\documentclass{aa}  

\usepackage{graphicx}
\usepackage{txfonts}
\usepackage{float}
\usepackage{dblfloatfix}
\usepackage{natbib}
\bibpunct{(}{)}{;}{a}{}{,} 
\usepackage{hyperref}
\hypersetup{colorlinks=true, urlcolor=blue, linkcolor=blue, citecolor=blue}
\usepackage{wrapfig}
\usepackage{comment}
\usepackage{svg}
\begin{document}

   \title{Linking planetary embryo formation to planetesimal formation I:\\ The impact of the planetesimal surface density in the terrestrial planet zone}
   \titlerunning{Linking planetary embryo formation to planetesimal formation}

   \author{Oliver Voelkel
          \inst{1}
          ,
          Rogerio Deienno
          \inst{2}
          ,
         Katherine Kretke
          \inst{2}
           ,
         Hubert Klahr
          \inst{1}
          }
    \authorrunning{O. Voelkel et al.}

   \institute{\inst{1} Max Planck Institute for Astronomy, Heidelberg, Königstuhl 17, 69117 Heidelberg, Germany\\
   \inst{2} Department of Space Studies, Southwest Research Institute,
Boulder, CO 80302\\
              \\
              \email{voelkel@mpia.de}
             }
  \abstract
   {
    The growth time scales of planetary embryos and their formation process are  imperative for our understanding on how planetary systems form and develop. They determine the subsequent growth mechanisms during the life stages of a circumstellar disk.
   }
   {
   We quantify the timescales and spatial distribution of planetary embryos via collisional growth and fragmentation of dynamically forming 100$\,$km sized planetesimals. In our study, the formation timescales of viscous disk evolution and planetesimal formation are linked to the formation of planetary embryos in the terrestrial planet zone. 
    }
   {
   We connect a one dimensional model for viscous gas evolution, dust and pebble dynamics and pebble flux regulated planetesimal formation to the N-body code LIPAD. Our framework enables us to study the formation, growth, fragmentation and evolution of planetesimals with an initial size of 100km in diameter for the first million years of a viscous disk.
   }
   {
   Our study shows the effect of the planetesimal surface density evolution on the preferential location and timescales of planetary embryo formation. A one dimensional analytically derived model for embryo formation based on the local planetesimal surface density evolution is presented. This model manages to reproduce the spatial distribution, formation rate and total number of planetary embryos at a fraction of the computational cost of the N-body simulations.
   }
   {
  The formation of planetary embryos in the terrestrial planet zone occurs simultaneously to the formation of planetesimals. The local planetesimal surface density evolution and the orbital spacing of planetary embryos in the oligarchic regime serve well as constraints to model planetary embryo formation analytically. Our embryo formation model will be a valuable asset in future studies regarding planet formation. 
   }

   \maketitle
%
%
\section{Introduction}
\subsection{Physical motivation}
\label{subsec:motivation}
The core accretion scenario is currently the most widely used theory for planet formation. It states that at first, planetary cores form in protoplanetary disks, which then continue to grow by various forms of accretion \citep{pollack1996formation}. It goes without saying that the formation of these planetary cores shapes the general picture of planet formation. To fully model the process of planet formation, one needs to track the different growth processes involved, beginning from dust coagulation, pebble and dust dynamics, the formation of planetesimals, the formation of planetary embryos and their subsequent growth until the circumstellar disk has vanished. A global model of planetesimal formation \citep{Lenz_2019}, that is regulated by the local pebble flux \citep{birnstiel2012simple} was introduced to a global model of planet formation \citep{EmsenhuberPrepA} in \cite{voelkel2020popsynth}. While this approach tracks the consistent formation and accretion of planetesimals on planetary embryos, the embryos themselves remain an ad hoc assumption. In this paper we investigate the formation of planetary embryos using N-body simulations \citep{levison2012lagrangian}, based on the evolution of the planetesimal surface density. Additionally, we construct an analytic, one dimensional, parameterized prescription of planetary embryo formation, that can be included into a global model of planet formation. In our companion paper, we add add the effect of pebble accretion on the formation of planetary embryos.
\\
Global models for planet formation that study planetary growth by solid accretion (\cite{mordasini2012extrasolar}, \cite{EmsenhuberPrepA}, \cite{bitsch2015growth}, \cite{ida2004toward} to mention just a few) generally begin with the initial presence of massive objects in the circumstellar disk, that are mostly referred to as embryos. Once an embryo has formed, it can grow by the accretion of solids and eventually the accretion of gas. While the accretion of gas on planets begins to be important at larger masses of around 10$\,$M$_{\oplus}$ \citep{pollack1996formation}, the presence of these 10$\,$M$_{\oplus}$ objects in the disk is the consequence of a previous phase of solid accretion on smaller embryos. For clarity, we define planetary embryos as objects of at least the mass of the earths moon ($M=0.0123\,$M$_{\oplus}$). These objects are massive enough to accrete planetesimals and pebbles from their surrounding orbits, but far from massive enough to effectively accrete gas.
\\
The growth of planetary embryos depends on the local disk environment, like e.g. the availability of planetesimals and pebbles. These quantities change over the course of the disks evolution and depend on the global evolution of the disk. Understanding where and when planetary embryos form, based on the circumstellar disks evolution is of vital relevance, as the evolution stage of the disk determines the subsequent growth of the embryos. 
 While the size range from lunar mass embryos to gas giant cores already spreads over roughly 4 orders of magnitude in mass, one has to keep in mind that these lunar mass embryos themselves are the product of long term planetesimal growth \citep{kokubo1998oligarchic,Kobayashi_2011,walsh2019planetesimals}. How, where and when these planetary embryos form out of much smaller planetesimals will be the main subject of this paper. 
\\
Despite all the uncertainties regarding the initial sizes that planetesimals were formed \citep{schlichting2013initial,schafer2017initial,walsh2017identification,morbidelli2009asteroids}; based in observational or theoretical arguments), here we will for simplicity assume planetesimals all formed with a diameter of 100 km (\cite{morbidelli2009asteroids}). Even though this planetesimal size is much larger than that inferred by other studies \citep{schlichting2013initial}, we find 5 orders of magnitude in mass between a lunar mass object and that of a 100$\,$km planetesimal.
Large planetesimals of 100 km are currently favored to explain the size distribution of asteroids and other minor bodies of the solar system \citep{morbidelli2009asteroids}. 100$\,$km also seems to be the most likely size in simulations of planetesimal formation \citep{Klahr2020, Johansen2009, abod2019mass} and as recent work suggests, this size is limited by diffusion \citep{Klahr2020}.  While 100$\,$km planetesimals from gravitational collapse are large in comparison to the small pebbles out of which they form, they are not massive enough to undergo pebble accretion. The formation of lunar mass objects from 100$\,$km planetesimals is therefore far from trivial and lays the foundation of subsequent planetary growth. 
\\
Forming massive planetary cores of 10$\,$M$_{\oplus}$ at larger distances to the star within the lifetime of a gaseous disk is currently challenging planetesimal accretion models \citep{johansen2019exploring}. A solution to this conundrum has appeared in the form of pebble accretion on distant planetary embryos \citep{KlahrBodenheimer2006,ormel2010effect,lambrechts2012rapid,bitsch2015growth}, \cite{Ndugu_2017}. This process describes the accretion of vastly smaller objects, that radially drift towards the star and is shown to be an effective planetary growth mechanism, even at larger distances up to the so called pebble isolation mass \citep{lambrechts2014separating}. 
\\
Similar to the case in which gas is accreted onto a 10$\,$M$_{\oplus}$ core, the accretion of pebbles also requires the presence of a massive body to effectively be accreted \citep{ormel2010effect}. The previously discussed planetesimal sizes of up to 100$\,$km are not believed to be large enough for significant pebble accretion. Assuming that pebble accretion can grow a lunar mass object over 4 orders of magnitude to the mass of a gas giant core therefore requires the ad hoc assumption of an initial planetary embryo at a given location. This approach is commonly used in planet formation studies that form gas giants from pebble accretion, but it lacks any description on the initial solid evolution of a circumstellar disk that would form the necessary embryo. While pebble accretion requires an active radial pebble flux, that is believed to decay faster than the presence of the gas disk due to radial drift \citep{birnstiel2012simple}, we face a similar conundrum as before. 
\\
Under which circumstances can a planetary embryo at a given radial distance form within the lifetime of a radial pebble flux? To answer this question, one needs a global study that models the formation of planetesimals from pebbles and track their following growth up to the size of lunar mass objects.
\subsection{Previous work}
\label{subsec:prev_work}
 Since lunar mass objects are not believed to form from the spontaneous collapse of a pebble cloud, there have been numerous studies that investigate the growth from planetesimals to planetary embryos in a circumstellar disk. 
Estimating the timescales of planet formation from a disk of planetesimals go back to \cite{safronov1969relative} and \cite{lissauer1987timescales}. Following up, it was shown that the growth of planetary embryos can be split in different growth phases, like runaway growth \citep{kokubo1996runaway} and eventually oligarchic \citep{kokubo1998oligarchic} once the embryo enhances the eccentricity of its surrounding planetesimals, effectively decreasing the accretion on the planet. Not only is the accretion of planetesimals suppressed in the runaway regime, but also arrange the embryos themselves around stable orbital separations when expressed in their mutual Hill radii (\cite{kokubo1998oligarchic}, \cite{walsh2019planetesimals}). While more has been done on the formation of planetary embryos, their growth timescales and orbital separation will be main subject to our study. 
\\
Planetary embryo formation depends on the spatial distribution of planetesimals within the circumstellar disk, as they are the building blocks of planetary embryos. Models for the viscous evolution of the gas suggest a shallow planetesimal surface density ($\Sigma_P$) profile of $\Sigma_P \propto r^{-0.9}$ \citep{shakura1973black}. The minimum mass solar nebula hypothesis suggests a steeper density profile of $\Sigma_P \propto r^{-1.5}$ (\cite{weidenschilling1977distribution}, \cite{hayashi}). However, if considering that planetesimal formation is proportional to the radial pebble flux, the surface density profile can be as steep as $\Sigma_P \propto r^{-2.1}$ \citep{Lenz_2019}. 
\\
The effect on planet formation of these different distributions under the assumption of initial embryo placement has recently been studied and suggests that the global planetesimal surface density distribution has major consequences for planet formation \citep{voelkel2020popsynth}. Therefore, studying the formation of planetary embryos based on the planetesimal surface density slope is the next logical step.
\subsection{The goal of this study}
Our goal is to determine the effect of the planetesimal surface density evolution on planetary embryo formation and derive an analytic recipe for planetary embryo formation. For that, we conduct N-body simulations and model the dynamical evolution, growth and fragmentation of planetesimals with an initial size of $d=100$km. Our study ranges from the initial gas and dust distribution, over pebble and planetesimal formation up to the finally formed planetary embryos within 0.5au and 5au of a protoplanetary disk around a solar type star. In order to make this possible, we have connected a one dimensional model for pebble flux regulated planetesimal formation \citep{Lenz_2019} with the N-body code LIPAD \citep{levison2012lagrangian}. This setup enables us to study the growth over multiple orders of magnitude in mass over $10^6$ years at a reasonable computational effort, allowing multiple simulations that cover a range of initial parameters. Based on analytic assumptions and numerical results, we present a one dimensional model for the formation of planetary embryos, as a function of the local planetesimal surface density evolution.
\\
In the following section, we will explain the physical models that we use in our study and our prescription on planetary embryo formation. (Sect. \ref{Sec:Pts_formation}). The connection between the one dimensional planetesmal formation model and LIPAD, as well as their explanation can be found in Sect. \ref{Sec:LIPAD_and_PLANETE}. Results and their discussion can be found in Sect. \ref{Sec:Numerical_Results} and Sect. \ref{Sec:Discussion}. Sect. \ref{Sec:Summary} contains a brief summary of our study and an outlook on how to proceed with the obtained results.
\section{Planetesimal $\&$ embryo formation}
\label{Sec:Pts_formation}
Our goal is to consistently model the growth timescales of planetary embryos from an initial disk of gas and dust. While this endeavour ranges over multiple orders of magnitude in mass, we have chosen to split it in two components. First we form planetesimals of 100$\,$km in diameter, using a one dimensional parameterized description while considering pebble flux regulated planetesimal formation. Following up on that we model the growth and fragmentation of the planetesimals in N-body simulations. Since both processes take place at the same time, it is necessary to connect our one dimensional parameterized model with the N-body simulation, as it is described in Sec. \ref{Sec:LIPAD_and_PLANETE}. Here we focus on the description of the one dimensional planetesimal formation and disk evolution model, as well as the equations of the following planetesimal growth.
\subsection{Disk evolution and planetesimal formation}
We have chosen to use a one dimensional viscous disk with an $\alpha$ prescription for turbulence \citep{shakura1973black}, in which we have added a two population model for solids \citep{birnstiel2012simple}. Based on the radial drift of the solids we form planetesimals with a parameterized efficiency. An exact description of the two population model can be found in \cite{birnstiel2012simple}. Here we will outline the basic equations and assumptions. The model uses a fixed mass relation between a smaller and a larger population of dust grains. These two populations are distinguished on whether particle growth is limited by radial drift or fragmentation respectively. Each time step solves one advection diffusion equation of the combined solid density
\begin{equation}
    \frac{\partial \Sigma_s}{\partial t} + 
    \frac{1}{r} \frac{\partial}{\partial r} 
    \left[ 
    r \left(
    \Sigma_s \bar{u} - D_{g} \Sigma_g \frac{\partial}{\partial r}
    \left(
    \frac{\Sigma_s}{\Sigma_g}
    \right)
    \right)
    \right]
    = 0
    \label{eq:advection_diffusion_equation}
\end{equation}
with $\Sigma_s$ the solid density, $\Sigma_g$ the gas density, $D_g$ the diffusion coefficient and $\bar{u}$ the weighted velocity of the two populations. The weighted velocity is given as 
\begin{equation}
    \bar{u} = (1-f_m(r)) \cdot u_0 + f_m(r) \cdot u_1
    \label{eq:weighted_velocity}
\end{equation}
where $f_m(r)$ is given as the aforementioned mass relation that separates the two populations with their corresponding velocities $u_0$ and $u_1$. The individual populations are then given as:
\begin{align}
 \Sigma_0 (r) &= \Sigma_s (r) \cdot (1-f_m(r)) 
    \label{twopop_dust_density}
    \\
    \Sigma_1 (r) &= \Sigma_s(r) \cdot f_m(r)
    \label{eq:twopop_pebble_density}
\end{align}
The mass relation $f_m$ was derived by fitting the two population model to more sophisticated simulations of dust coagulation by \cite{Birnstiel_2010}. The values that showed the best results are given as 
\begin{align}
    f_m = \left\{\begin{array}{ll} 
        0.97, & \text{drift limted case} \\
        0.75, & \text{fragmentation limited case}
        \end{array}\right. 
        \label{eq:f_m_parameter}
\end{align}
The decision on whether a particle is within $\Sigma_0$  or $\Sigma_1$ is done by its Stokes number \citep{birnstiel2012simple}. $\Sigma_0$ contains particles with a small Stokes number ($St \ll 1$). The motion of these particles is coupled to the motion of the gas. $\Sigma_1$ contains larger particles with $St \ge 1$, which are no longer coupled to the gas. In the following we will refer to $\Sigma_0$ as dust and $\Sigma_1$ as pebbles.
Planetesimals are formed based on the radial drift of the solid material in our disk. A detailed description of the planetesimal formation model can be found in \cite{Lenz_2019}. For our purpose we assume planetesimals to form with an initial size of 100$\,$km in diameter. This choice is supported by numerical simulations of planetesimal formation by \cite{Klahr2020} and observations of asteroid and kuiper belt objects (\cite{morbidelli2009asteroids}, \cite{schafer2017initial}, \cite{walsh2017identification}). 
\\
The formation of planetesimals as described in \cite{Lenz_2019} occurs in trapping zones in which disk instabilities can trigger planetesimal formation. These zones are distributed within the whole disk. 
The formation rate of planetesimals is then given proportional to the radial pebble flux and can be written as
\begin{align}
    \dot{\Sigma}_{\text{p}}(r) = \frac{\epsilon}{d(r)} \frac{\dot{M}_{\text{peb}}}{2 \pi r}
    \label{eq:planetesimal_formation}
\end{align}
with $\epsilon$ the formation efficiency, $d(r)$ the radial separation of pebble traps and $r$ the radial distance to the star. We have chosen $d(r)$ to be 5 gas pressure scale heights and $\epsilon = 0.05$ in our simulations. $\dot{M}_{\text{peb}}$ is the radial pebble flux, which in our model is defined as 
\begin{align}
  \dot{M}_{\text{peb}}   := 2 \pi r \sum_{
  \text{St}_\text{min} \leq \text{St} \leq \text{St}_\text{max} 
  } 
  {
  | v_\text{drift} (r,\text{St}) | \Sigma_\text{s} (r,\text{St} )
  }
  \label{pebble_flux}
\end{align}
The pebble flux regulated model for planetesimal formation results in a steeper radial planetesimal surface density profile ( $\Sigma_p \propto r^{-2.1}$, \cite{Lenz_2019}) as suggested by the minimum mass solar nebula hypothesis ($\Sigma_P \propto r^{-1.5}$) or the gas surface density profile of a viscously evolving disk ($\Sigma_P \propto r^{-0.9}$). Due to the fact we do not specify the physical process that form planetesimals (e.g., streaming instability, Kelvin Helmholts, etc.), this one dimensional planetesimal formation description can be considered model independent. The formation of planetesimals is regulated by the local pebble flux. The latter is regulated by dust coagulation and disk evolution \citep{birnstiel2012simple}. This approach enables us to connect the timescales of the dynamical pebble evolution of the disk with the timescales of planetesimal formation.
In our study, we chose to focus on three planetesimal surface density profiles, while applying the formation rate from the pebble flux regulated model, as it connects the viscous timescales of the disk with the formation of planetesimals.
\subsection{Planetesimal growth and embryo formation}
\label{Sec:Planetesimal_growth}
In the following we will describe the one dimensional analytical model that determines where and when lunar mass planetary embryos will be formed (based on the local planetesimal surface density evolution). The model connects analytic growth rates with the orbital seperation of planetary embryos in the oligarchic regime. The mass of the largest object at an orbital distance $r$ to the star at a time $t$ is given as $M_{p}(r,t)$.
Once planetesimals have fomred at a time $t_0$ at a distance $r$, we introduce
\begin{align}
    M_{p}(r,t_{0}) = M_{100\,km}
\end{align} We set the initial mass to that of a 100$\,$km in diameter planetesimal with a solid density of $\rho_s = 1.0\,$g/cm$^3$. During the evolution of the planetesimal disk, we integrate the mass growth rate of $M_{p}$ within a swarm of planetesimals in every timestep. The local mass growth rate is then given as \citep{lissauer1993planet}
\begin{align}
    \frac{d M_{p}(r,t)}{d t} 
    = 
    \frac{\sqrt{3}}{2} \Sigma_P(r,t) \cdot \Omega(r) \pi r^2 
    \left(
    1 + \frac{v^2_{\text{esc}}(M_p,r)}{v^2_{\infty}(r,t)}
    \right)
    \label{eq:mass_growth}
\end{align}
with $\Omega$ as the orbital keppler frequency, $v_{\text{esc}}$ the escape velocity of an object with mass $M_p$ and $v_{\infty}$ the mean dispersion velocity within the swarm of planetesimals. We choose $v_{\infty}(r) = e(r,t) \cdot v_{k}(r)$ with $e(r,t)$  as the local mean planetesimal eccentricity in our analytical model computation. $v_{k}(r)$ is the kepplerian velocity at an orbital distance $r$.
Eq. \ref{eq:mass_growth} is integrated in every timestep with the updated values for $\Sigma_P$, $v_{\text{esc}}$ and $v_{\infty}$, hence new planetesimals form over time. Once $M_p$ has reached the minimum mass of a planetary embryo $M_{emb}$ (which in our study is given as a lunar mass) at a distance $r$, we determine this to be the location at which a lunar mass planetary embryo can be formed. We do not track the subsequent evolution of the embryo. Our approach is solely designed to estimate the local timescales involved to form an embryo mass object within an evolving planetesimal disk. The eccentricity for the analytical model computation is given as $e(r,t) = 5 \cdot 10^{-4}( 1 + r^{0.8})$, which results in a good fit to the numerical simulations. It is known that the size of planetesimals has a significant effect on the accretion rate \citep{fortier2007oligarchic}. The planetesimal size appears in $v_{\infty}$, $v_{\text{esc}}$ and $M_{p}(r,t=0)$. Our model runs in Sect. \ref{Sec:Numerical_Results} considers all planetesimals, including $M_{p}(r,t=0)$ to be 100 km in diameter. Eq. \ref{eq:mass_growth} however is still valid for different planetesimal sizes, by adapting $v_{\infty}$, $v_{esc}$ and $M_{p}(r,t=0)$.
\\
Our one dimensional embryo formation model can be described by two criteria. The first criterion refers to the necessary growth time at a distance $r$ as a function of the planetesimal surface density evolution. The second criterion concerns the orbital seperation to already present embryos. Criterion I for the embryo formation model can be written as:
\begin{align}
    M_p(r,t) \ge M_{\text{emb}}
    \label{eq:criterion_I}
\end{align}
The second criterion for the formation of a planetary embryo at $r_i$ is the orbital separation to other planetary embryos at $r_j$. As suggested by numerical studies by \cite{kokubo1998oligarchic,Kobayashi_2011,walsh2019planetesimals} we find an orbital separation of planetary embryos in the oligarchic growth regime of $\Delta r_{\text{orbit}} \sim $10-20$R_{\text{Hill}}$. We choose a randomized Gaussian distribution for the orbital separation around 17 $R_{\text{Hill}}$ with a standard deviation of $\sigma_{\Delta r} = 2.5 R_{\text{Hill}}$ in our analytic model runs. The mass for the computation of the Hill Radius is always given as the mass of the embryos that have already been placed. Criterion II is then given as: 
\begin{align}
    \Delta r_{\text{orbit} i,j } \ge \Delta r_{\text{min}}
    \label{eq:criterion_II}
\end{align}{}
where $\Delta r_{\text{orbit} i,j }$ is the orbital distance of an embryo at $r_i$ to an embryo at $r_j$. $\Delta r_{\text{min}}$ is chosen from the Gaussian. The embryos that are formed with the one dimensional analytic model are compared to the results of the N-body simulations in Sect. \ref{subsec:toy_model_comparison}.
\section{LIPAD and the growth of planetesimals}
\label{Sec:LIPAD_and_PLANETE}
LIPAD (Lagrangian Integrator for Planetary Accretion and Dynamics; \cite{levison2012lagrangian}), is a particle-based (i.e., Lagrangian) code. LIPAD was developed to follow the collisional, accretional, and dynamical evolution of a large number of meter- to kilometer-sized objects through the entire growth process to become planets, making it ideal for our study. A detailed description, as well as an extensive suite of tests, of LIPAD can be found in \cite{levison2012lagrangian}. In addition, LIPAD has been succesfuly employed in previous studies of planet formation, as well as collisional evolution of meter- to kilometer-sized planetesimals interacting with planet/protoplanets \citep{kretke2014challenges,levison2015growing,walsh2016terrestrial,walsh2019planetesimals,deienno2019energy,deienno2020collisional}.

LIPAD uses the concept of tracer particles to represent a large number of small bodies with roughly the same orbit and size. Tracers are characterized by three numbers: the physical radius, the bulk density, and the constant total mass of the disk particles that it will represent. 

Collisional routines are employed to determine when collisions between tracers will happen. In this event, following a probabilistic outcome based on a fragmentation law by \cite{benz1999catastrophic}, tracers can be assigned new physical radii. Therefore, a distribution of tracers in LIPAD will represent the size distribution of the evolving planetesimal population. The interaction among tracers is resultant from statistical algorithms for viscous stirring, dynamical friction, and collisional damping. 

Large enough tracers can be promoted to become planetary embryos. Planetary embryos interact among themselves, as well as with Tracers, via normal N-body routines \citep{duncan1998multiple}.

LIPAD also has a prescription of the gaseous nebula from \cite{hayashi}. This gas disk provides aerodynamic drag, eccentricity, and inclination damping on every object. 
\subsection{Planetesimal formation in LIPAD}
\label{Subsec:Panetesimal_formation_in_LIPAD}
\begin{figure}
    \centering
    \includegraphics[width=1.0\linewidth]{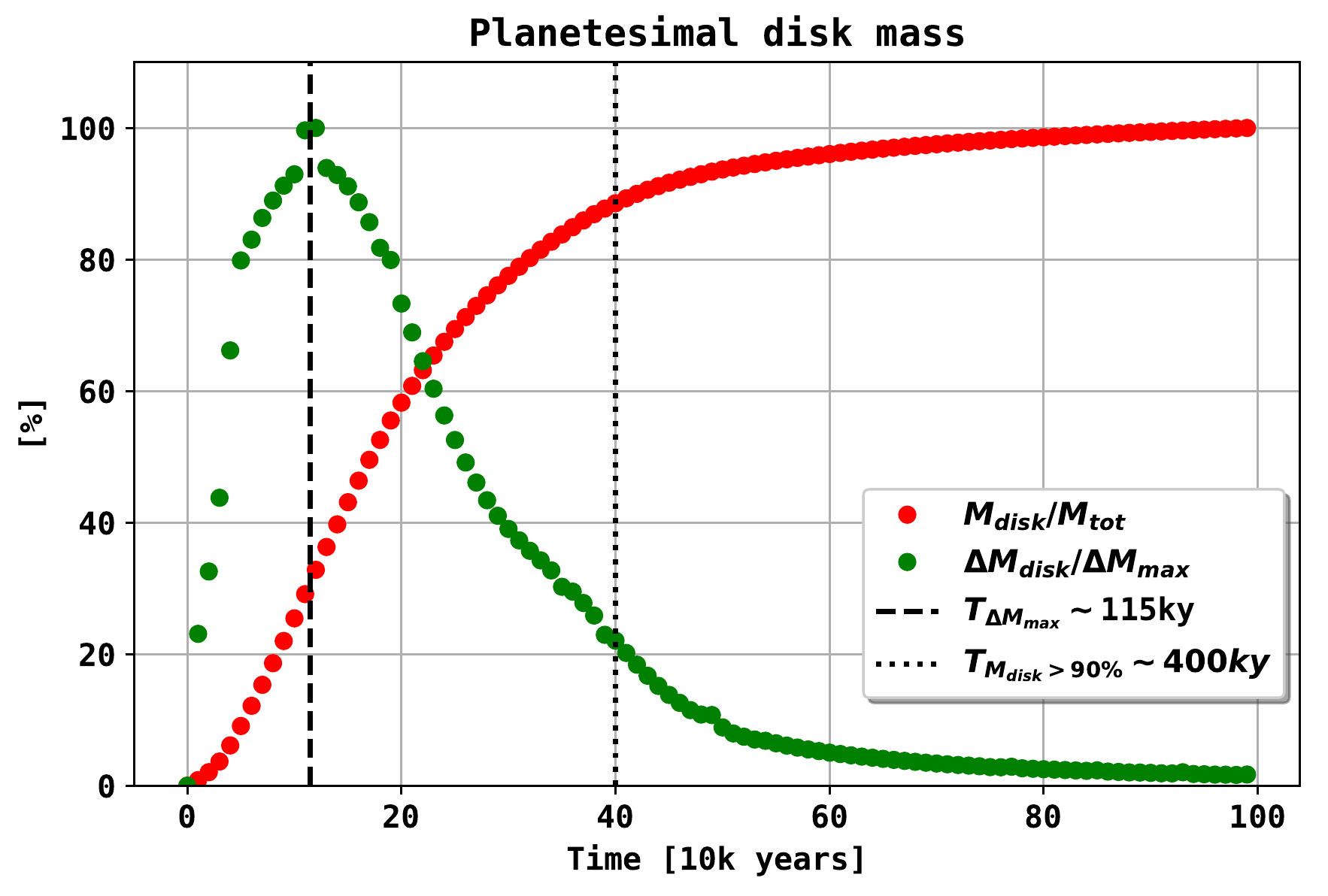}
    \caption{Qualitative change of the planetesimal disk mass $M_{disk}$ (red dots), normalized by the total disk mass after $10^6$ years that we use in the analytic setups. The green dots indicate the disk mass increase every $10^4$ years $\Delta M_{disk}$, normalized by the maximum mass change $\Delta M_{max}$. }
    \label{fig:formation_rate}
\end{figure}

We aim to investigate different total masses of planeteimals and surface density profiles, while taking their formation timescales into account. For that purpose we apply the formation rate from our one dimensional model and scale it to the total masses after $10^6$ years between 0.5$\,$au and 5$\,$au. The formation of planetesimals as described in Sect. \ref{Sec:Pts_formation} scales linearly with the planetesimal formation efficiency $\epsilon$, which is why we choose the same qualitative formation rate for our various setups. The normalized disk mass change can be seen in Fig. \ref{fig:formation_rate}. 
Our planetesimal formation model uses a surface density distribution to describe planetesimals in the disk, whereas LIPAD uses tracer particles. For that matter we transform our surface density into a discrete number of tracer particles. We initially define a total number of tracer particles $N_{Tracer}\mathcal{O}(\approx 10^4)$ to be generated in the simulation within $10^6$ years. To get the mass of the individual tracers $M_{Tracer}$ we use the final mass that is in planetesimals $M_{Pts}$ after $10^6$ years:
\begin{equation}
    M_{Tracer} = \frac{M_{Pts}}{N_{Tracer}}
\end{equation}
The domain of the one dimensional surface density is split into individual rings of mass. Every $10^4$ years we add new planetesimal tracers to the LIPAD simulation according to the formation of the planetesimal surface density $\Delta M_{disk}$. Each of those newly formed tracers is assigned a heliocentric distance that is chosen randomly between the inner and outer edge of the ring in which it formed. Doing this in every time step and every ring, we ensure that the overall heliocentric distribution of planetesimal tracers in the LIPAD simulations will match the density slope and planetesimal distribution of the one dimensional model. LIPAD then continues with the newly included planetesimal tracers additional to the previously included objects that by then have grown and fragmented until the next group of tracers is included. Using this setup we connect the timescales of pebble growth and drift, the formation of planetesimals and their simultaneous growth. The qualitative mass change of the individual setups can be seen in Fig. \ref{fig:formation_rate}. 
The peak of the planetesimal formation rate occurs at $T_{\Delta M_{max}}\sim$ 115ky. Since the formation of planetesimals requires the presence of a radial pebble flux, we can file conclusions from the planetesimal formation rate on the remaining pebbles. About 90 $\%$ of planetesimals have formed within 400$\,$ky of our setup
\section{Numerical results}
\label{Sec:Numerical_Results}
In the following we will present the results of nine different setups in which we vary the total mass within  0.5$\,$au to 5$\,$au and the surface density slope with which planetesimals enter the simulation. The total masses after $10^6$ years are 6$\,M_{\oplus}$, 13$\,M_{\oplus}$ and 27$\, M_{\oplus}$ and for each we vary the density slope with $\Sigma_p \propto r^{-1.0}$, $\Sigma_p \propto r^{-1.5}$ and $\Sigma \propto r^{-2.0}$ respectively. The planeteismal formation rate for these analytic setups is shown in Fig. \ref{fig:formation_rate}. We focus on the mass and semimajor axis evolution of planetary embryos in LIPAD (Fig. \ref{Fig:Emb_form_LIPAD_6_ME} - Fig. \ref{Fig:Emb_form_LIPAD_27_ME}, Sect. \ref{subsec:mass_semimajor_axis_evol}). The embryo mass occurences are shown in Fig. \ref{Fig:Histogramms}, Sect. \ref{subsec:mass_occurrences}. The LIPAD results are compared to the analytic model for embryo placement in Fig. \ref{Fig:Time_Semi}, Sect. \ref{subsec:toy_model_comparison}. Following up on this we display the cumulative number of embryos formed (Fig. \ref{Fig:Cumulative_number}, Sect. \ref{Sec:Cumulative_number}), their orbital separation (Fig. \ref{Fig:Orbital_Seperation}, Sect. \ref{Sec:orbital_seperation}) and the active embryo number as well as the total mass in embryos (Fig. \ref{Fig:Active_number}, Sect. \ref{Sec:active_number}).
\subsection{Mass and semimajor axis evolution}
\label{subsec:mass_semimajor_axis_evol}
\begin{figure*}[]
\label{Subsection:Comparison}
\centering
\includegraphics[width=1.0\linewidth]{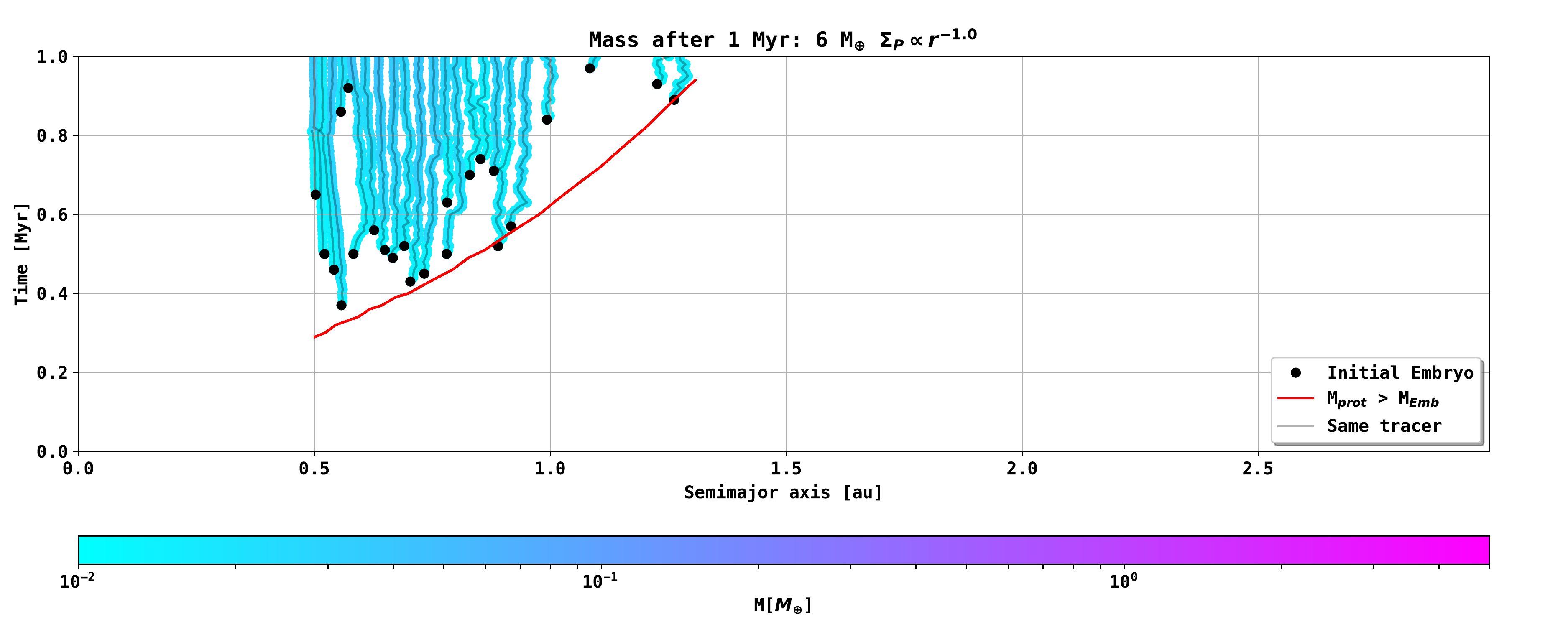} \\
\includegraphics[width=1.0\linewidth]{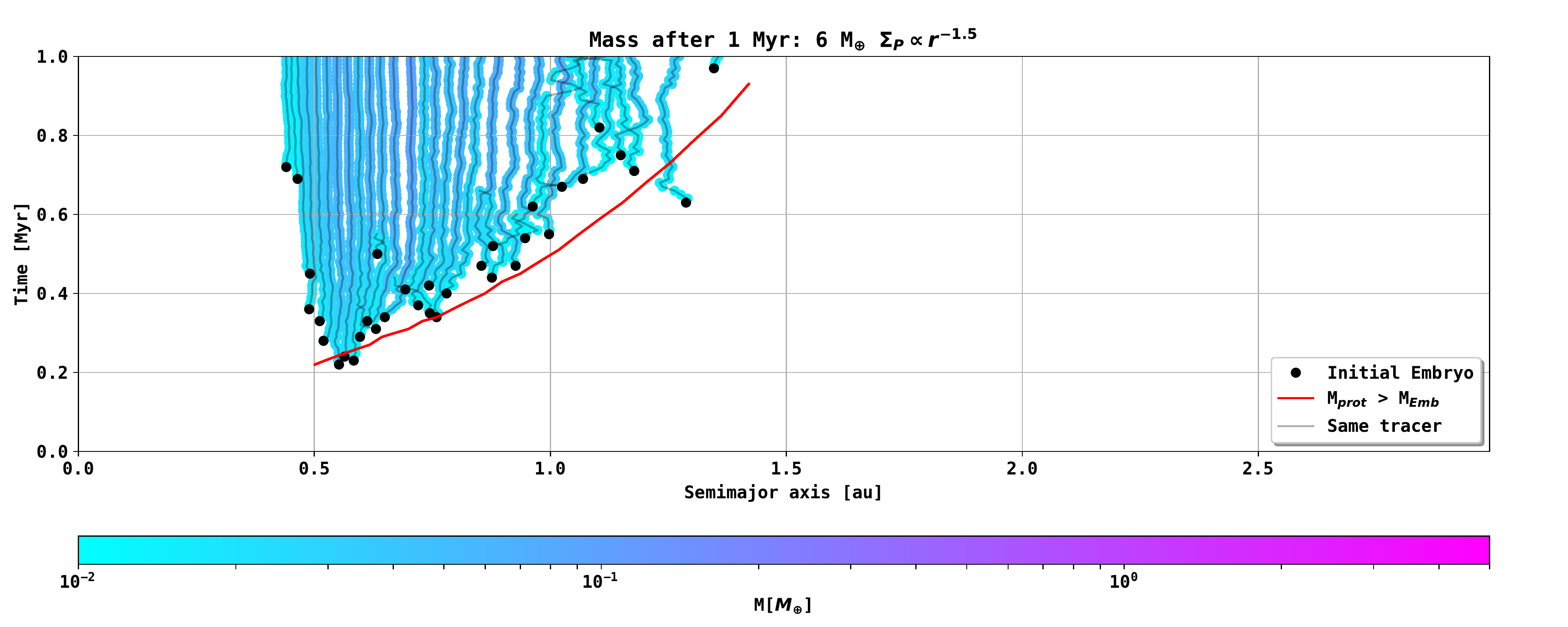} \\
\includegraphics[width=1.0\linewidth]{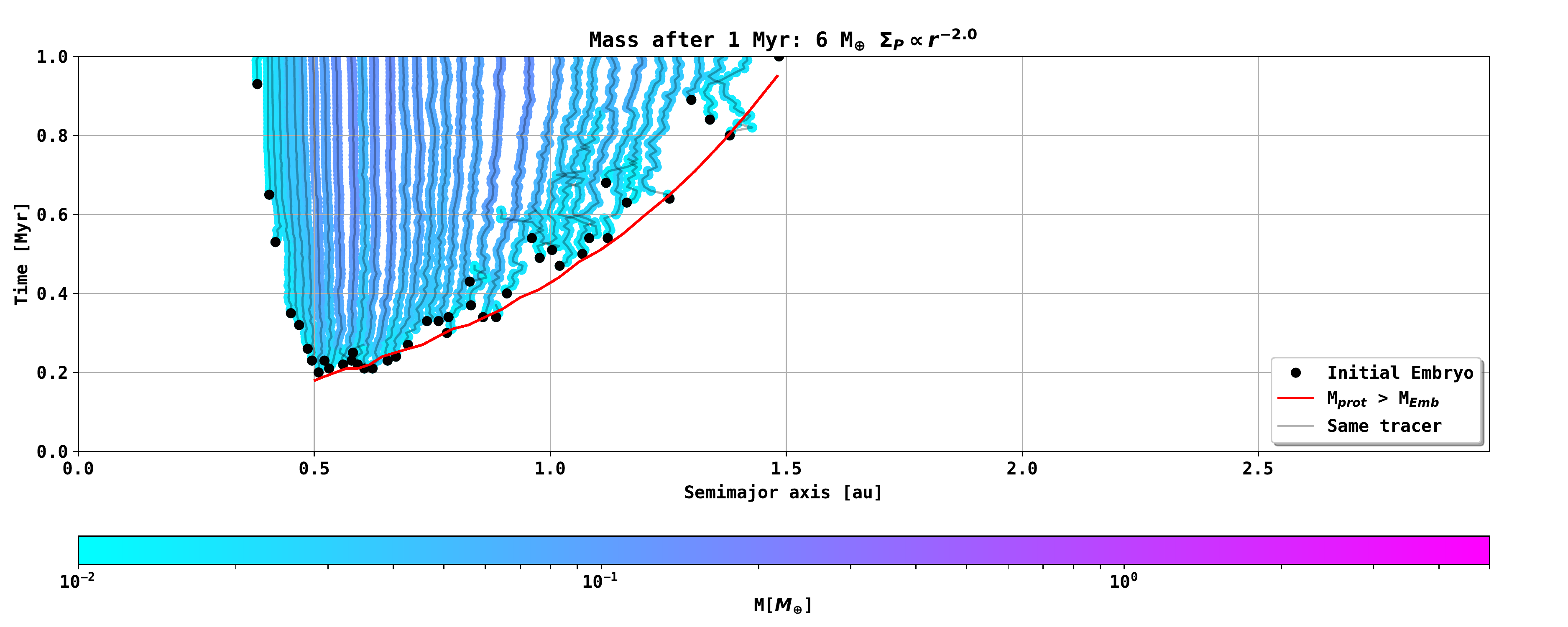} \\
\caption{\small  Time over Semimajor axis evolution of the N-body simulation in LIPAD. The Time and location at which an object has first reached lunar mass is indicated by the black dots in the plot. The subsequent growth of the embryo is tracked and connected with the grey lines (its mass is given by the colorbar). The mass after 1 million years in planetesimals is 6 M$_{\oplus}$ in these runs. The planetesimal surface density slope is varied ($\Sigma_P \propto r^{-1.0}$, $\Sigma_P \propto r^{-1.5}$ , $\Sigma_P \propto r^{-2.0}$ ). The red line indicates where M$_{prot}$ surpasses the mass of a lunar mass planetary embryo in the analytical model from Sect. \ref{Sec:Planetesimal_growth}, assuming the same evolution of the planetesimal surface density that is given to the N-body simulation.
}
\label{Fig:Emb_form_LIPAD_6_ME}
\end{figure*}

\begin{figure*}[]
\label{Subsection:Comparison}
\centering
\includegraphics[width=1.0\linewidth]{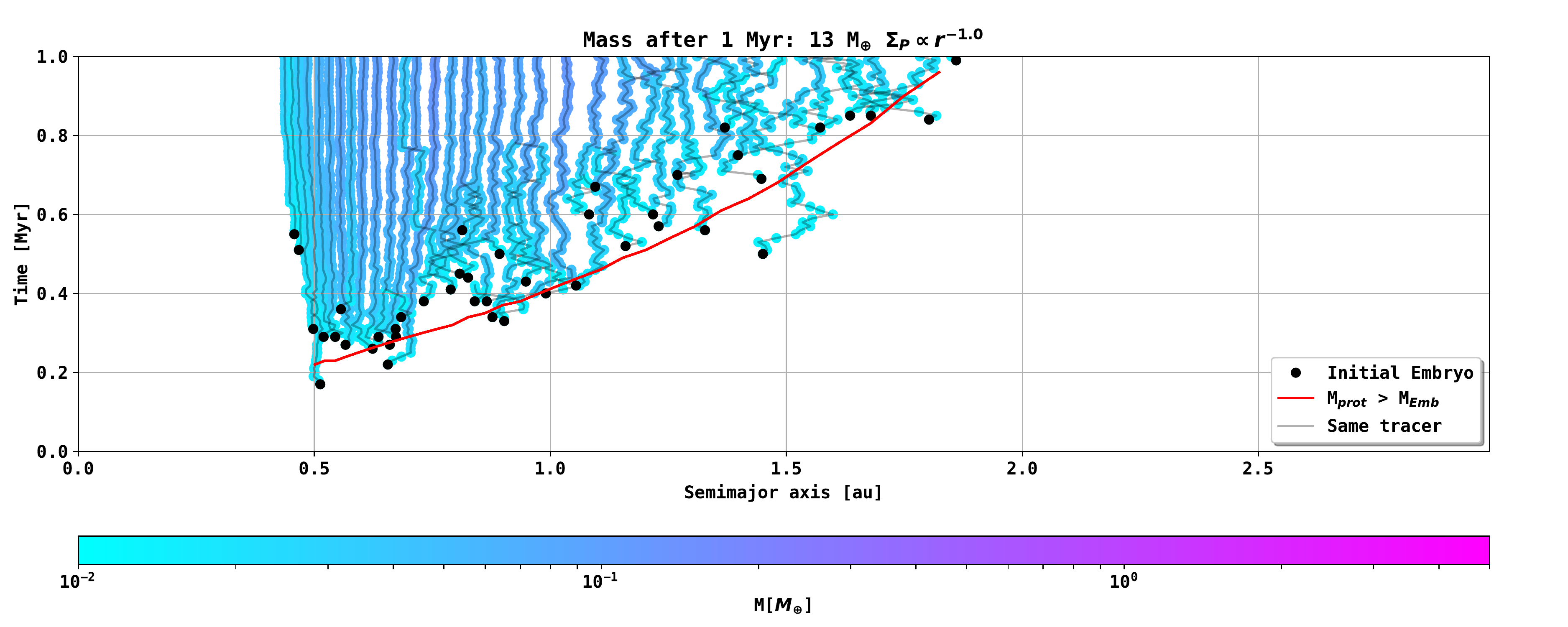} \\
\includegraphics[width=1.0\linewidth]{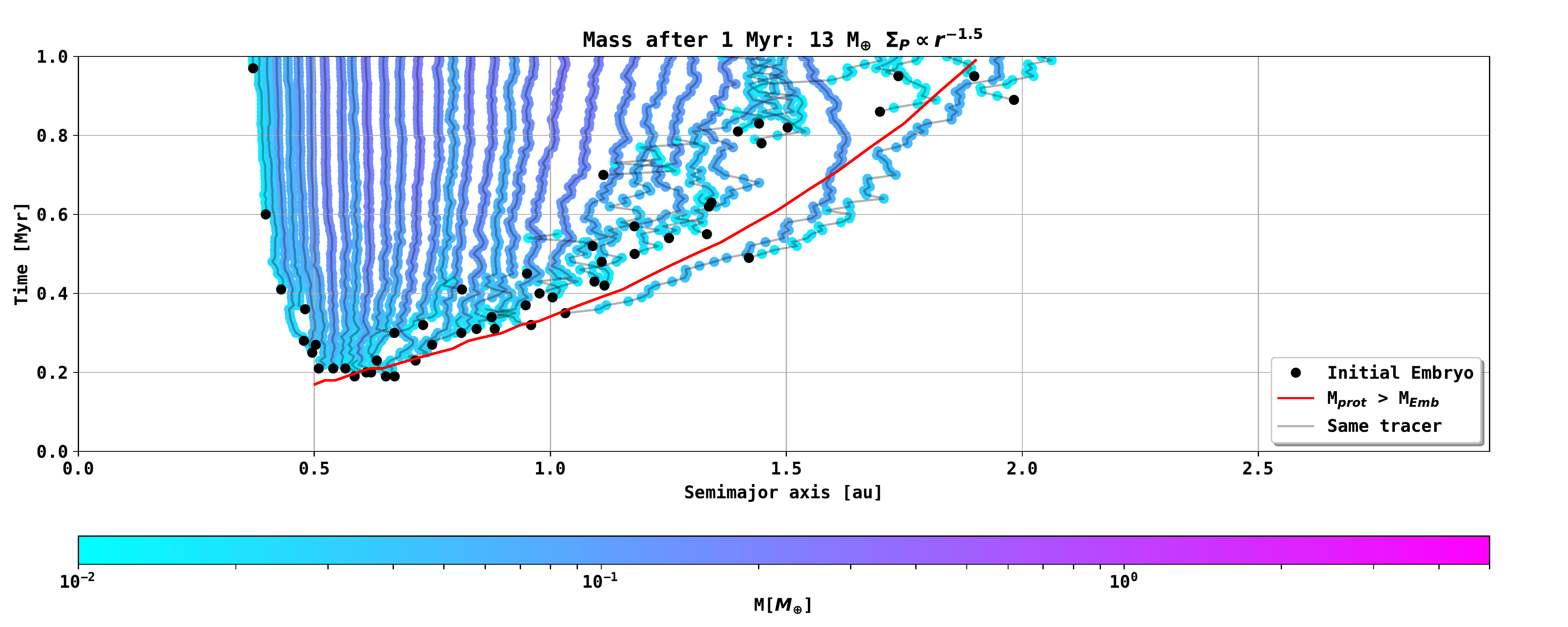} \\
\includegraphics[width=1.0\linewidth]{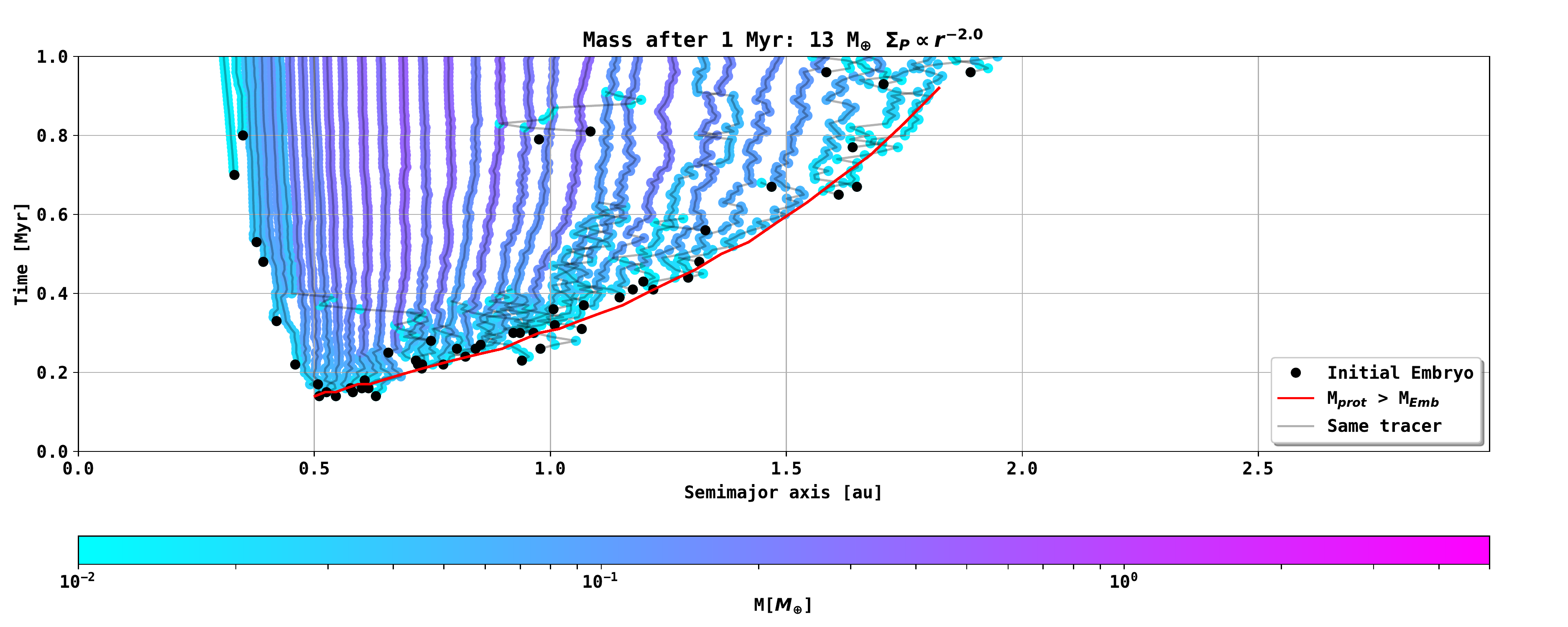} \\
\caption{\small  Time over Semimajor axis evolution of the N-body simulation in LIPAD. The Time and location at which an object has first reached lunar mass is indicated by the black dots in the plot. The subsequent growth of the embryo is tracked and connected with the grey lines (its mass is given by the colorbar). The mass after 1 million years in planetesimals is 13 M$_{\oplus}$ in these runs. The planetesimal surface density slope is varied ($\Sigma_P \propto r^{-1.0}$, $\Sigma_P \propto r^{-1.5}$ , $\Sigma_P \propto r^{-2.0}$ ). The red line indicates where M$_{prot}$ surpasses the mass of a lunar mass planetary embryo in the analytical model from Sect. \ref{Sec:Planetesimal_growth}, assuming the same evolution of the planetesimal surface density that is given to the N-body simulation.
}
\label{Fig:Emb_form_LIPAD_13_ME}
\end{figure*}

\begin{figure*}[]
\label{Subsection:Comparison}
\centering
\includegraphics[width=1.0\linewidth]{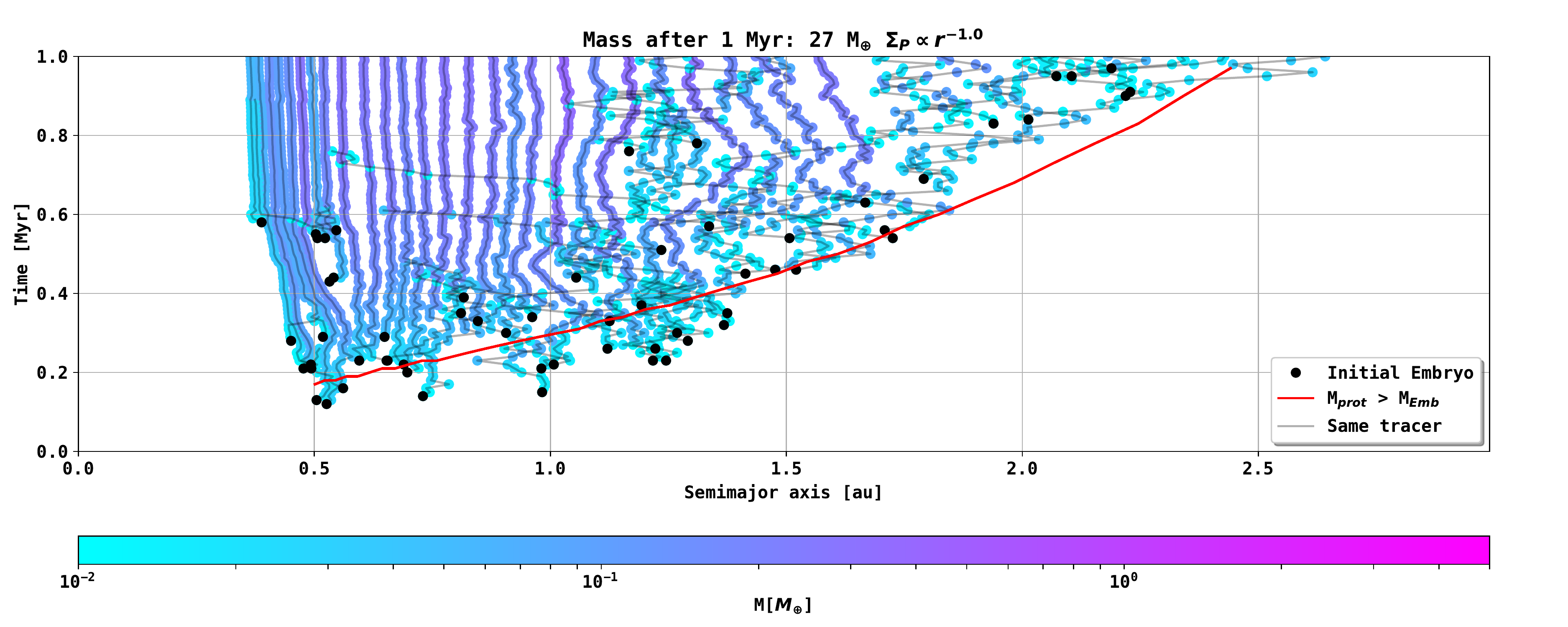} \\
\includegraphics[width=1.0\linewidth]{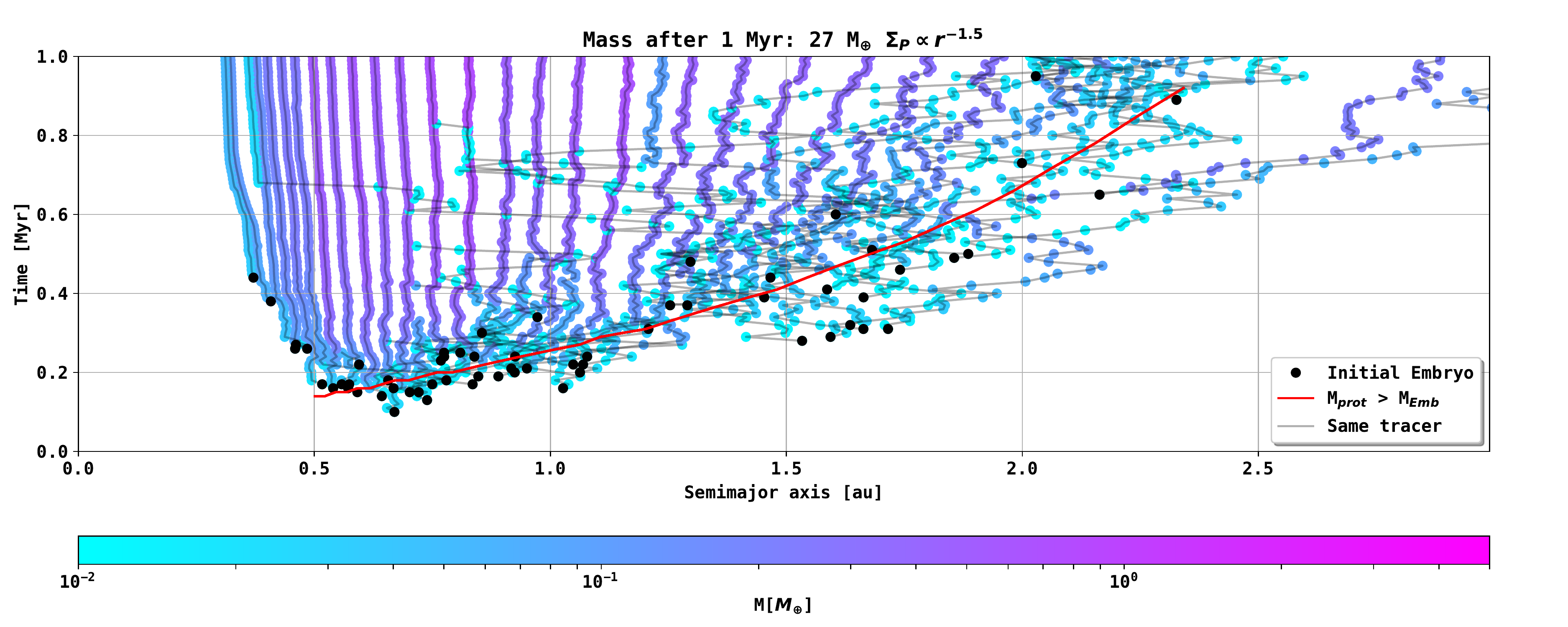} \\
\includegraphics[width=1.0\linewidth]{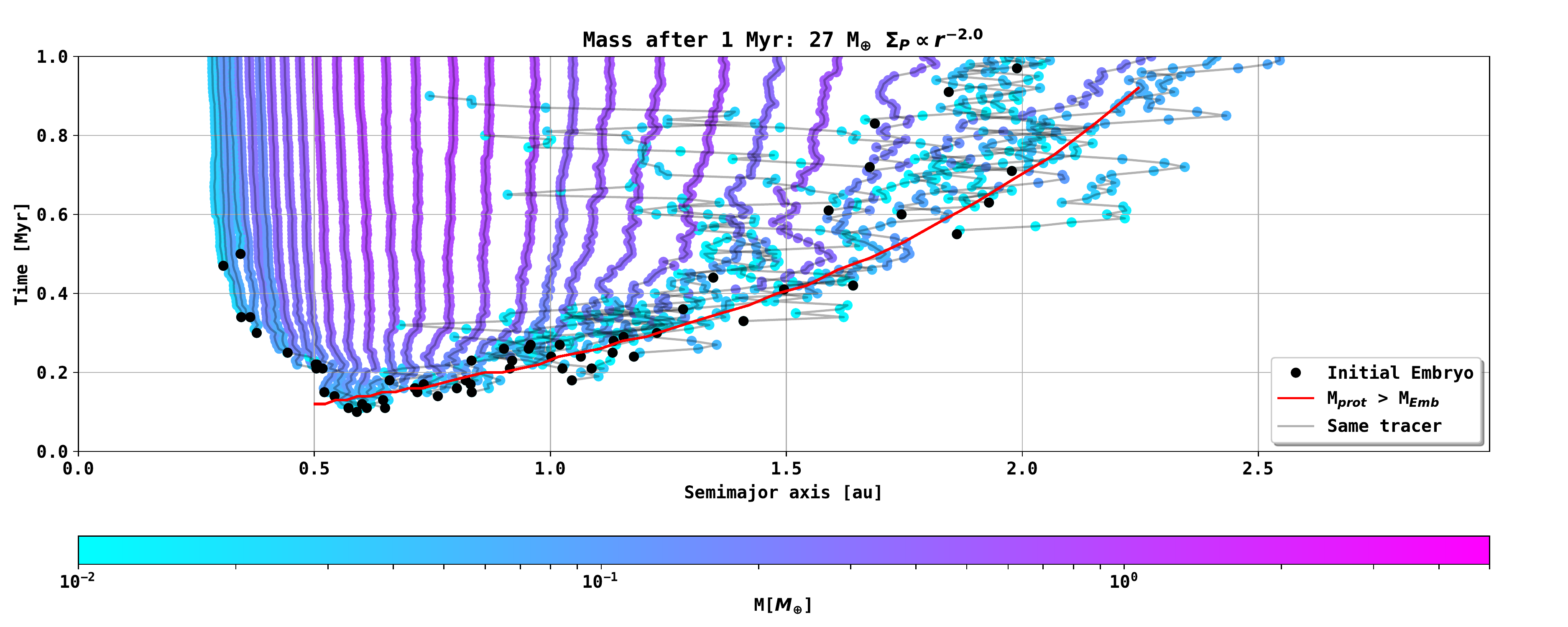} \\
\caption{\small  Time over Semimajor axis evolution of the N-body simulation in LIPAD. The Time and location at which an object has first reached lunar mass is indicated by the black dots in the plot. The subsequent growth of the embryo is tracked and connected with the grey lines (its mass is given by the colorbar). The mass after 1 million years in planetesimals is 27 M$_{\oplus}$ in these runs. The planetesimal surface density slope is varied ($\Sigma_P \propto r^{-1.0}$, $\Sigma_P \propto r^{-1.5}$ , $\Sigma_P \propto r^{-2.0}$ ). The red line indicates where M$_{prot}$ surpasses the mass of a lunar mass planetary embryo in the analytical model from Sect. \ref{Sec:Planetesimal_growth}, assuming the same evolution of the planetesimal surface density that is given to the N-body simulation.
}
\label{Fig:Emb_form_LIPAD_27_ME}
\end{figure*}

Fig. \ref{Fig:Emb_form_LIPAD_6_ME} - Fig. \ref{Fig:Emb_form_LIPAD_27_ME} show the evolution of the N-body system within 1 Myrs. We show the time and semimajor axis evolution of objects that were classified as planetary embryos in the LIPAD simulation. The classification of an embryo occurs after a tracer particle represents a single object of lunar mass. The tracer is then promoted to a planetary embryo and is treated as a single N-body object with an initial lunar mass. The subsequent growth of a given embryo is represented by the color bar and its semimajor axis evolution by a grey line. The occurrence of when a tracer is promoted to an embryo is shown as black dots. 
\\
Since embryos can collide and eventually merge during their evolution we make a distinction between active embryos and initial embryos. The number of initial embryos are the events in which tracers have been promoted to planetary embryos (number of black dots) and the number of active embryos is the number of embryos at a given time $t$. The red line in the plots refers to the analytic model. It indicates where $M_{P}$ has surpassed a lunar mass (Criterion I), when assuming the same analytic planetesimal surface density evolution as in the N-body simulation. The red line is shown only for reference, comparing the N-body simulation with the analytical result. Even though all planetesimals that enter the LIPAD simulation have an initial semimajor axis of above 0.5$\,$au, we find embryo formation within 0.5$\,$au as well. This is due to dynamical interactions/scattering of the LIPAD tracer particles that lead to a nonzero planetetsimal distribution wihtin the edge of its original formation. Since this effect is not taken into account in the analytical model density distribution, we cannot see a change of the red line within 0.5$\,$au. This effect also has to be considered when comparing the cumulative number of initial embryos (see Fig. \ref{Fig:Cumulative_number}). Finally our results show that the more massive disks (Fig. \ref{Fig:Emb_form_LIPAD_27_ME}) form embryos earlier at close distance. Additionally, embryos in massive disks can form at larger heliocentric distances, than in their less massive counterparts (Fig. \ref{Fig:Emb_form_LIPAD_6_ME}).
\subsection{Embryo mass occurrences}
\label{subsec:mass_occurrences}
\begin{figure*}[]
\centering
\begin{minipage}{.33\textwidth}
  \centering
  \includegraphics[width=1.0\linewidth]{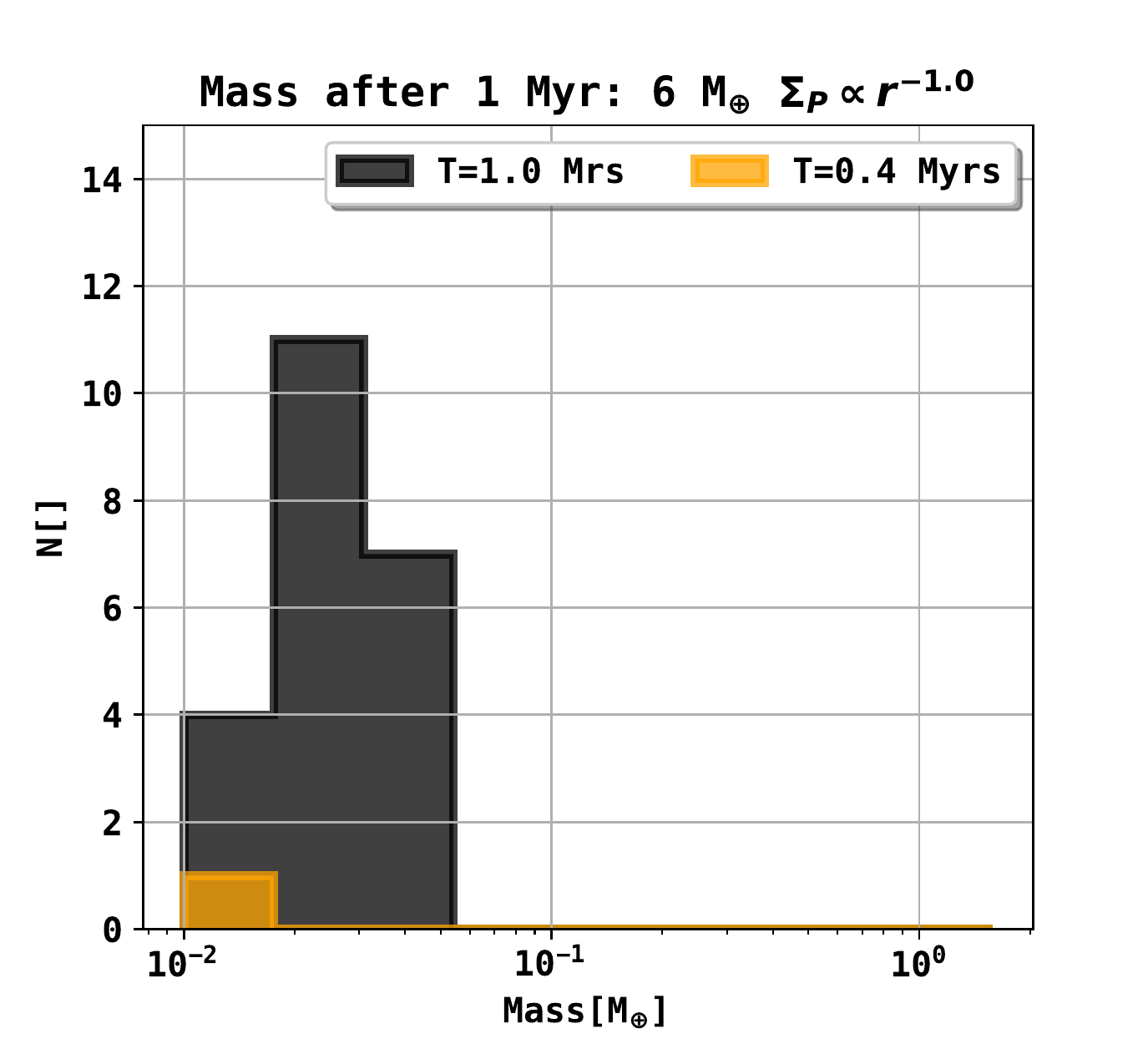}
\end{minipage}
\begin{minipage}{.33\textwidth}
  \includegraphics[width=1.0\linewidth]{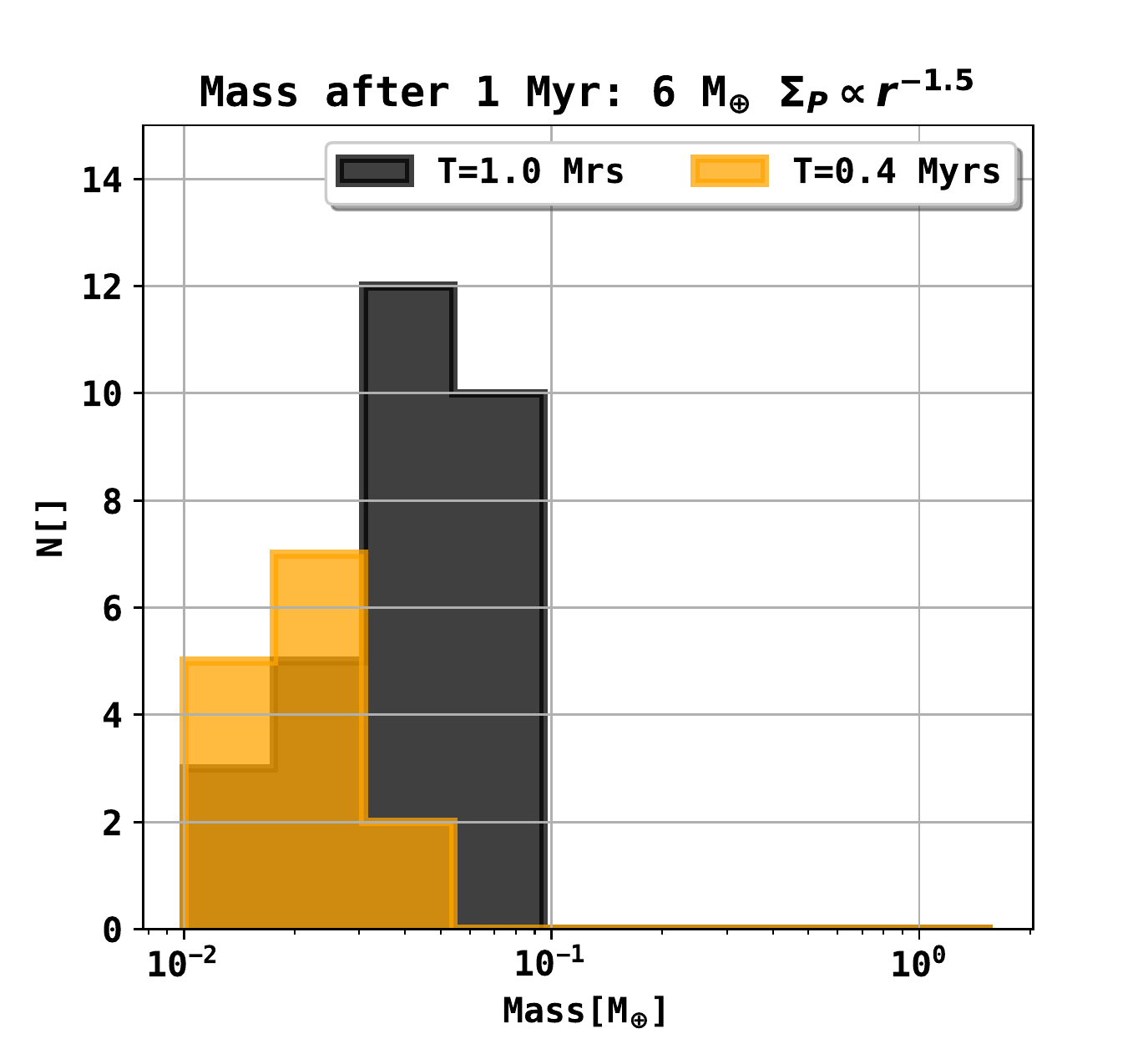}
\end{minipage}%
\begin{minipage}{.33\textwidth}
  \includegraphics[width=1.0\linewidth]{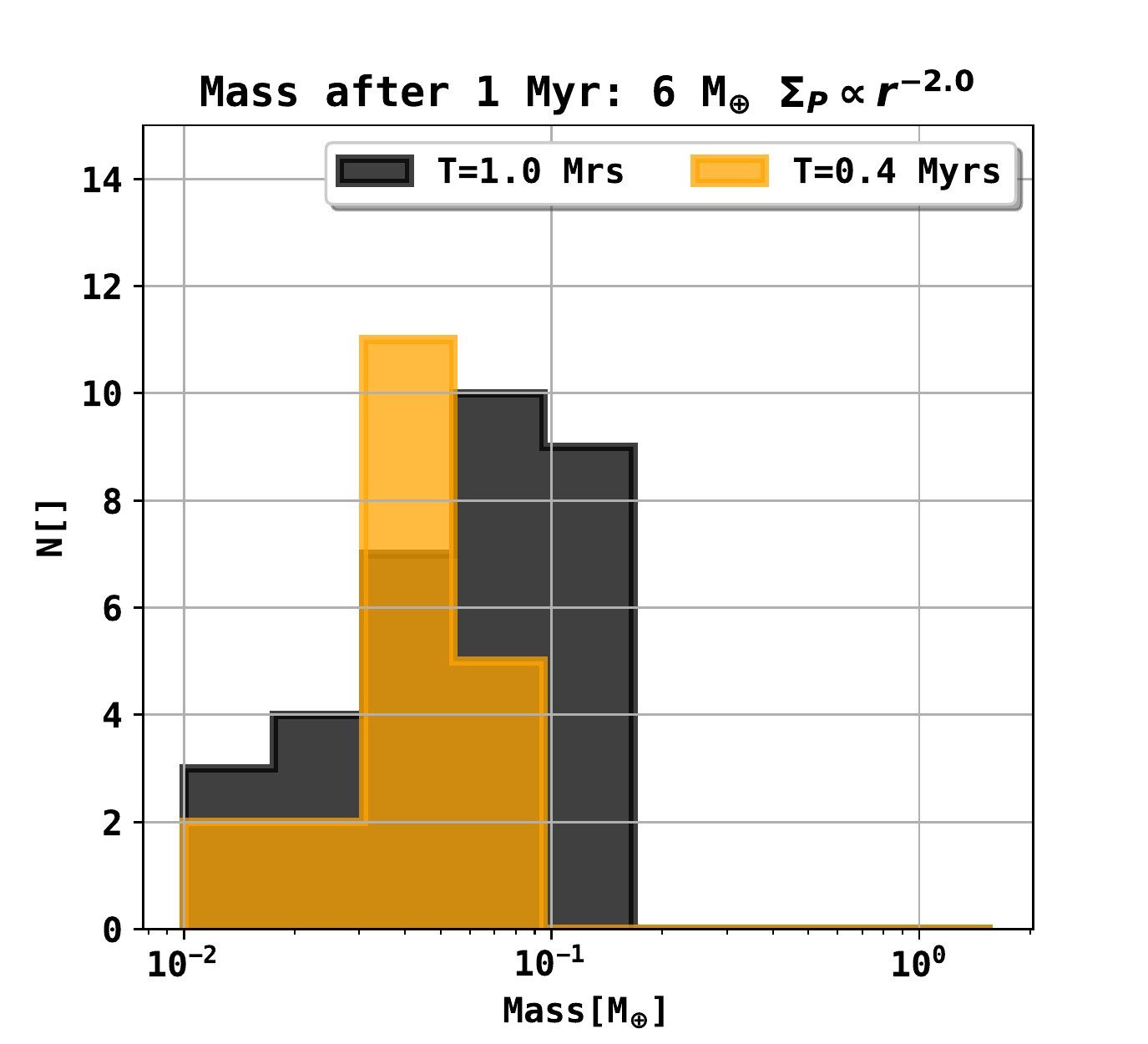}
\end{minipage}%
\\
\begin{minipage}{.33\textwidth}
  \centering
  \includegraphics[width=1.0\linewidth]{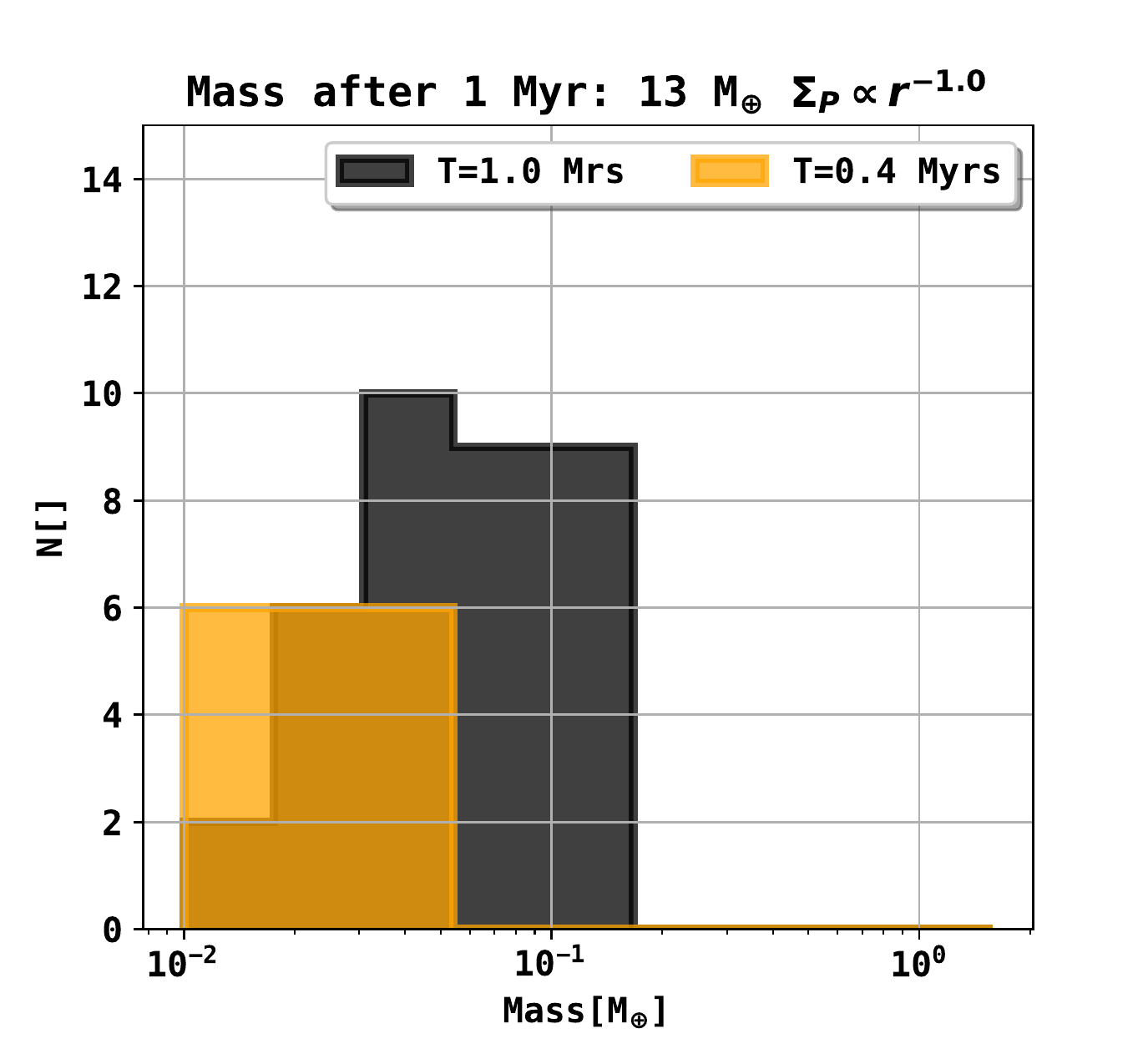}
\end{minipage}
\begin{minipage}{.33\textwidth}
  \includegraphics[width=1.0\linewidth]{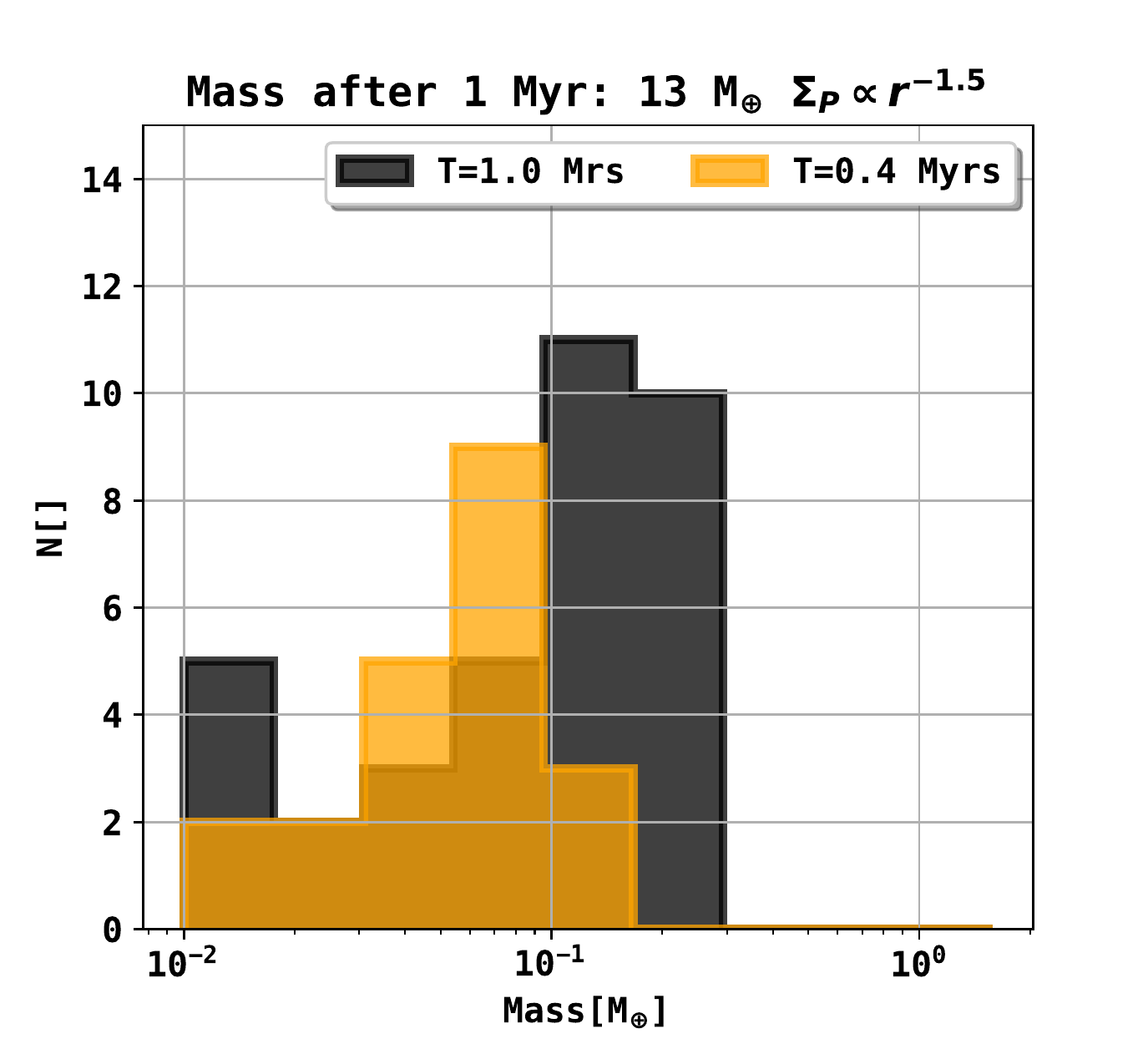}
\end{minipage}%
\begin{minipage}{.33\textwidth}
  \includegraphics[width=1.0\linewidth]{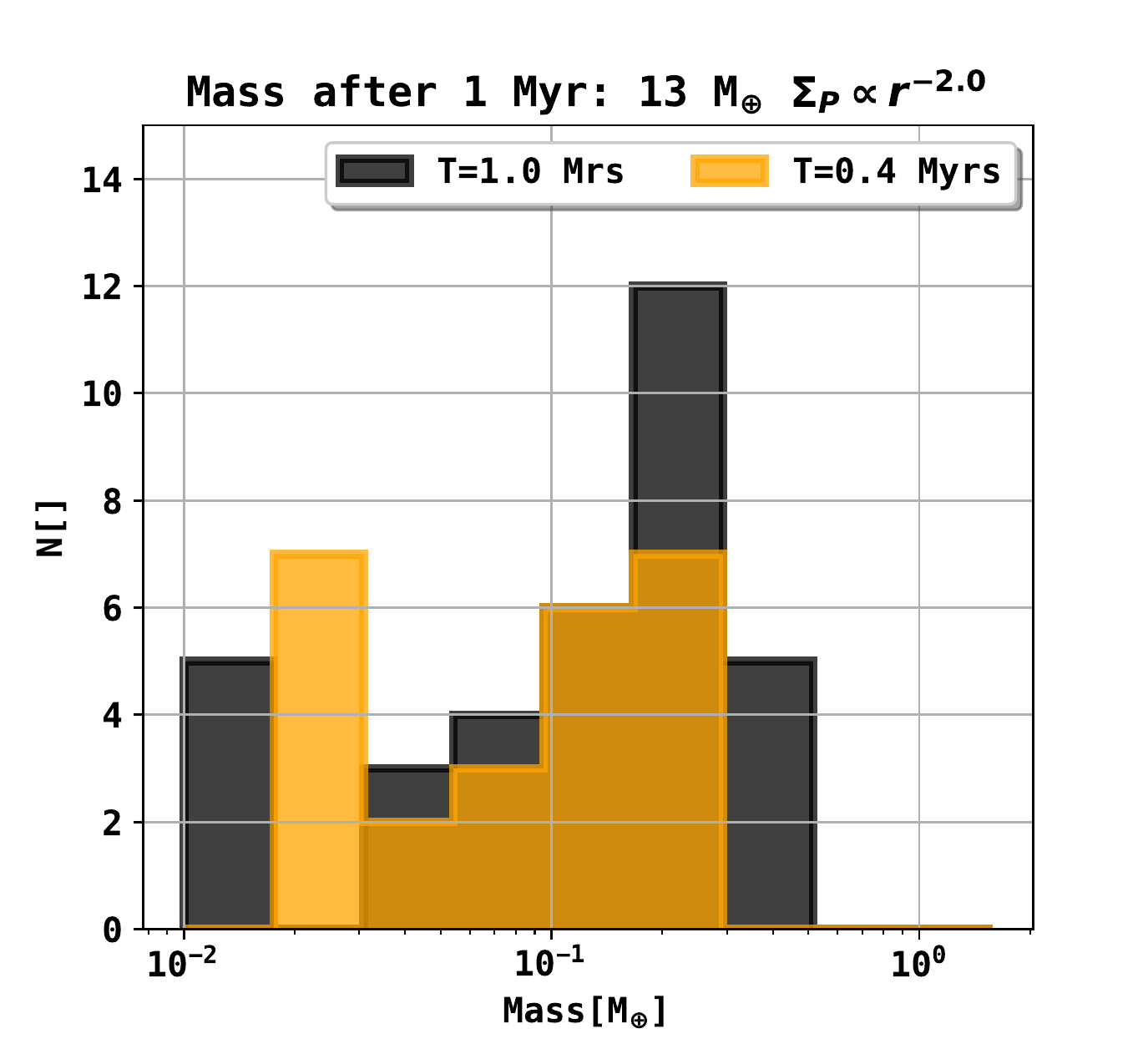}
\end{minipage}%
\\
\begin{minipage}{.33\textwidth}
  \centering
  \includegraphics[width=1.0\linewidth]{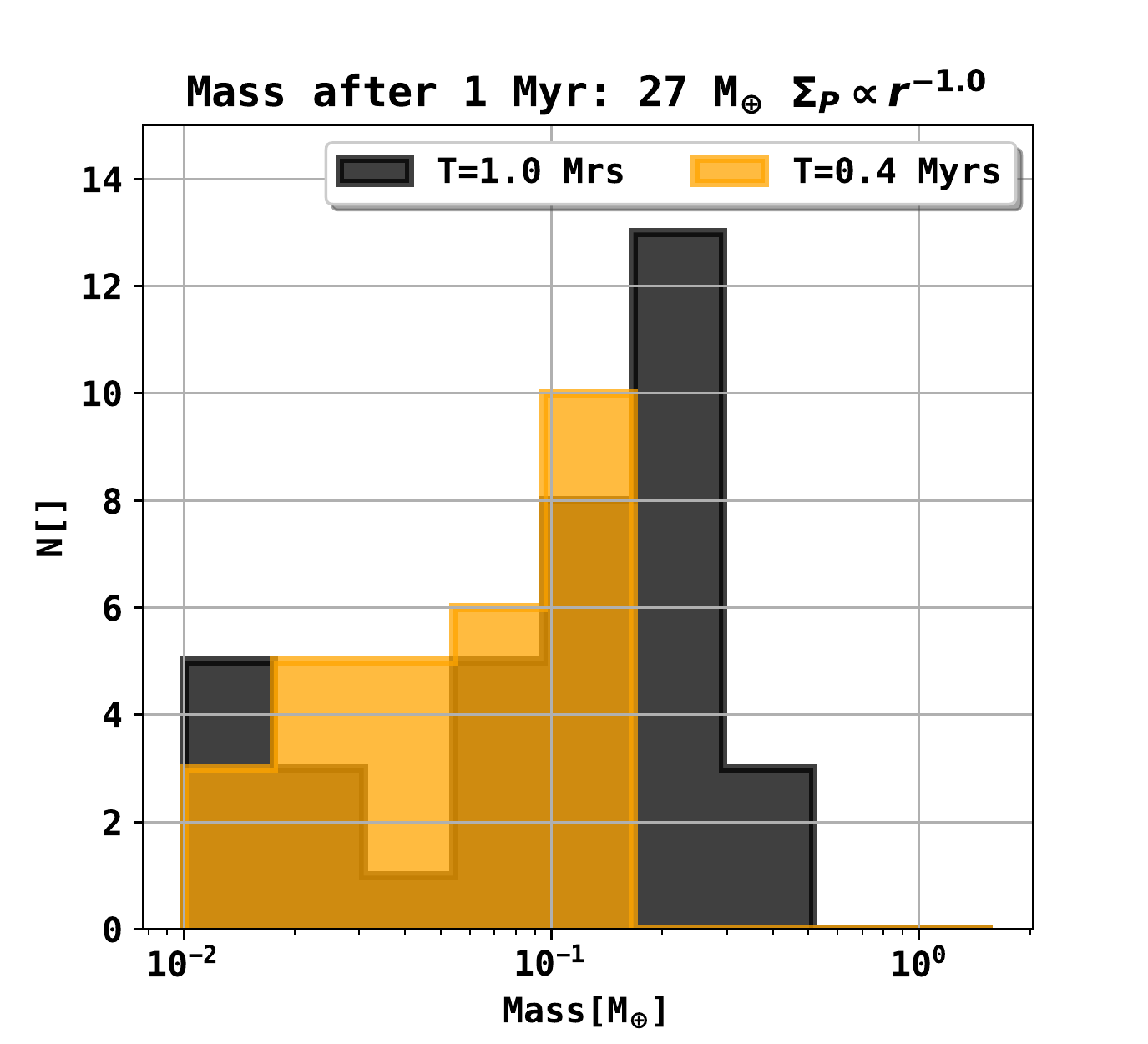}
\end{minipage}
\begin{minipage}{.33\textwidth}
  \includegraphics[width=1.0\linewidth]{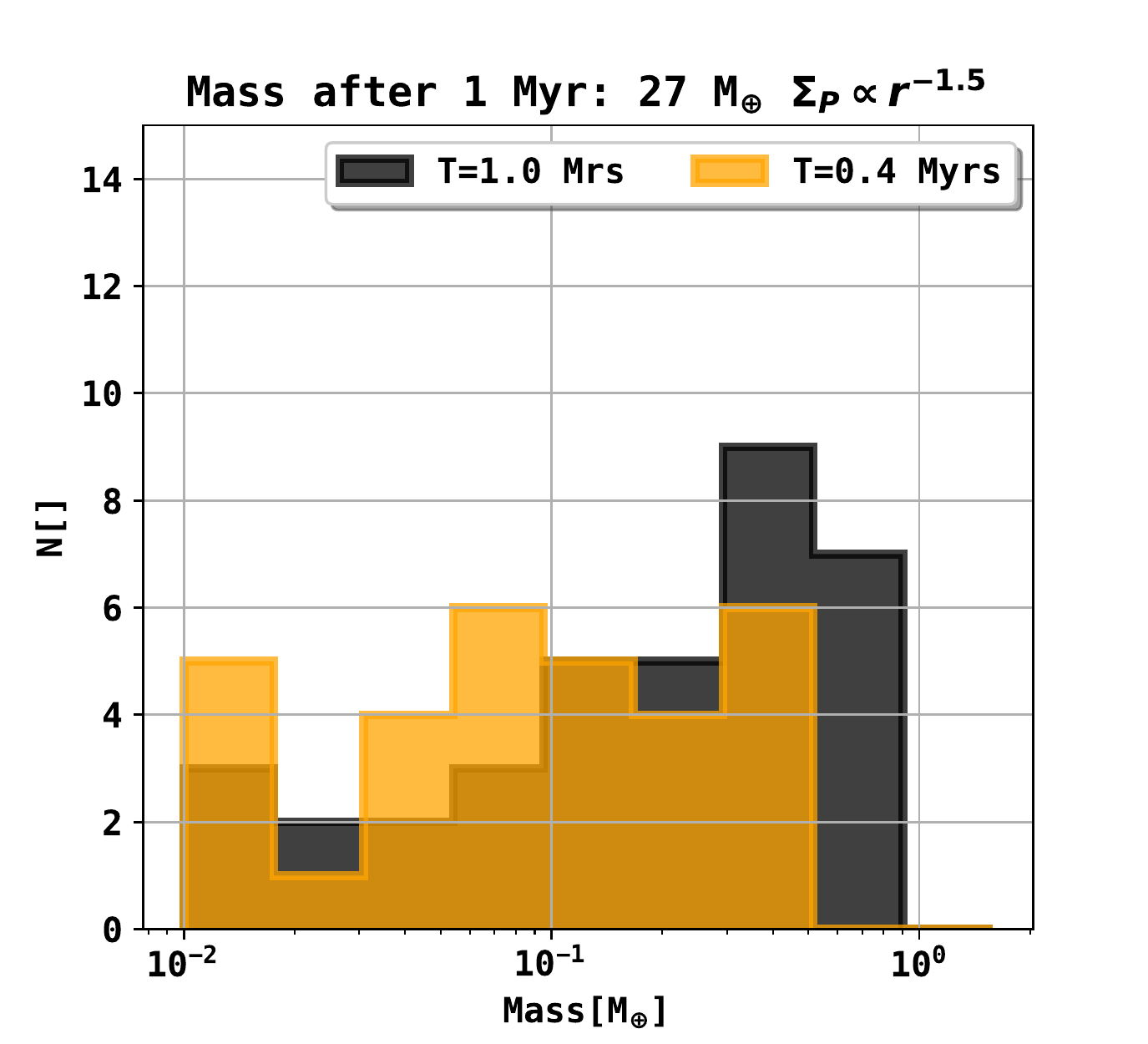}
\end{minipage}%
\begin{minipage}{.33\textwidth}
  \includegraphics[width=1.0\linewidth]{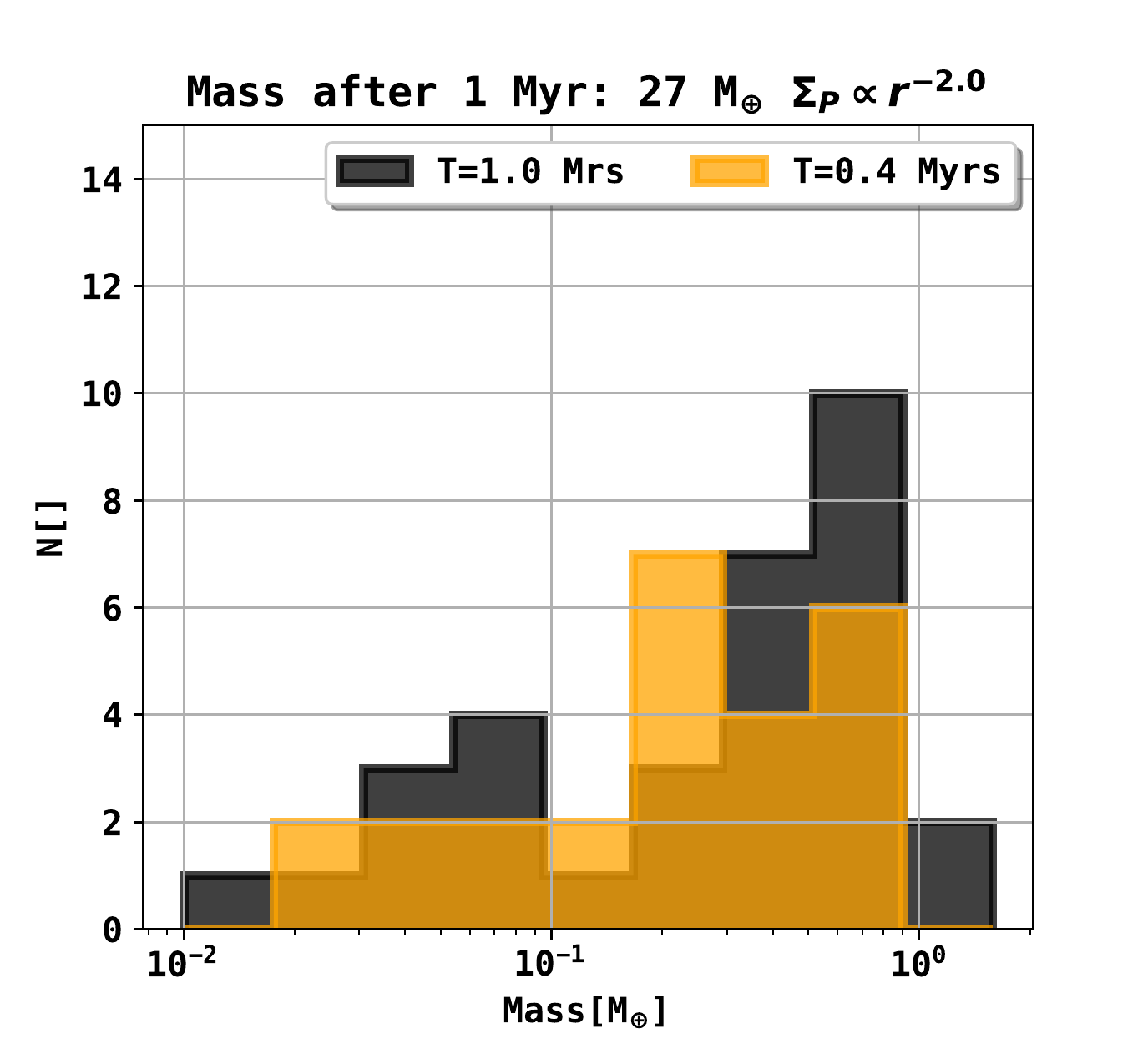}
\end{minipage}%
\caption{\small  Number of embryos within a given mass bin at T = 400$\,$ky ($T_{M_{disk}> 90\%}$) and T = 1$\,$Myrs from the simulations of Fig. \ref{Fig:Emb_form_LIPAD_6_ME} - Fig. \ref{Fig:Emb_form_LIPAD_27_ME}.
}
\label{Fig:Histogramms}
\end{figure*}

Fig. \ref{Fig:Histogramms} shows the number of embryo masses at $T_{M_{disk}>90\%}$(400$\,$ky) and at 1 Myr from the simulations of Fig. \ref{Fig:Emb_form_LIPAD_6_ME} - Fig. \ref{Fig:Emb_form_LIPAD_27_ME}. Most embryos in each simulation are found in the higher mass end of their simulation. Embryos that have low masses ($\approx 0.0123\,$M$_{\oplus}$) are less abundant than embryos that share the highest possible masses in the system. There is no single embryo growing substantially larger than the others in the system, in agreement with standard oligarchic growth models \citep{kokubo1998oligarchic}. 
\subsection{Comparison with the analytical model}
\label{subsec:toy_model_comparison}
\begin{figure*}[]
\label{Subsection:Comparison}
\centering
\begin{minipage}{.33\textwidth}
  \centering
  \includegraphics[width=1.0\linewidth]{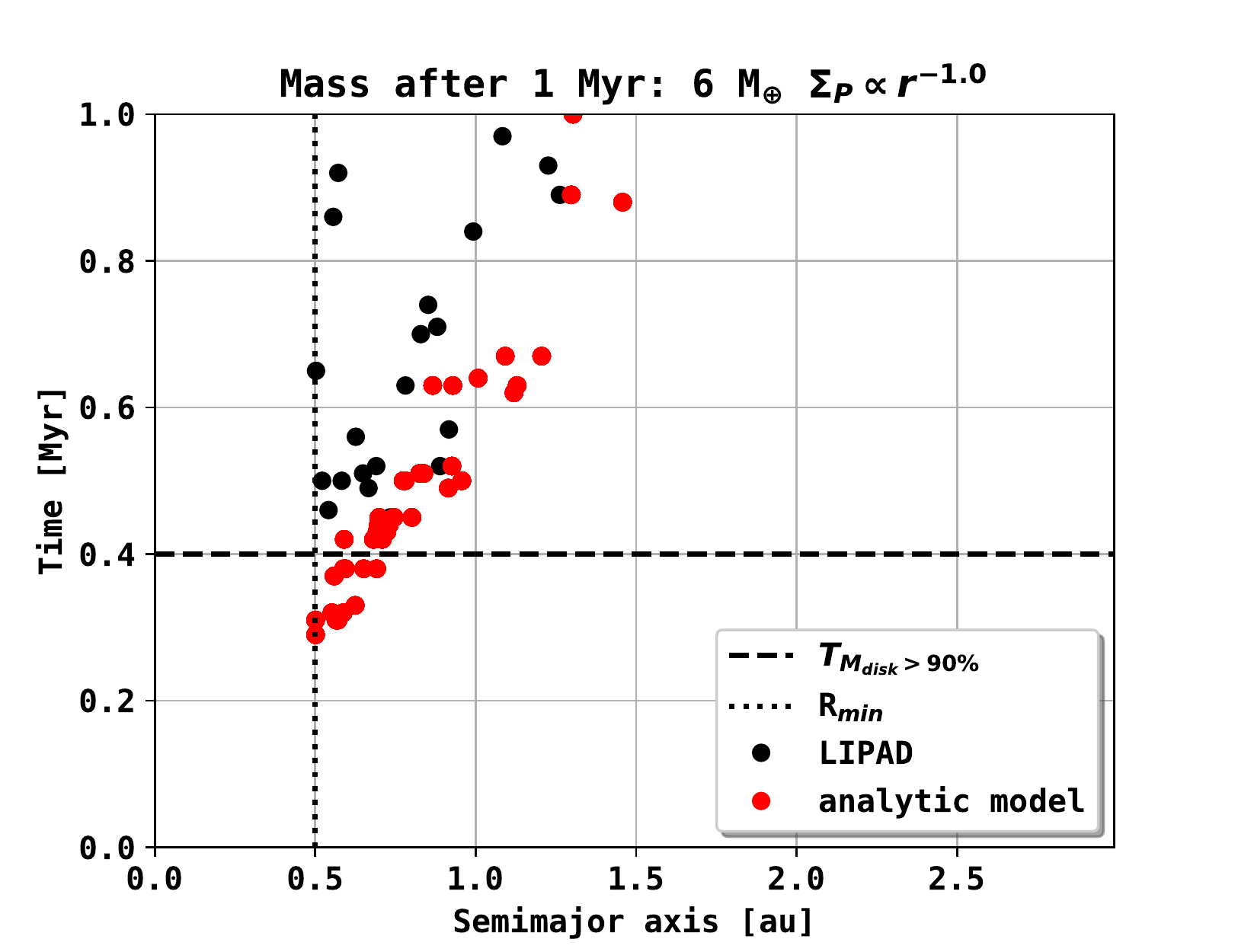}
\end{minipage}
\begin{minipage}{.33\textwidth}
  \includegraphics[width=1.0\linewidth]{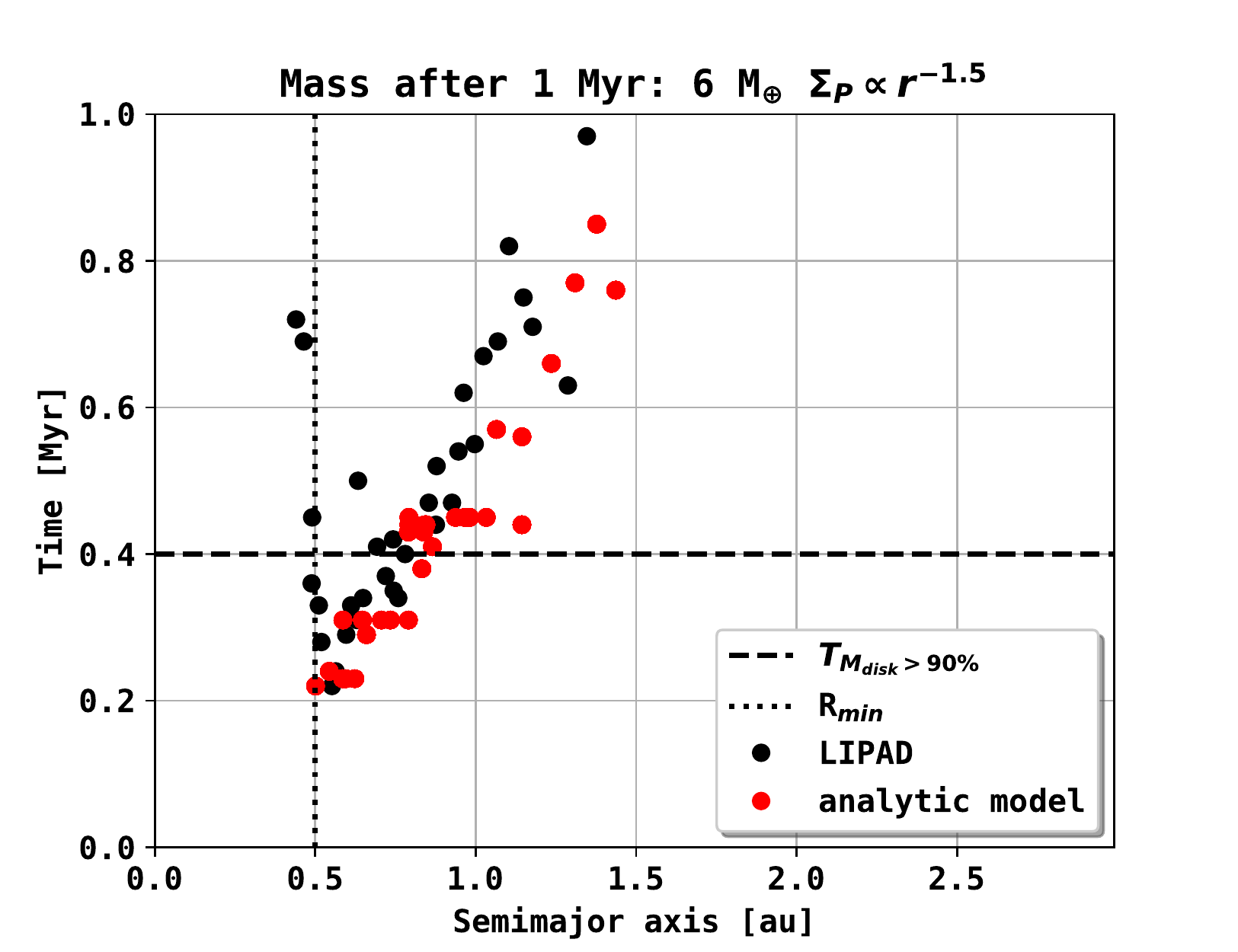}
\end{minipage}%
\begin{minipage}{.33\textwidth}
  \includegraphics[width=1.0\linewidth]{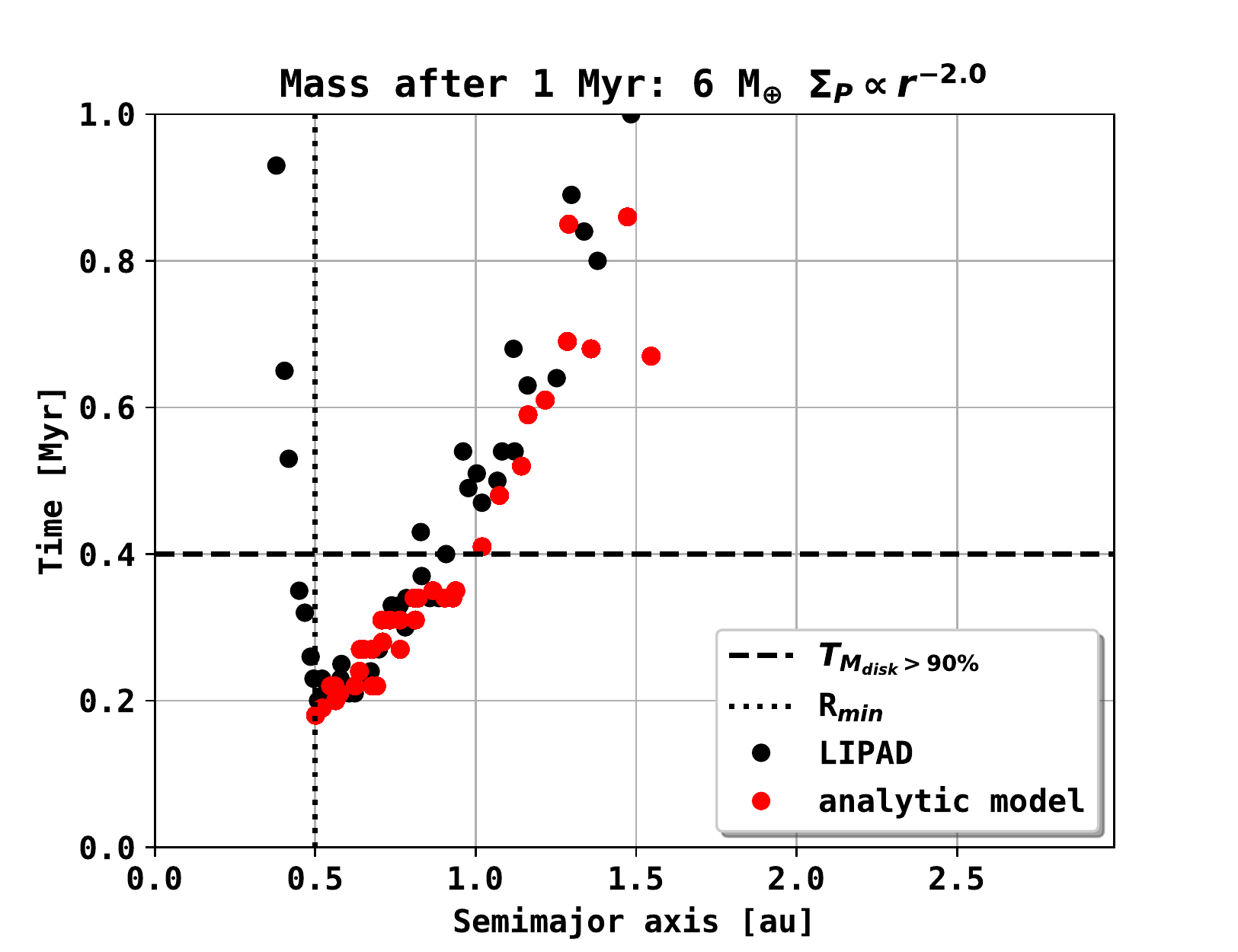}
\end{minipage}%
\\
\begin{minipage}{.33\textwidth}
  \centering
  \includegraphics[width=1.0\linewidth]{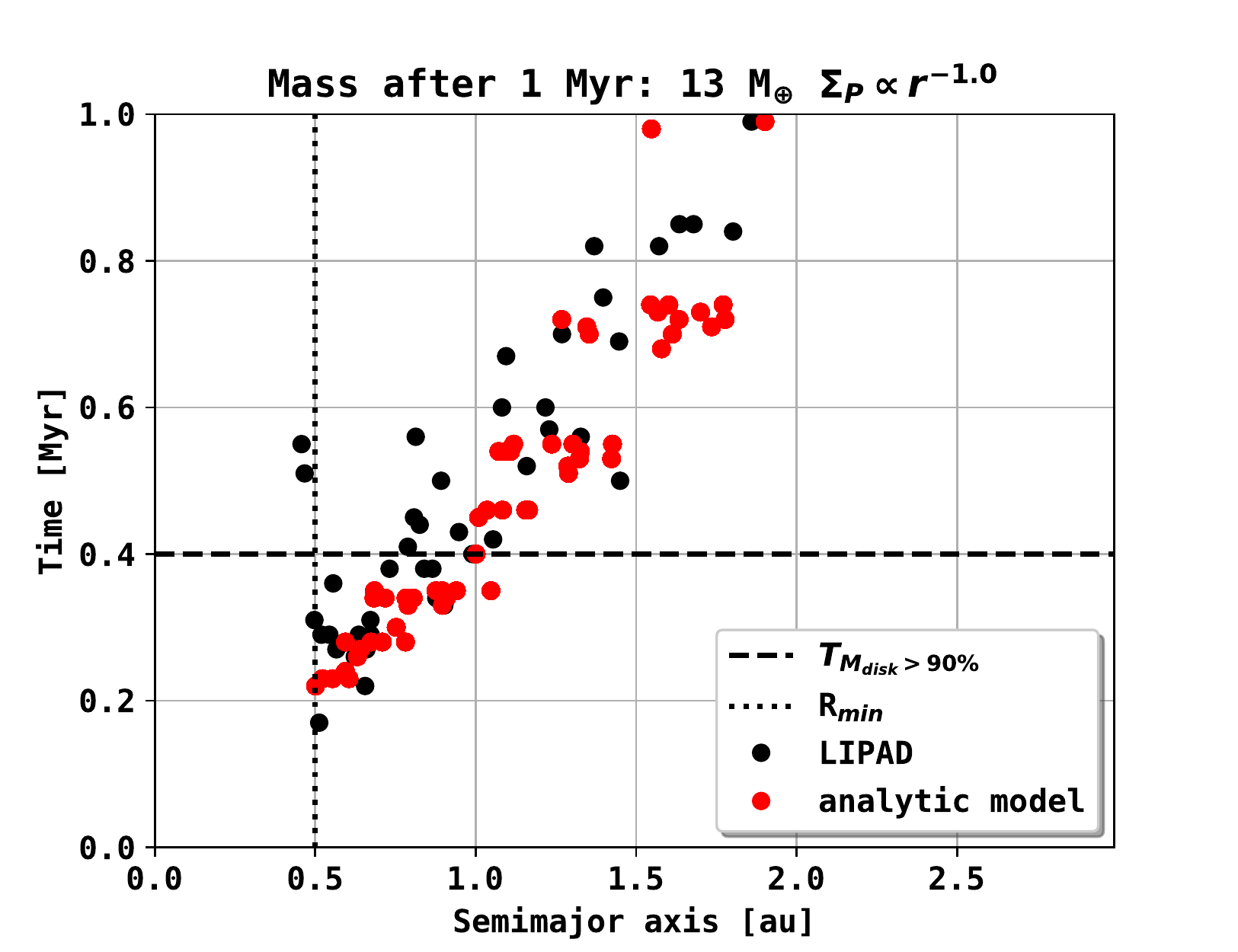}
\end{minipage}
\begin{minipage}{.33\textwidth}
  \includegraphics[width=1.0\linewidth]{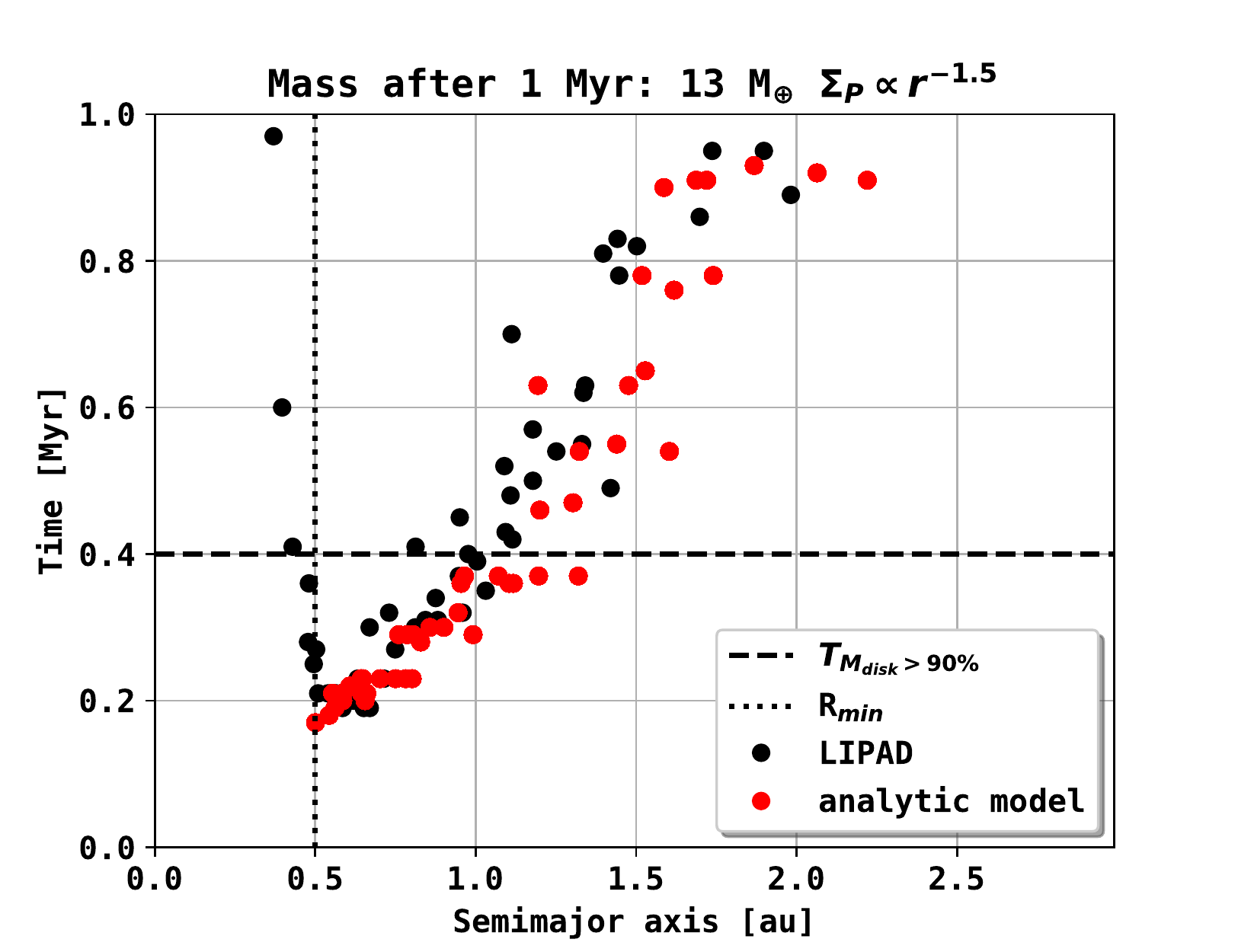}
\end{minipage}%
\begin{minipage}{.33\textwidth}
  \includegraphics[width=1.0\linewidth]{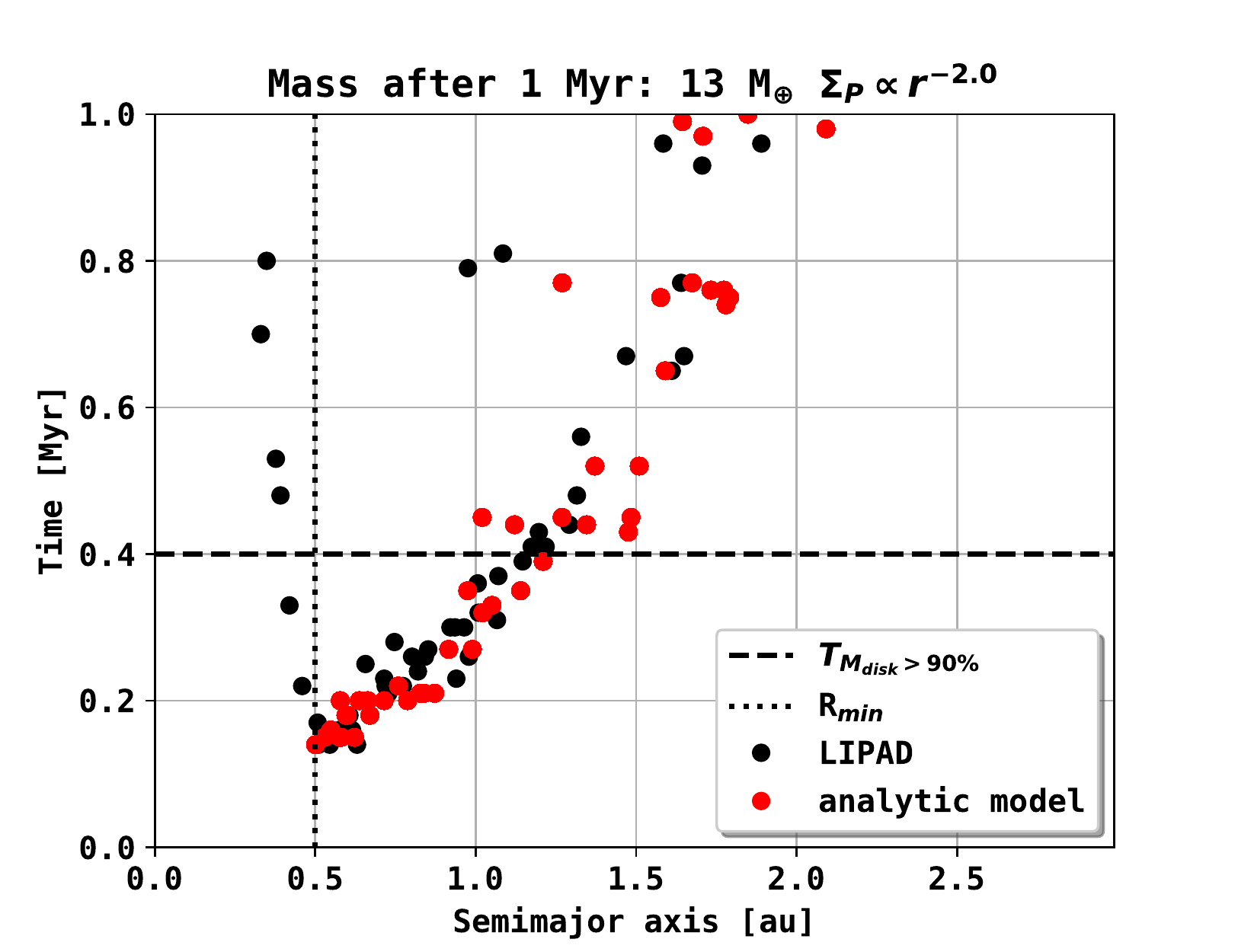}
\end{minipage}%
\\
\begin{minipage}{.33\textwidth}
  \centering
  \includegraphics[width=1.0\linewidth]{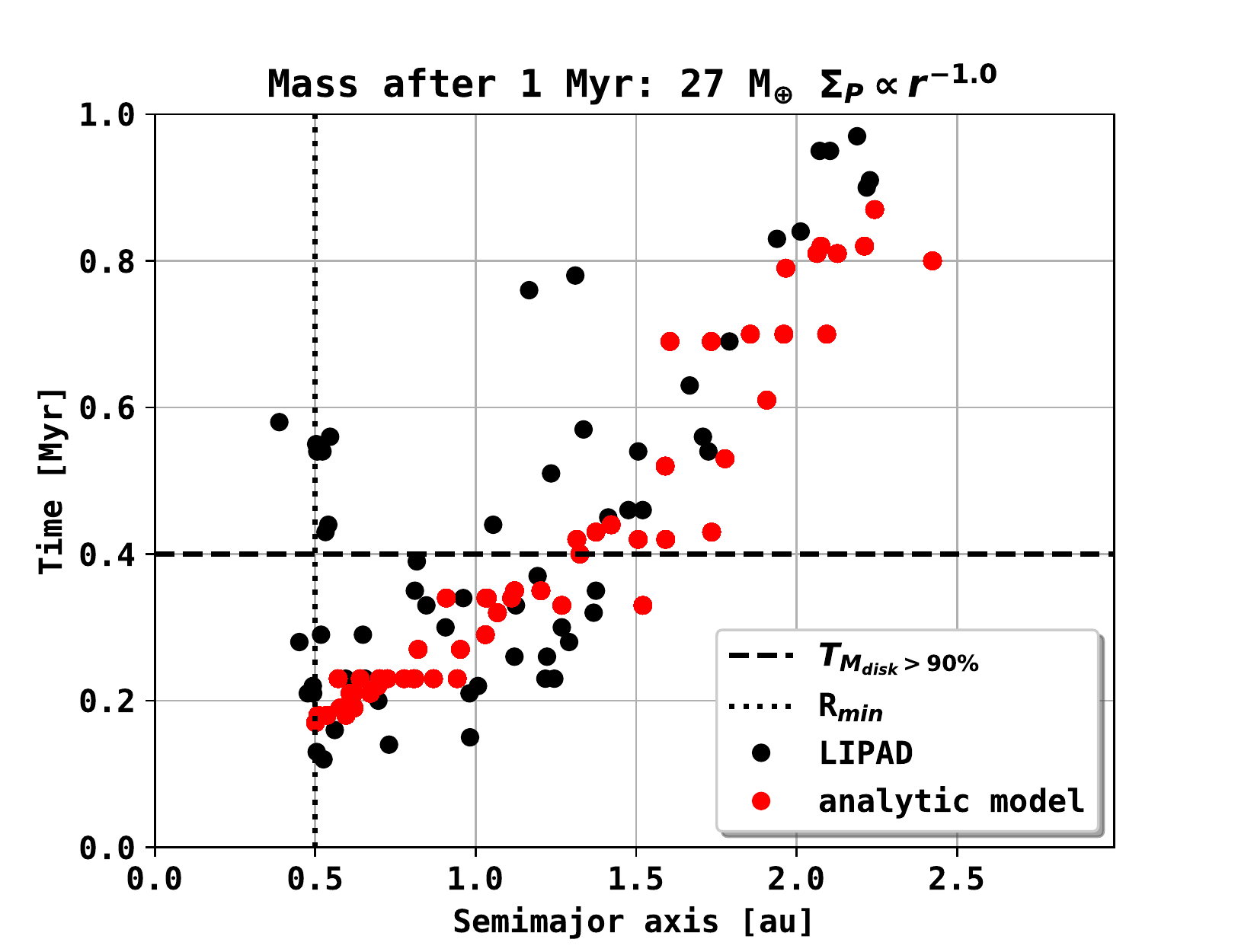}
\end{minipage}
\begin{minipage}{.33\textwidth}
  \includegraphics[width=1.0\linewidth]{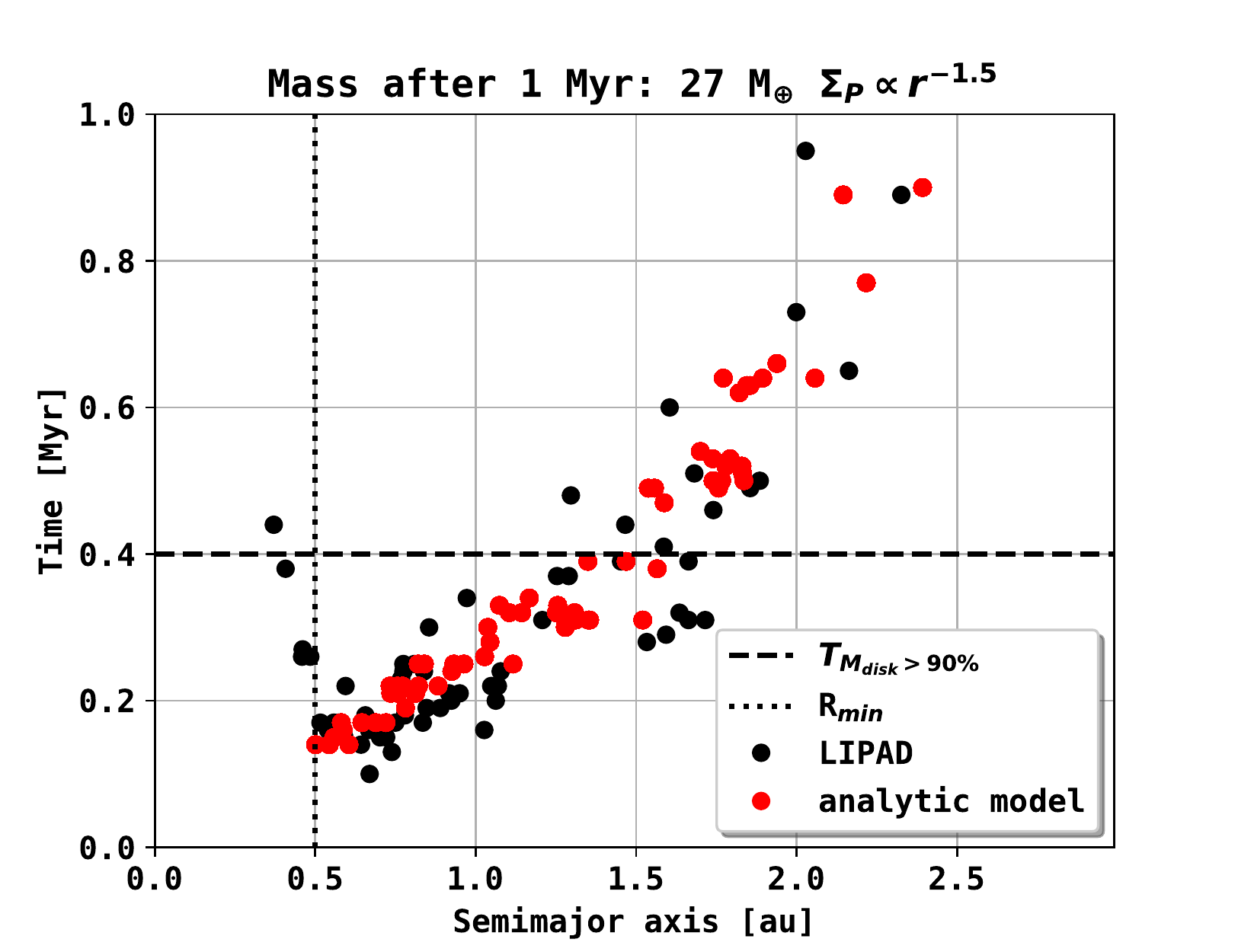}
\end{minipage}%
\begin{minipage}{.33\textwidth}
  \includegraphics[width=1.0\linewidth]{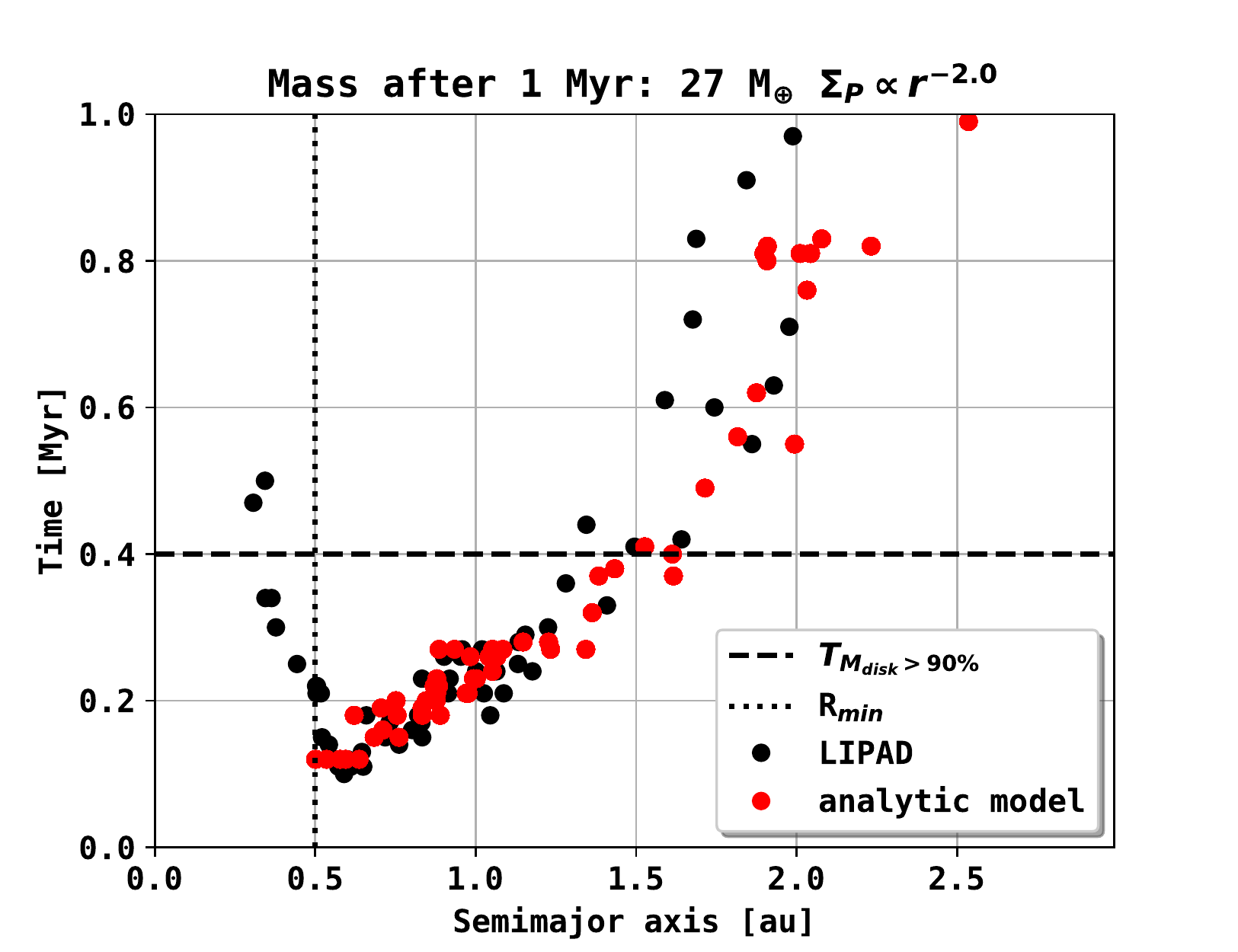}
\end{minipage}%
\caption{\small  Analytical model for embryo formation and embryo formation in LIPAD. The black dots indicate at what time and at which location an object reached planetary embryo mass in the LIPAD simulations. The red dots indicate the time for each distance from the star at which a planetary embryo is placed using our analytical model. The orbital seperation as input to the analytical model is given as 17 R$_{Hill}$ with a randomization of 2.5 R$_{Hill}$ The inner edge of planetesimal formation in the LIPAD runs was chosen to be at 0.5$\,$au for numerical performance. We vary the total mass in planetesimals after $10^6$ years from 0.5$\,$au to 5$\,$au (6$M_{\oplus}$,13$M_{\oplus}$,27$M_{\oplus}$) and the planetesimal surface density slope ($\Sigma_P \propto r^{-1.0}$, $\Sigma_P \propto r^{-1.5}$, $\Sigma_P \propto r^{-2.0}$).
}
\label{Fig:Time_Semi}
\end{figure*}

Fig. \ref{Fig:Time_Semi} shows the time and location at which an object has reached the mass of a planetary embryo for both the LIPAD simulation and the analytical model from Sect. \ref{Sec:Planetesimal_growth} (Eq. \ref{eq:mass_growth}, \ref{eq:criterion_I}, \ref{eq:criterion_II}). The red dots refer to analytical model with the same planetesimal surface density increase as in the LIPAD simulations, whereas the black dots are those from the N-body simulation shown in Fig. \ref{Fig:Emb_form_LIPAD_6_ME} - Fig. \ref{Fig:Emb_form_LIPAD_27_ME}. We show the inner edge of planetesimal formation in LIPAD at 0.5$\,$au and give the time by which the planetesmal disk mass has reached $90\%$ of its total value ($T_{M_{disk}> 90\%}  = 400$ky). The randomization of the semimajor axis in our analytical model is given by 2.5$R_{Hill}$ as explained in Sec. \ref{Subsec:Panetesimal_formation_in_LIPAD}. We find that the overall time and semimajor axis distribution of the analytical model is in well agreement with the larger N-body simulations from LIPAD. The randomization of the semimajor axis does well in reproducing the stochastic nature of the N-body process, as well as the analytic growth equation does for the time it takes until embryo formation (based on the local planetesimal surface density evolution at a given distance from the star) is possible.
\\
The innermost embryos (0.5$\,$au to 1$\,$au) in every setup form well below $T_{M_{disk}> 90\%}$, but no embryo outside 2$\,$au forms within $T_{M_{disk}> 90\%}$. The implications on possible pebble accretion from this behavior are discussed in Sect. \ref{SubSec:Implications_pebble_acc}.
\subsection{Cumulative number}
\label{Sec:Cumulative_number}
\begin{figure*}[]
\centering
\begin{minipage}{.33\textwidth}
  \centering
  \includegraphics[width=1.0\linewidth]{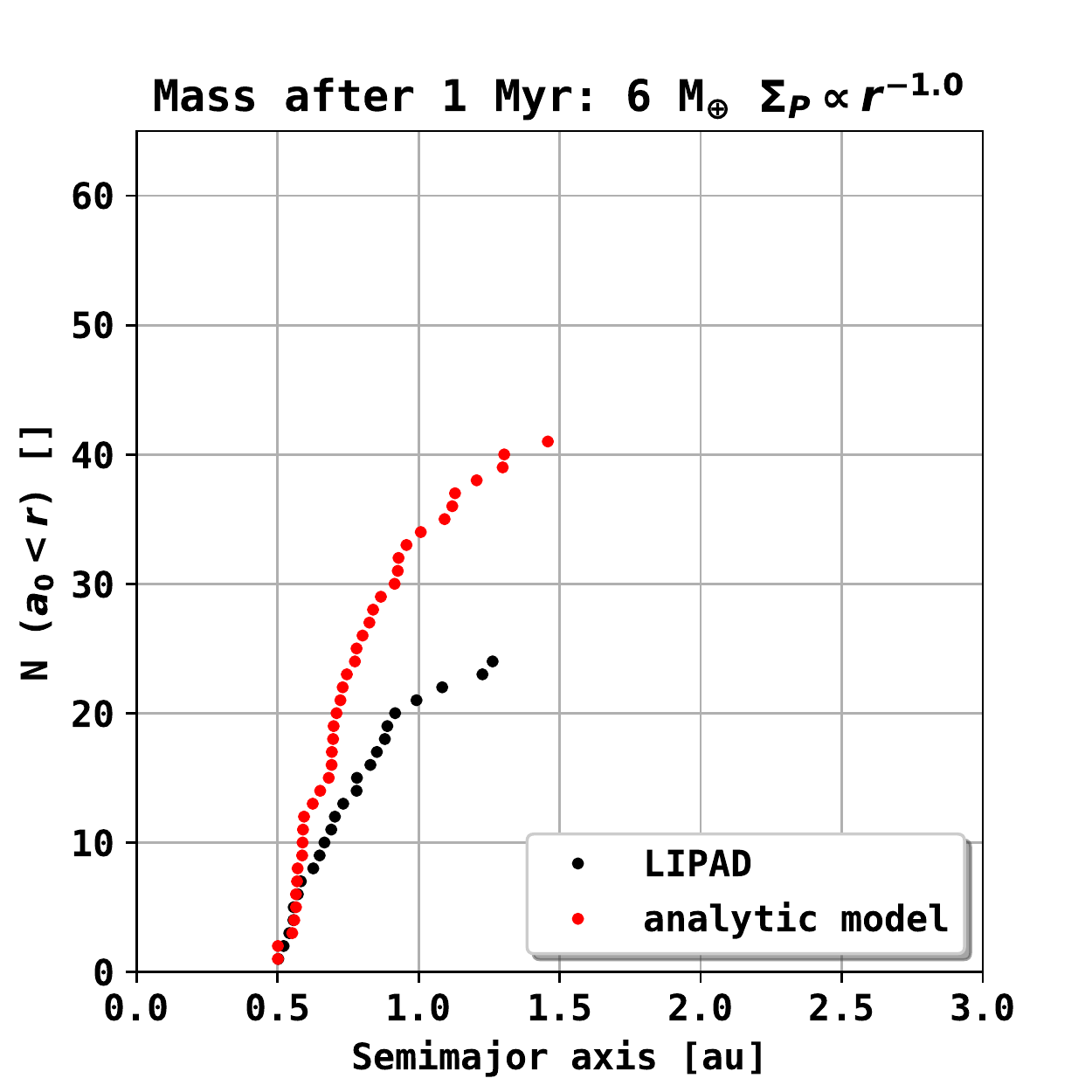}
\end{minipage}
\begin{minipage}{.33\textwidth}
  \includegraphics[width=1.0\linewidth]{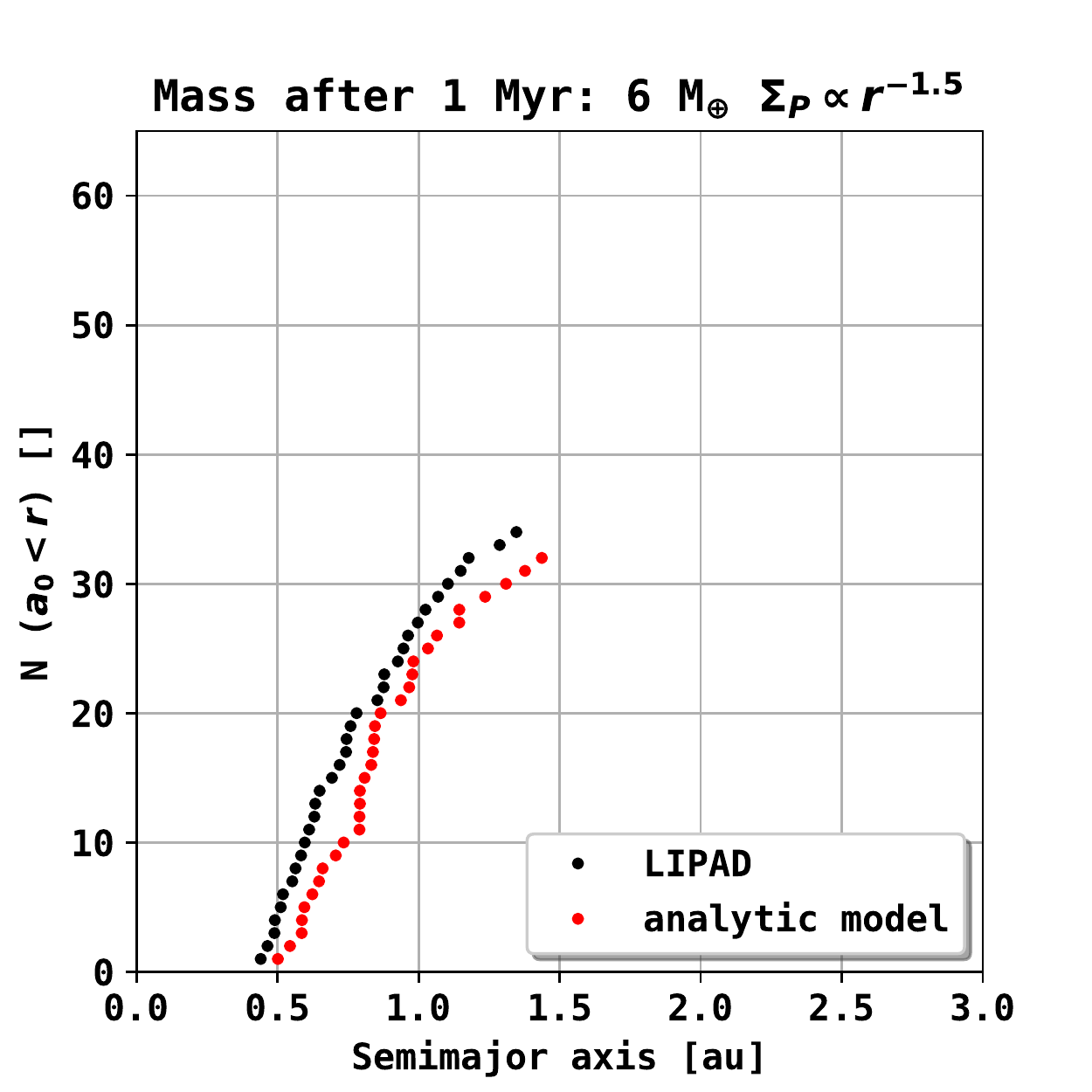}
\end{minipage}%
\begin{minipage}{.33\textwidth}
  \includegraphics[width=1.0\linewidth]{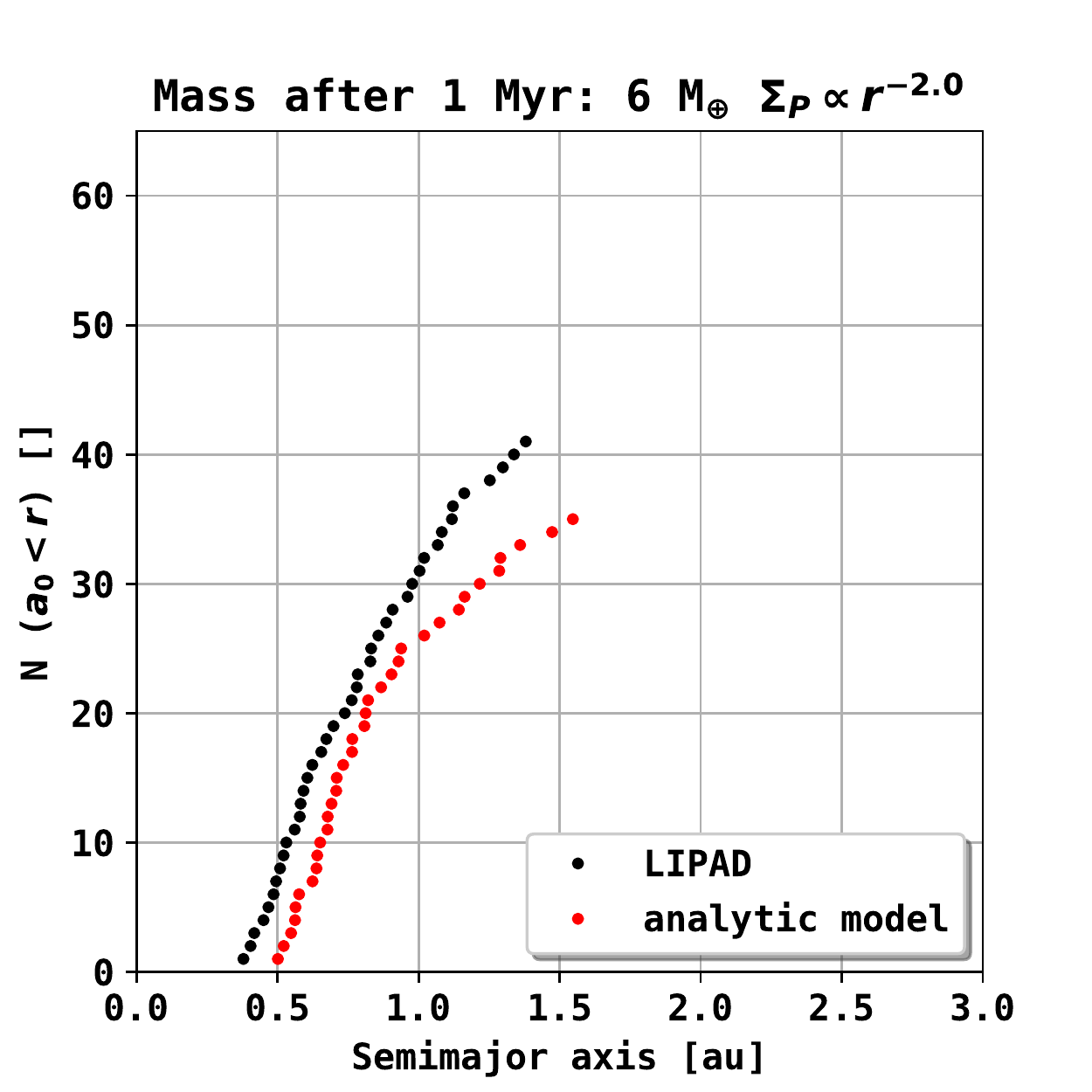}
\end{minipage}%
\\
\begin{minipage}{.33\textwidth}
  \centering
  \includegraphics[width=1.0\linewidth]{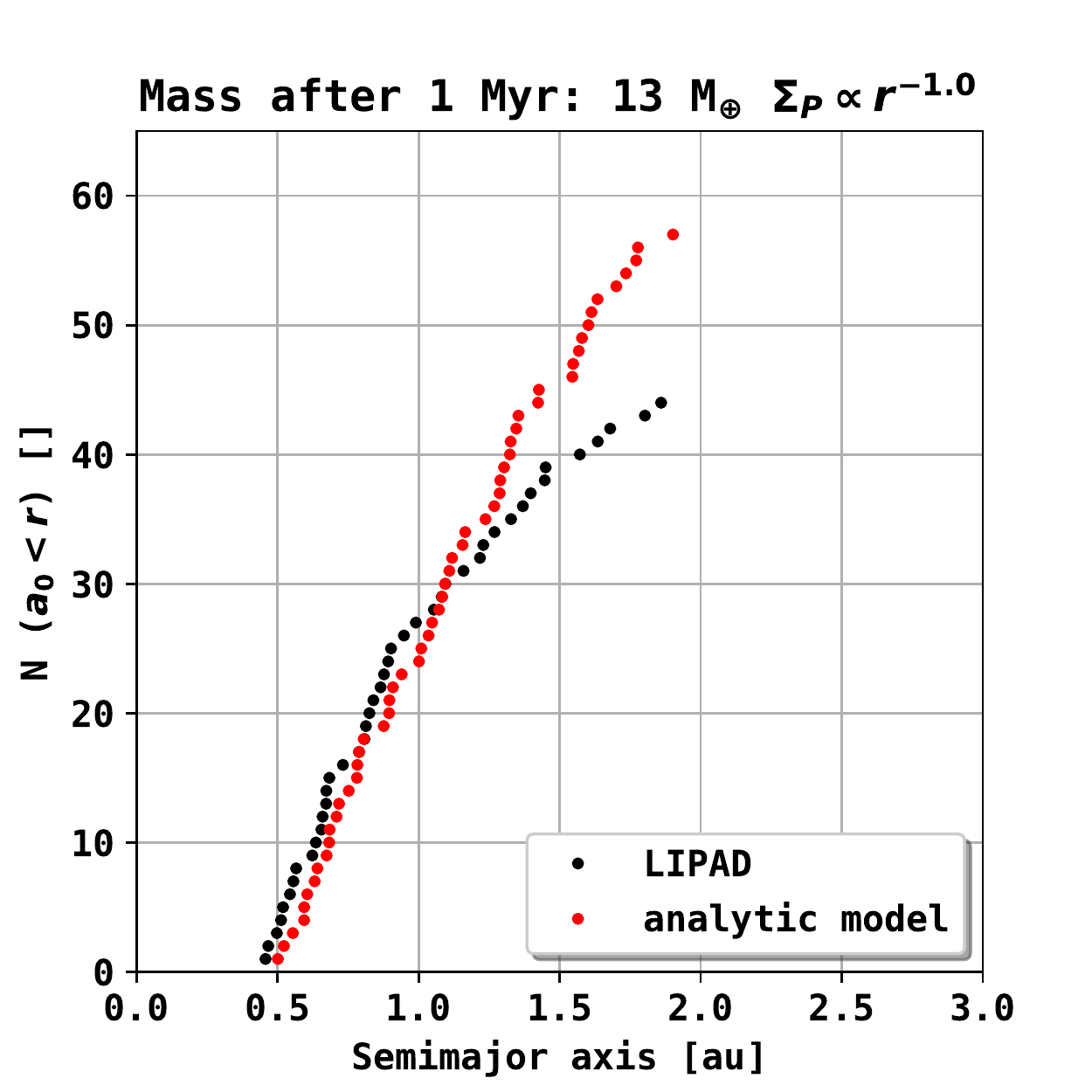}
\end{minipage}
\begin{minipage}{.33\textwidth}
  \includegraphics[width=1.0\linewidth]{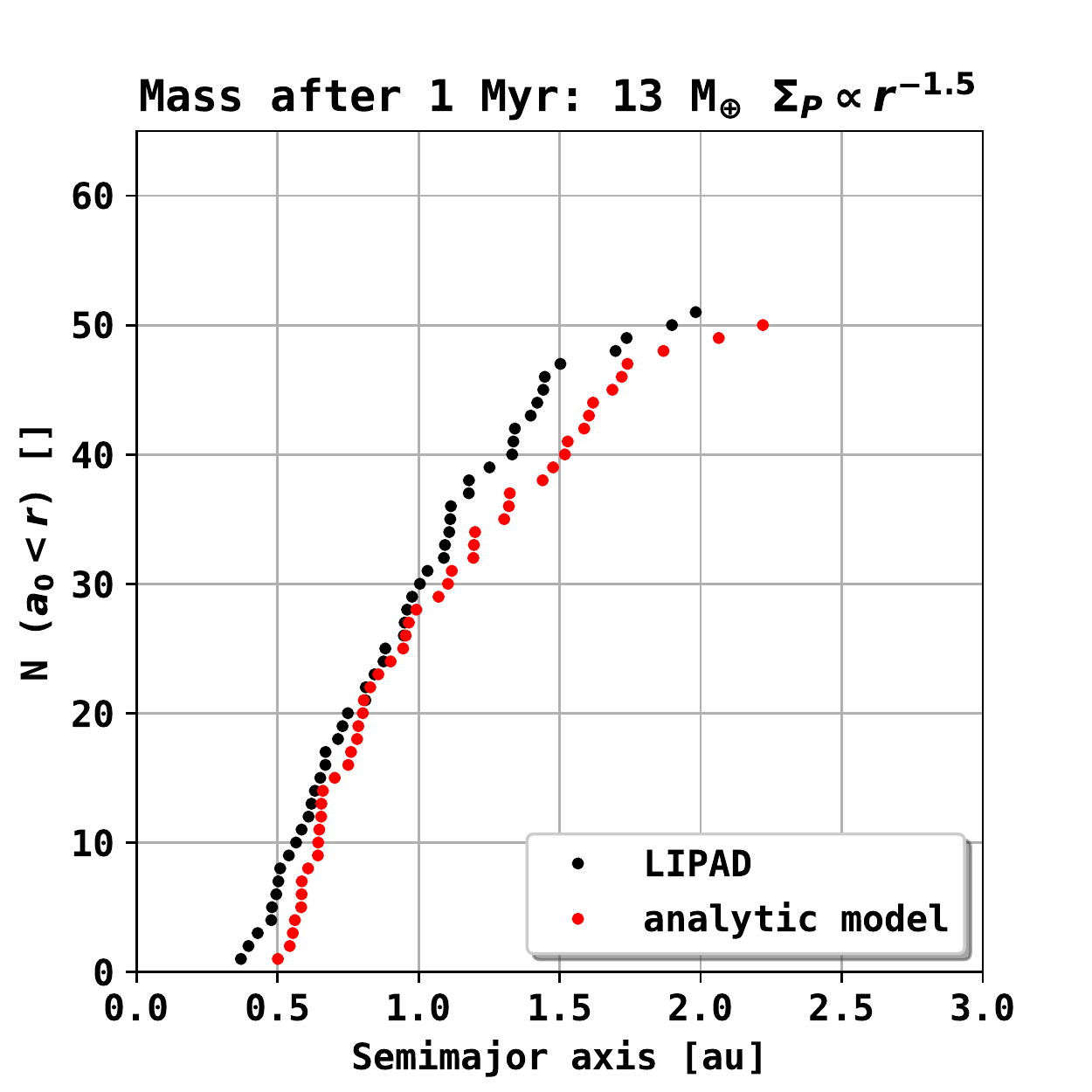}
\end{minipage}%
\begin{minipage}{.33\textwidth}
  \includegraphics[width=1.0\linewidth]{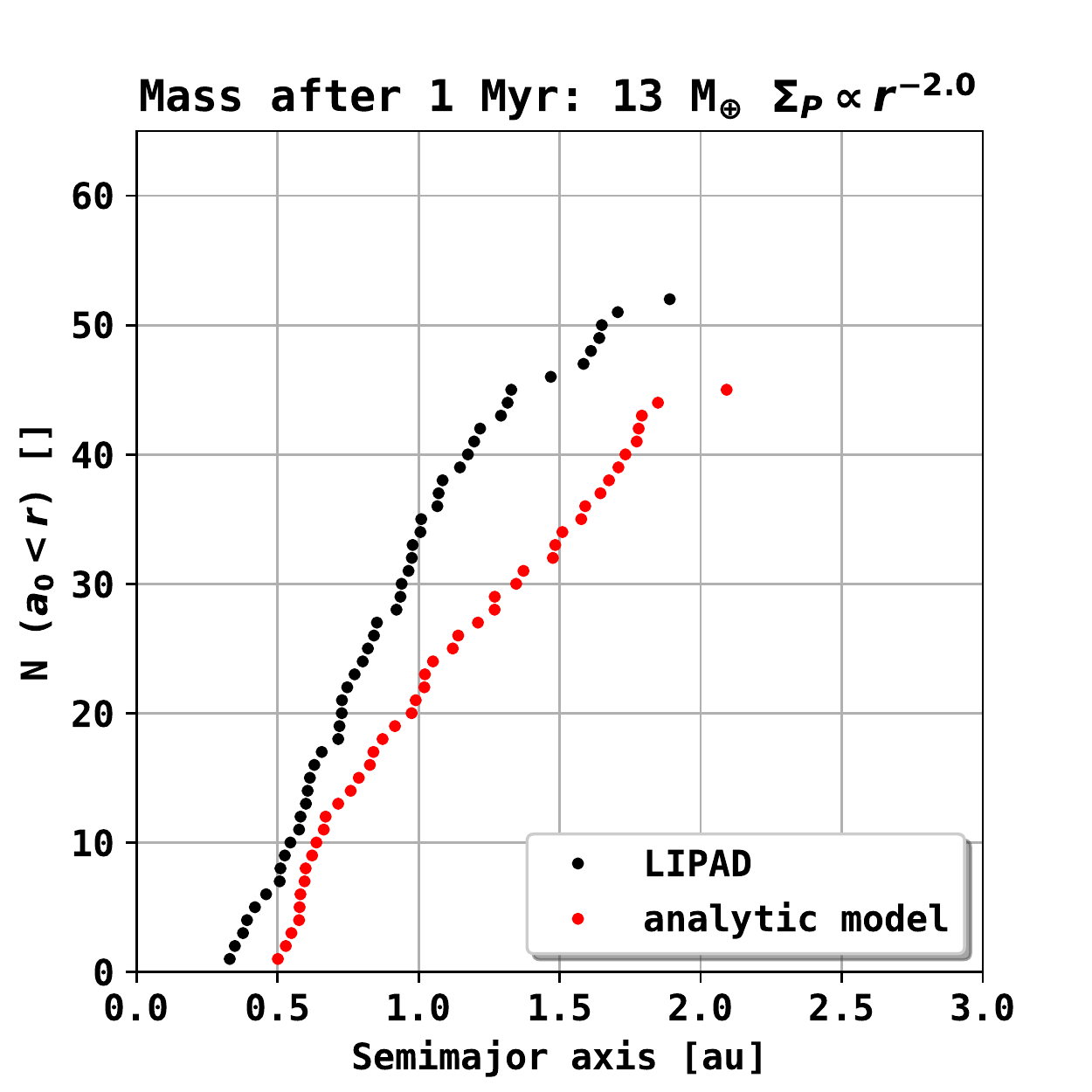}
\end{minipage}%
\\
\begin{minipage}{.33\textwidth}
  \centering
  \includegraphics[width=1.0\linewidth]{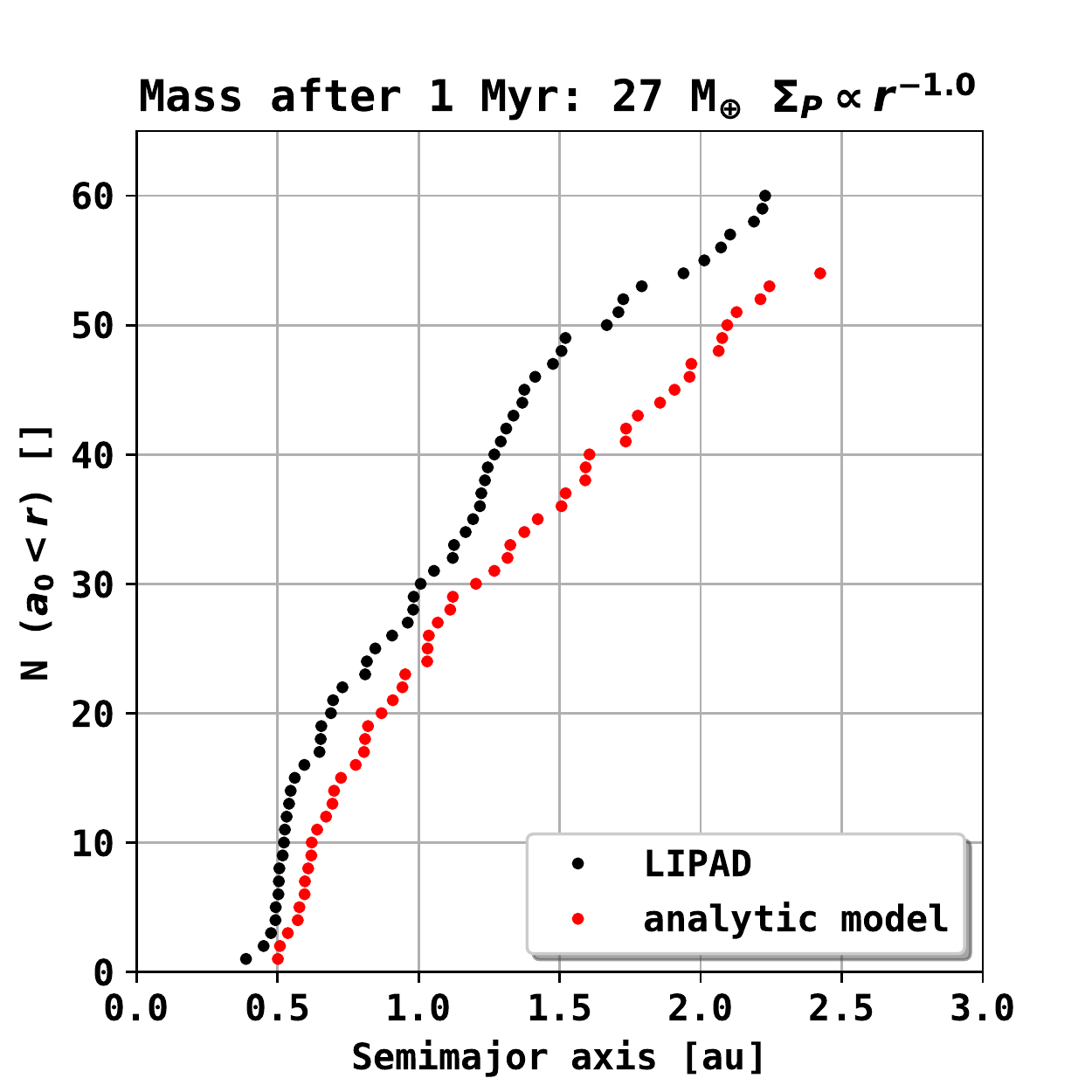}
\end{minipage}
\begin{minipage}{.33\textwidth}
  \includegraphics[width=1.0\linewidth]{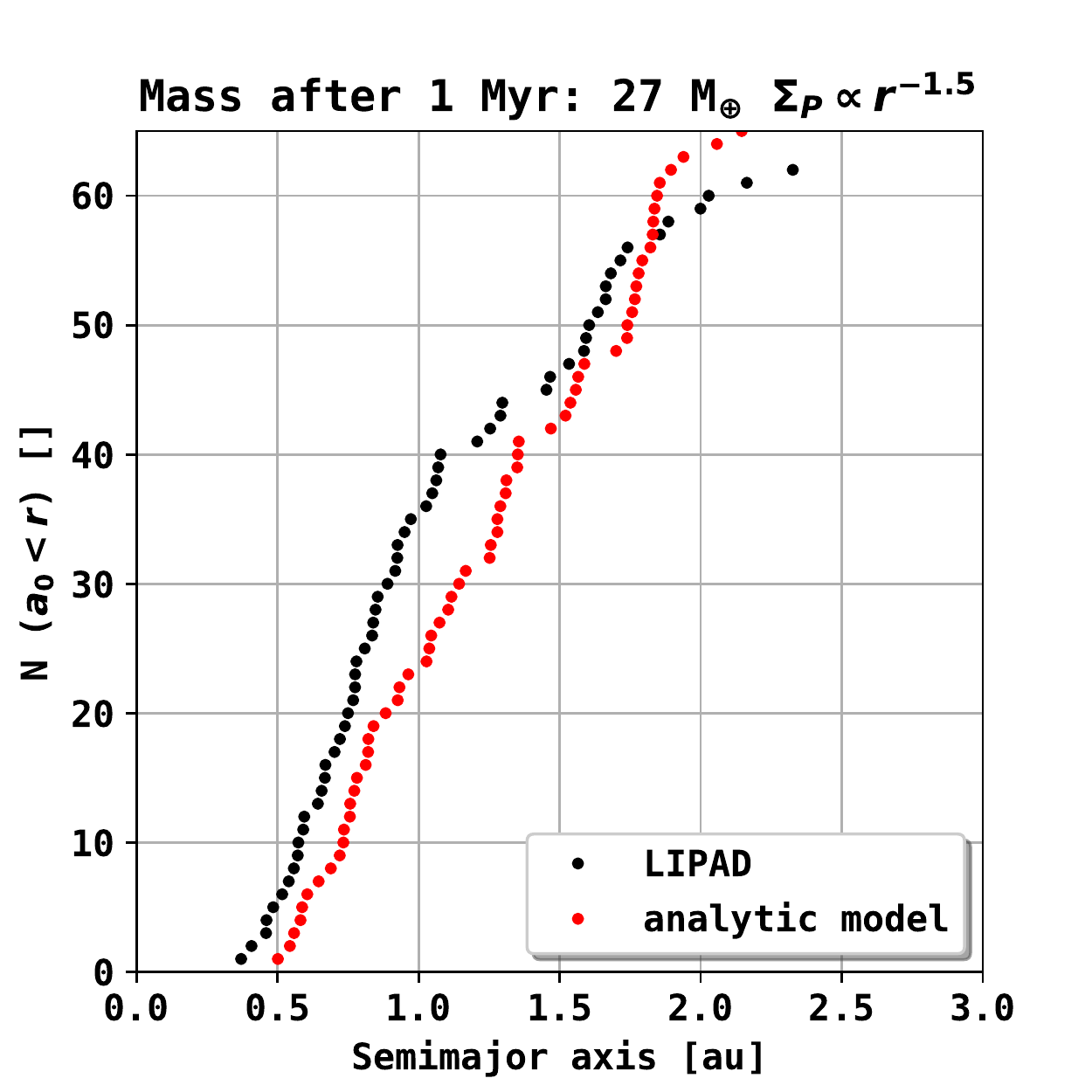}
\end{minipage}%
\begin{minipage}{.33\textwidth}
  \includegraphics[width=1.0\linewidth]{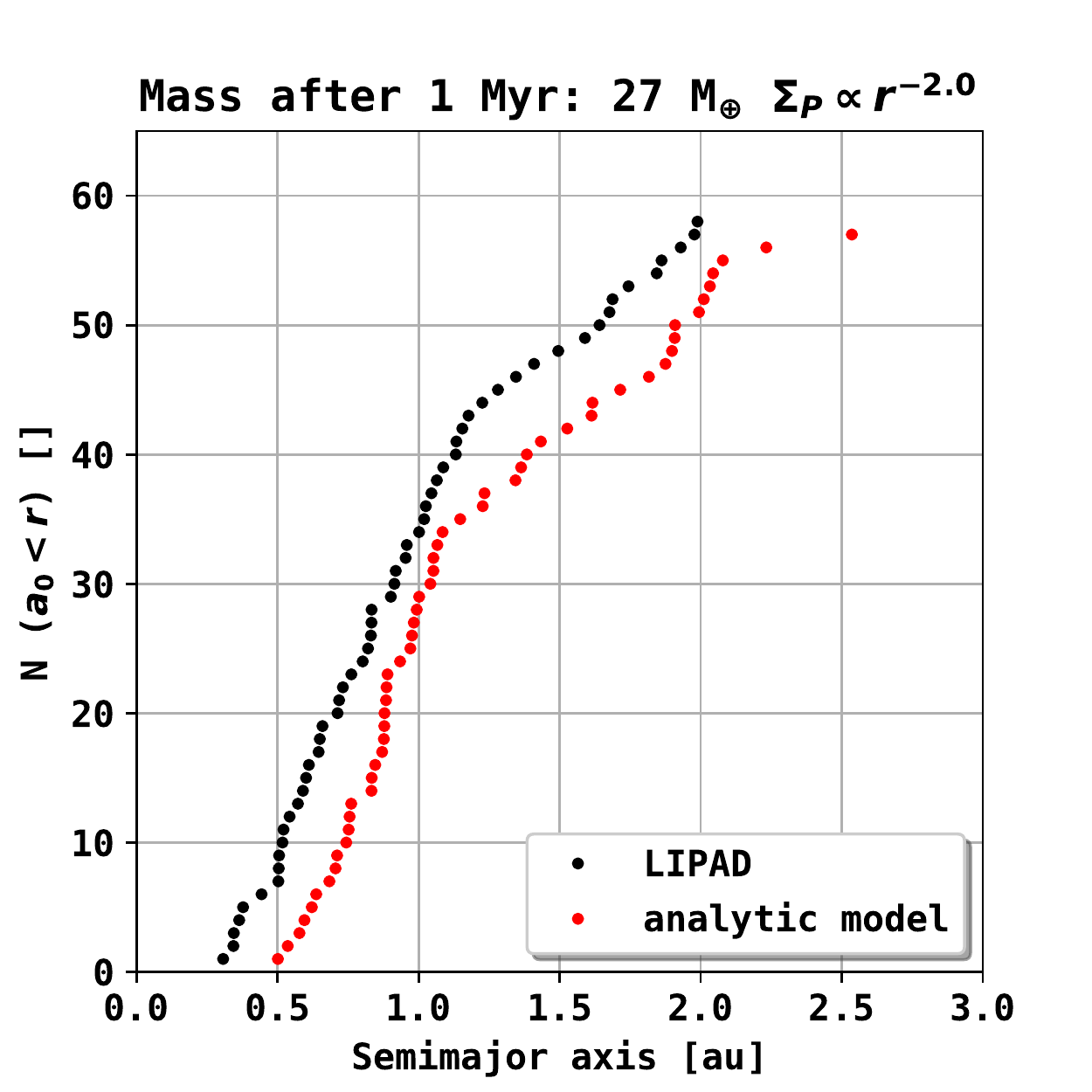}
\end{minipage}%
\caption{\small  Cumulative number of embryos formed during the LIPAD runs from Fig. \ref{Fig:Emb_form_LIPAD_6_ME} - Fig \ref{Fig:Emb_form_LIPAD_27_ME} (black dots). The red dots show the cumulative number of embryos that would be placed according to the analytical model from Sect. \ref{Sec:Planetesimal_growth}.
}
\label{Fig:Cumulative_number}
\end{figure*}

Fig. \ref{Fig:Cumulative_number} shows the cumulative number of initial embryos from the LIPAD simulation and the analytic model. The orbital separation of the embryos in the analytic model scales linearly with their Hill radius, which again scales linearly with their distance to the star. The cumulative number of planetary embryos in the analytic model therefore scales logarithmic with distance. Since the orbital separation of embryos in the N-body simulation converges to the same amount of Hill radii, we also find a logarithmic trend in the cumulative number of embryos formed in LIPAD. The total number of initial embryos is related to the total mass in planetesimals after 1 Myrs. The reason for this is that in more massive disks, embryos can form at larger distances. 
\\
The N-body simulations show embryo formation within 0.5$\,$au, which is not possible in the analytic model, since the planetesimal formation within 0.5$\,$au is neglected. The innermost embryos that form in the N-body simulation are therefore due to planetesimals that moved within 0.5$\,$au due to their dynamcial interactions. 
\\
 This spatial area of embryo formation is well defined by Criterion I, see Fig. \ref{Fig:Emb_form_LIPAD_6_ME} - Fig. \ref{Fig:Emb_form_LIPAD_27_ME}. The number of embryos within this area can be determined using Criterion II by setting their orbital separations.
\subsection{Orbital separation}
\label{Sec:orbital_seperation}
\begin{figure*}[]
\label{Subsection:Comparison}
\centering
\begin{minipage}{.33\textwidth}
  \centering
  \includegraphics[width=1.0\linewidth]{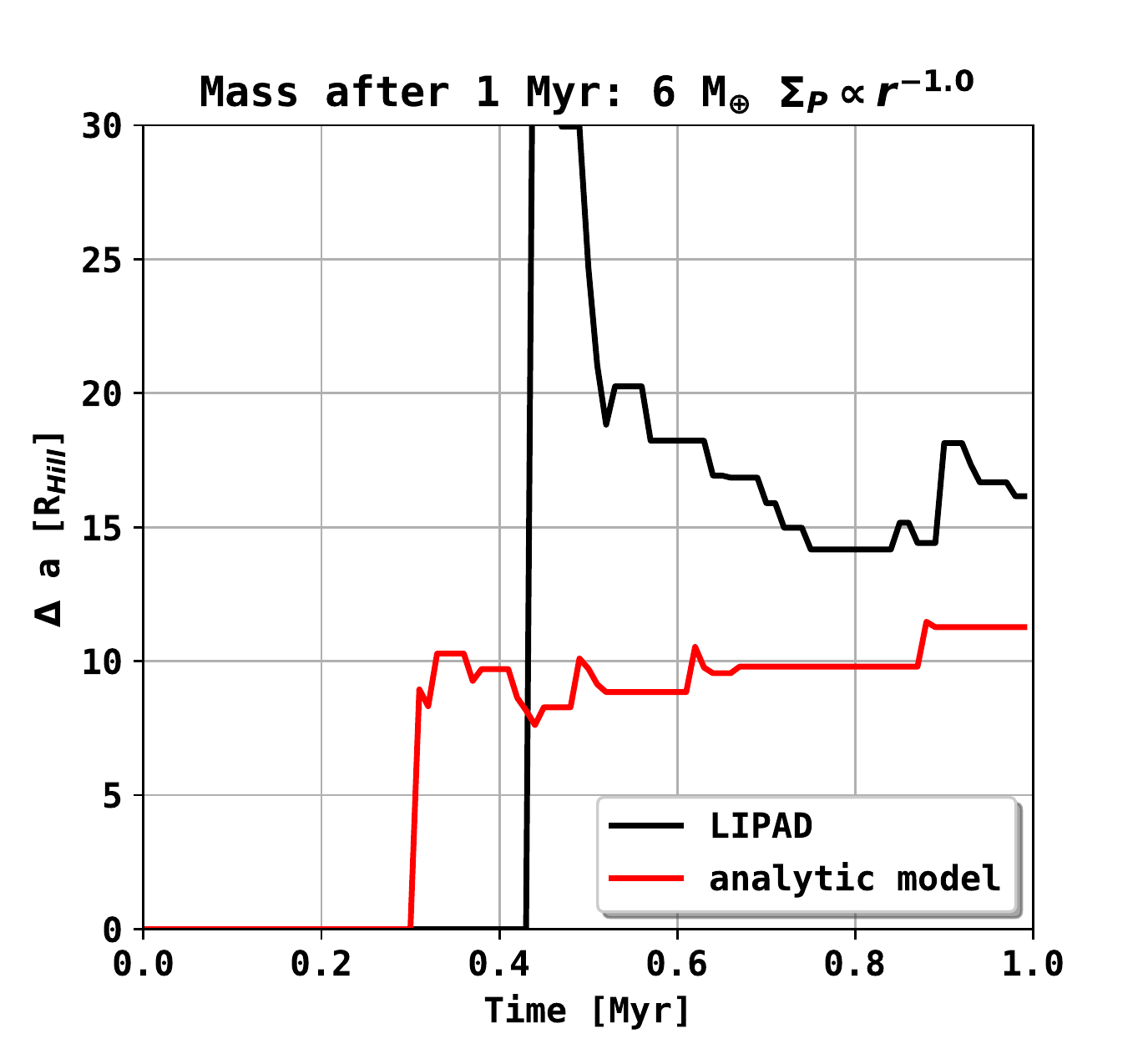}
\end{minipage}
\begin{minipage}{.33\textwidth}
  \includegraphics[width=1.0\linewidth]{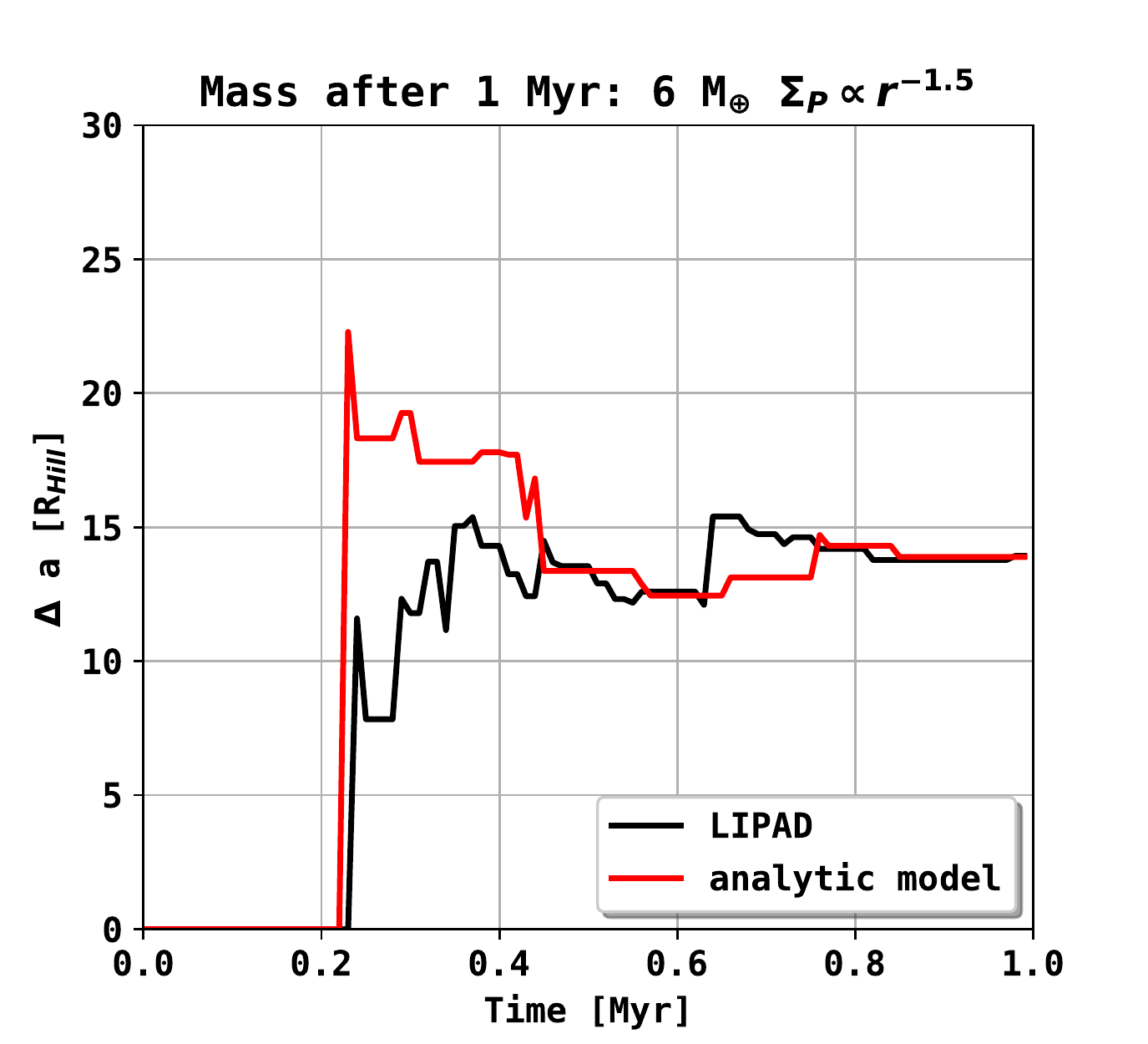}
\end{minipage}%
\begin{minipage}{.33\textwidth}
  \includegraphics[width=1.0\linewidth]{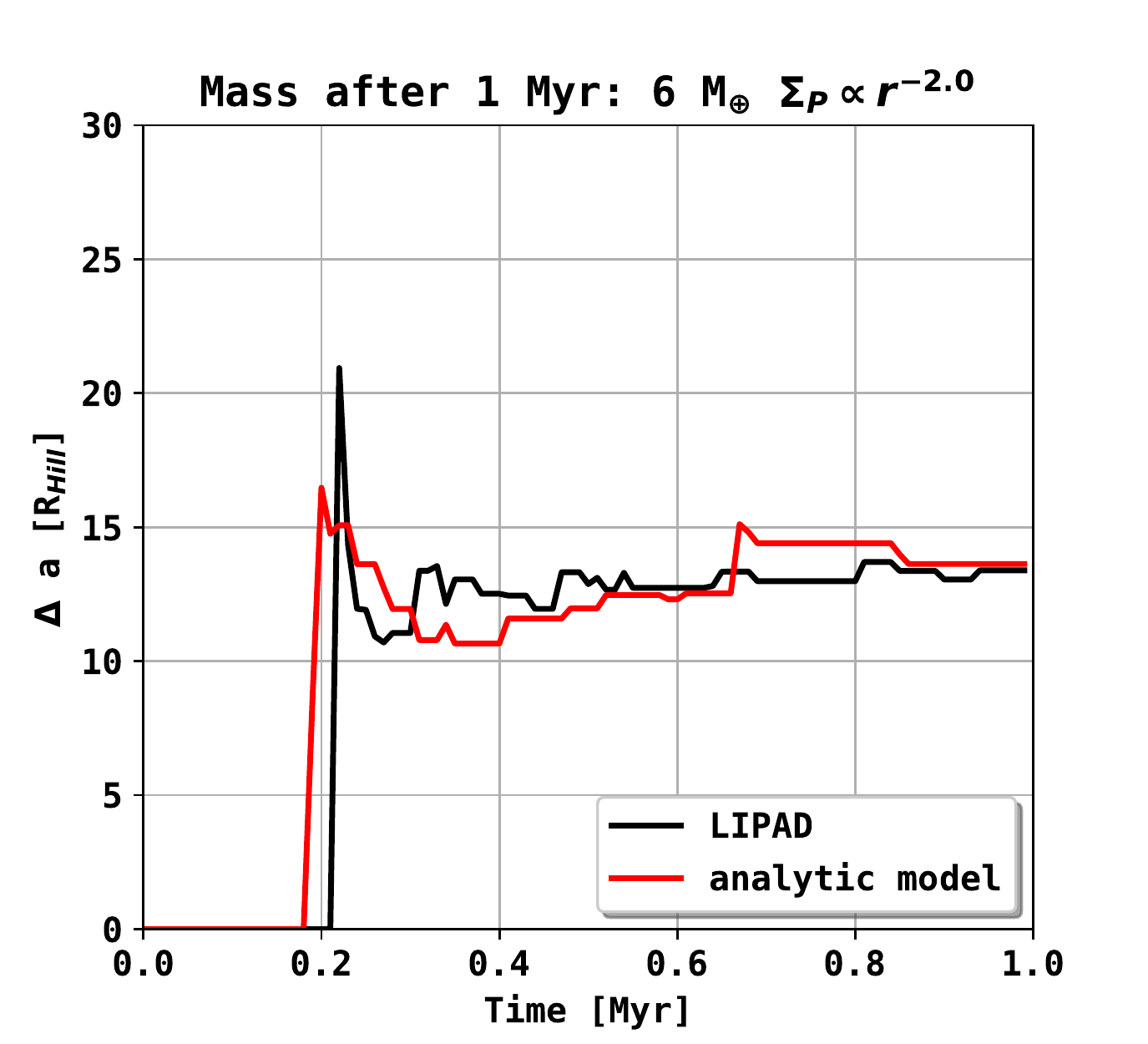}
\end{minipage}%
\\
\begin{minipage}{.33\textwidth}
  \centering
  \includegraphics[width=1.0\linewidth]{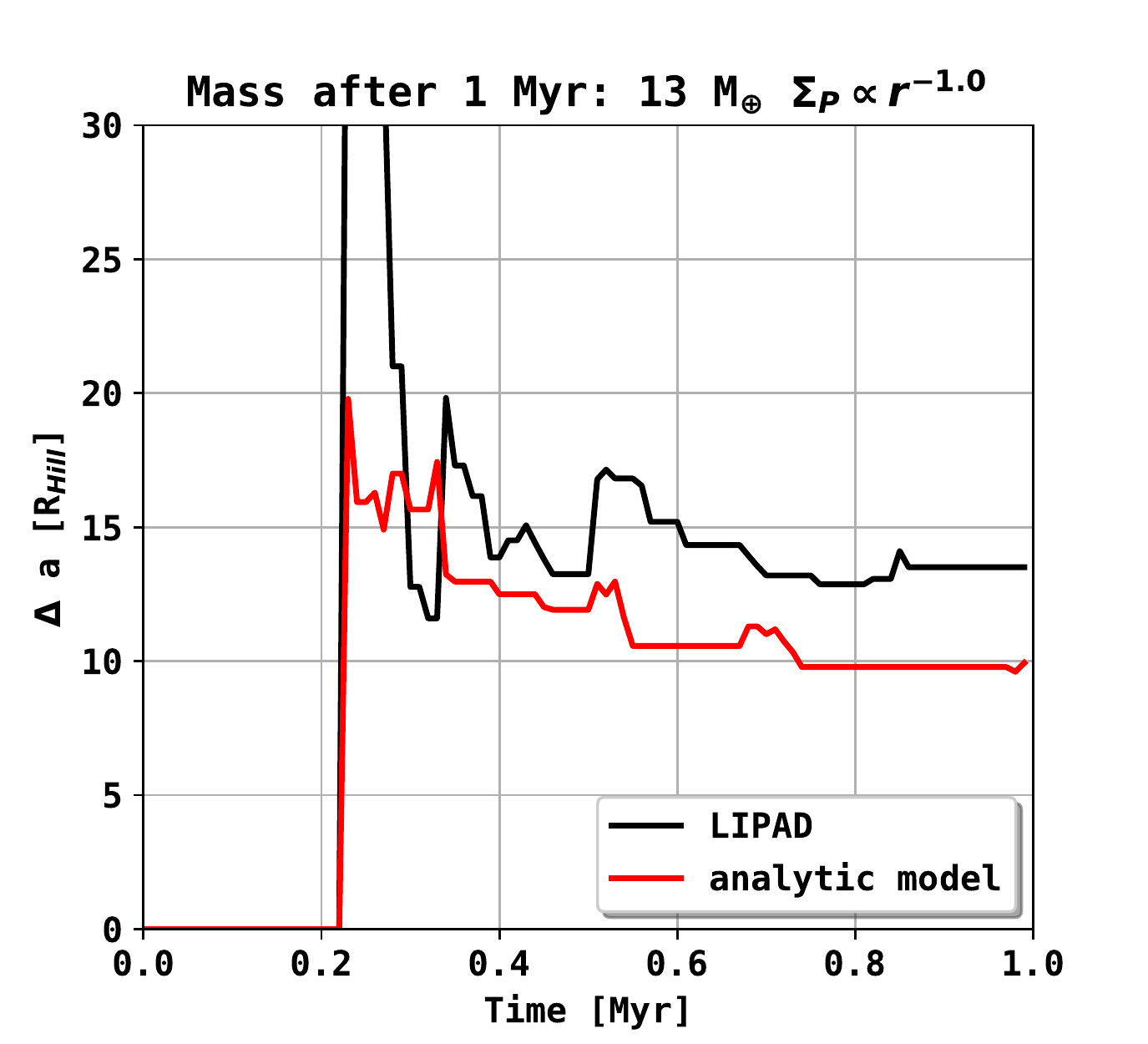}
\end{minipage}
\begin{minipage}{.33\textwidth}
  \includegraphics[width=1.0\linewidth]{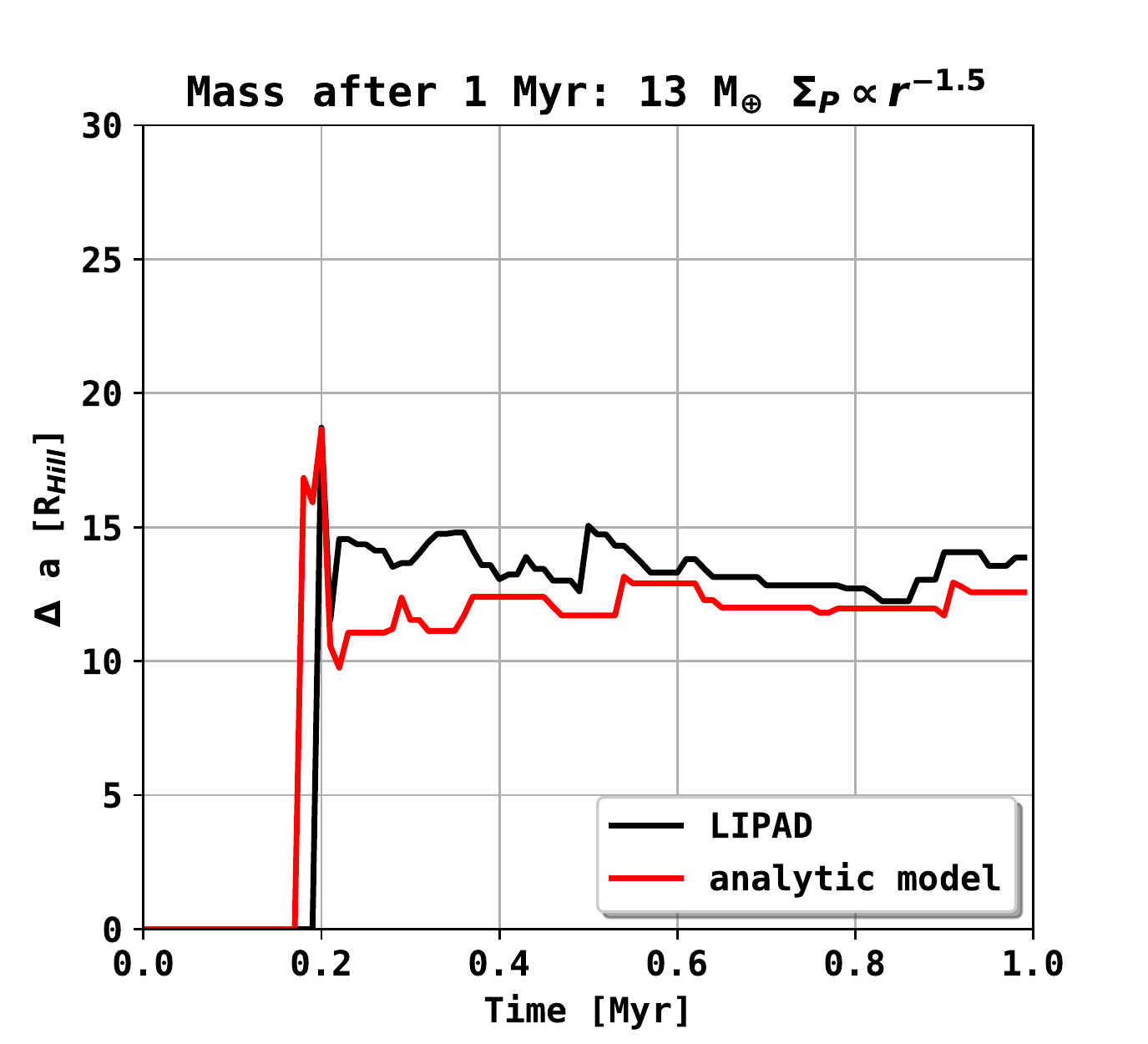}
\end{minipage}%
\begin{minipage}{.33\textwidth}
  \includegraphics[width=1.0\linewidth]{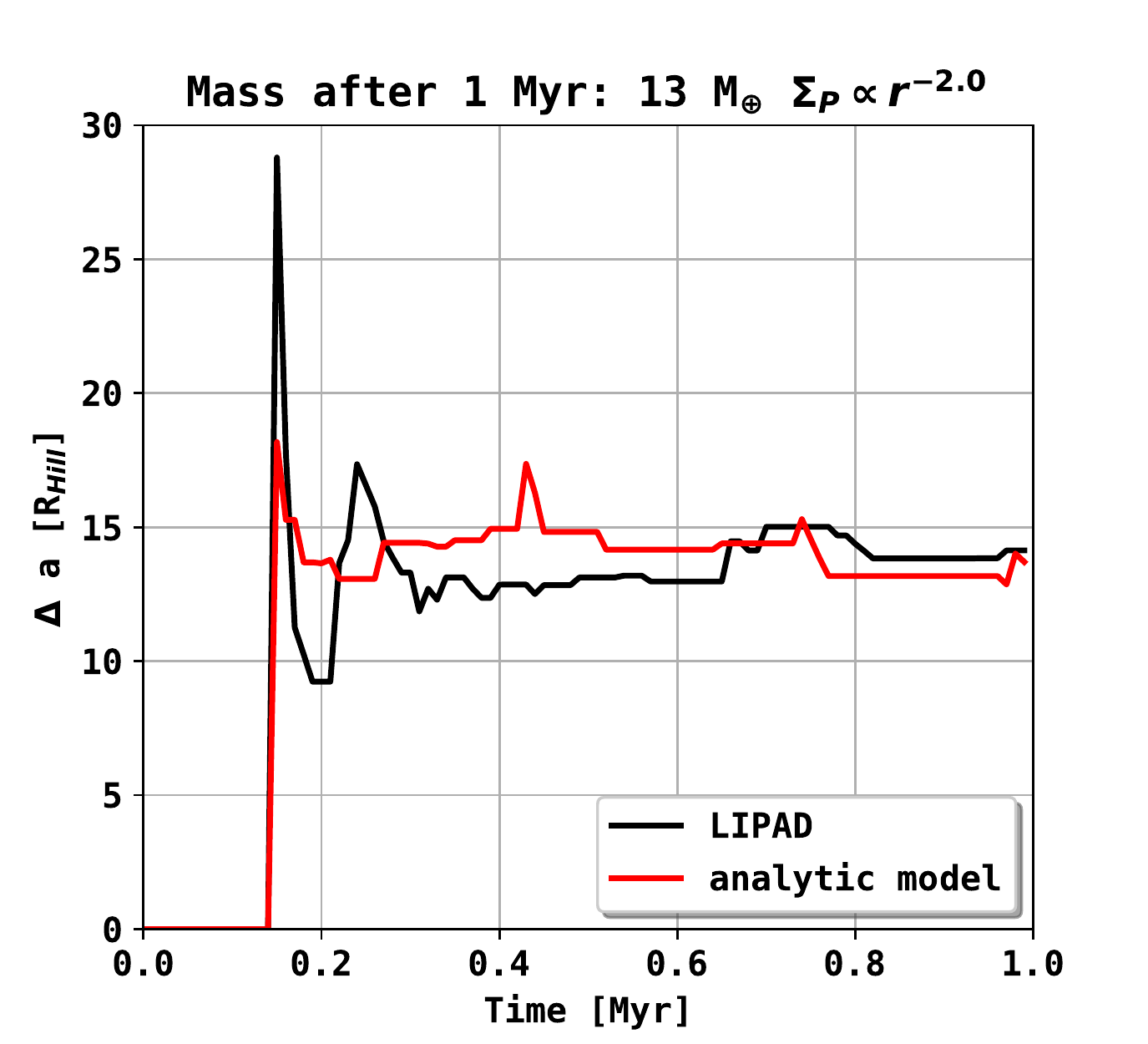}
\end{minipage}%
\\
\begin{minipage}{.33\textwidth}
  \centering
  \includegraphics[width=1.0\linewidth]{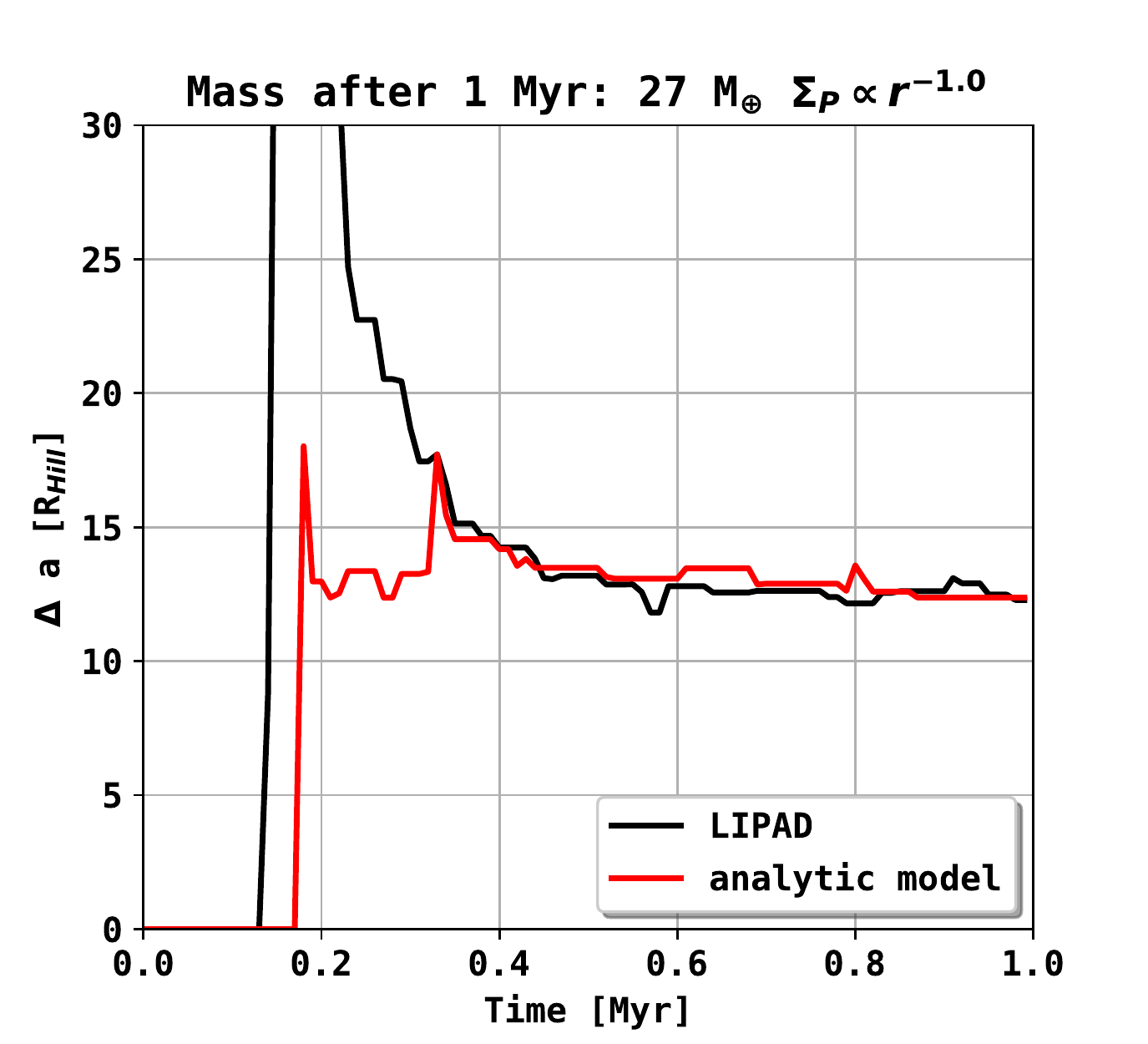}
\end{minipage}
\begin{minipage}{.33\textwidth}
  \includegraphics[width=1.0\linewidth]{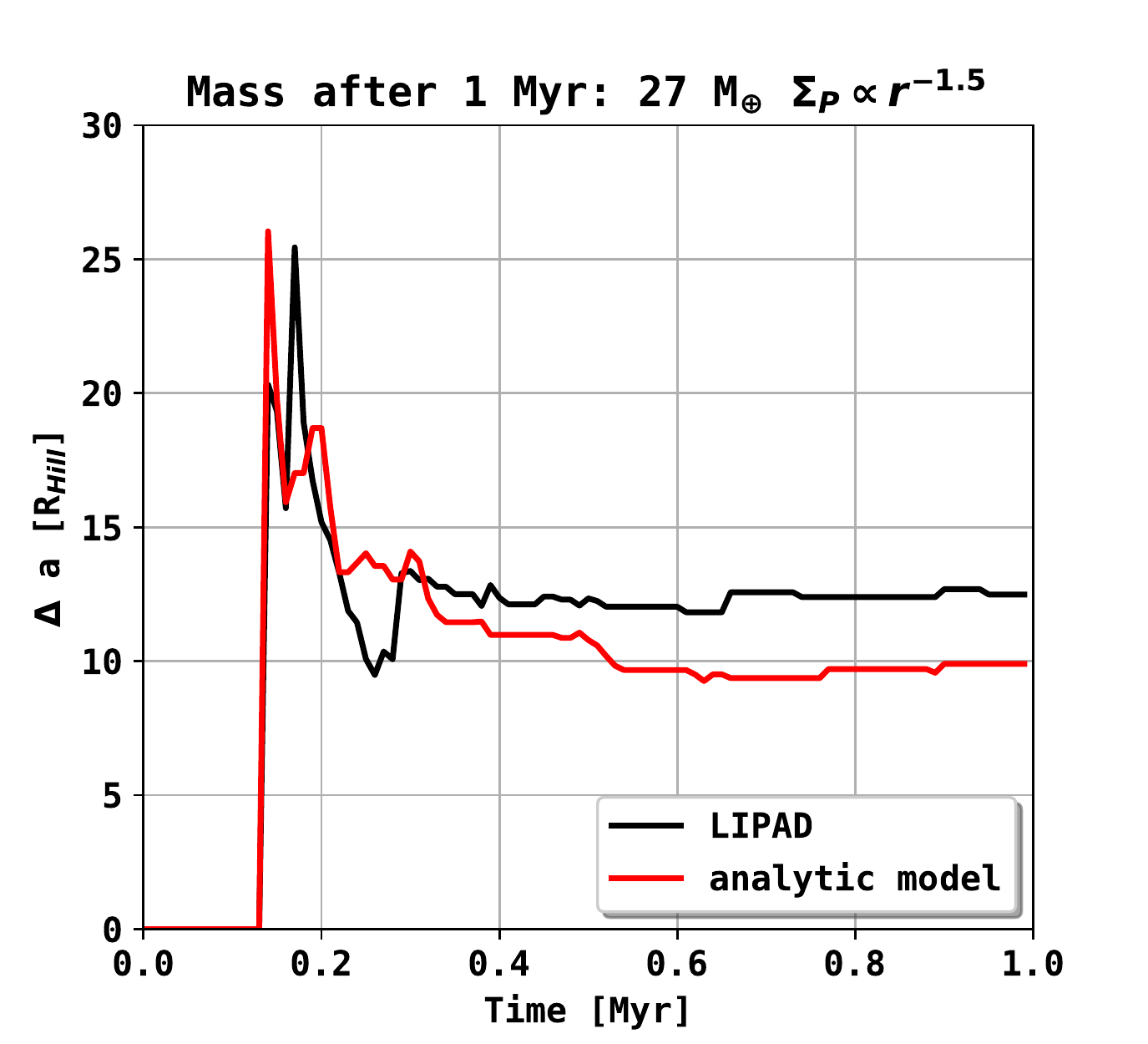}
\end{minipage}%
\begin{minipage}{.33\textwidth}
  \includegraphics[width=1.0\linewidth]{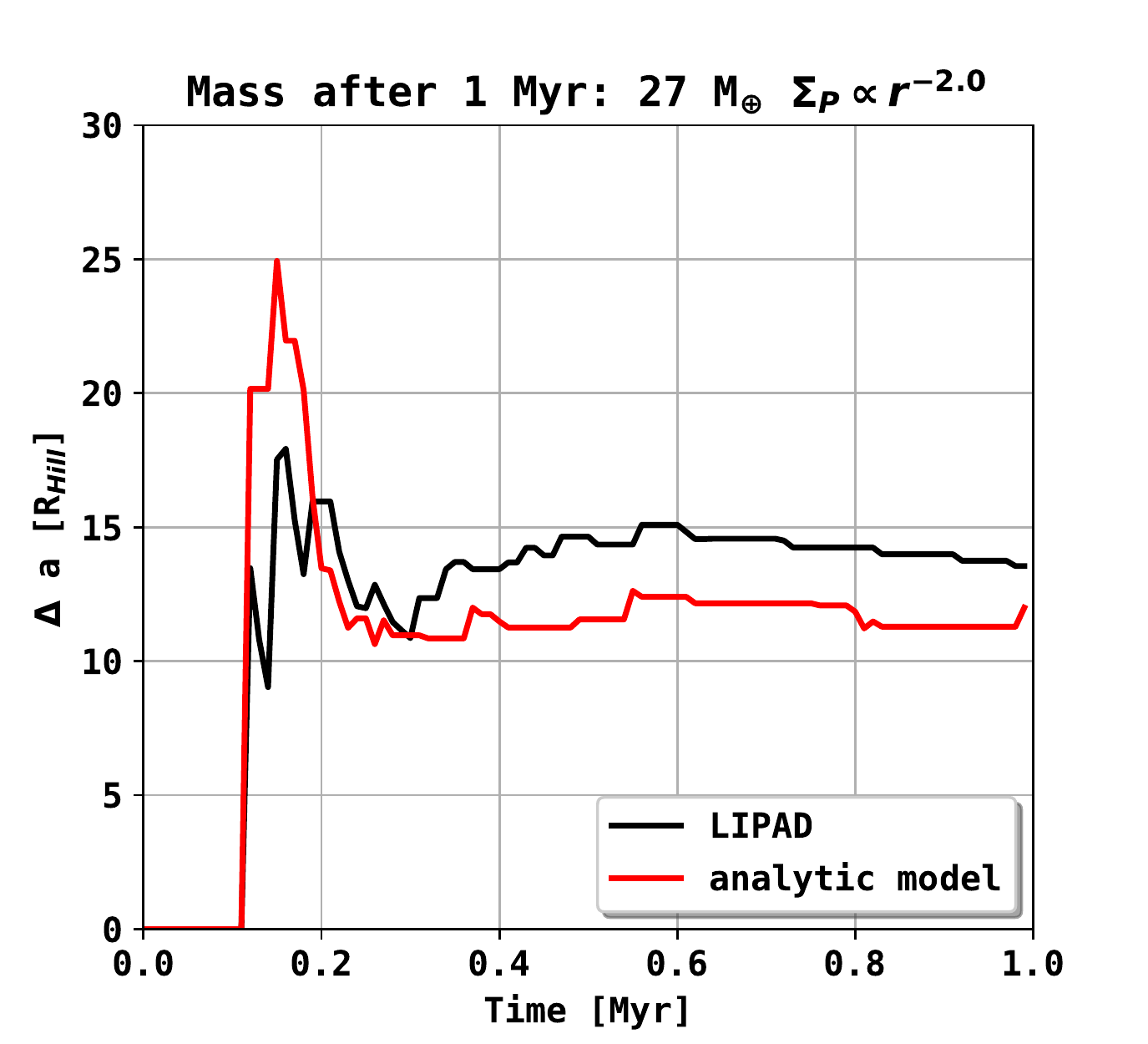}
\end{minipage}%
\caption{\small  Mean orbital separation of the initial embryos from the LIPAD runs and the analytical model embryos over time. The input parameters for the analytical model are given as 17 R$_{Hill}$ with a randomization of 2.5 R$_{Hill}$.
}
\label{Fig:Orbital_Seperation}
\end{figure*}

Fig. \ref{Fig:Orbital_Seperation} shows the time evolution of the average orbital separation of initial embryos for the LIPAD simulation and the analytical model, see Fig. \ref{Fig:Time_Semi}. The mean orbital separation of all systems converges to a value around 13-15 $R_{Hill}$ after 200-400$\,$ky. The orbital separation is a free parameter from Criterion II of the analytical model that we have chosen to fit the numerical results from our N-body simulations. In combination with Criterion I this allows us to predict the number, spatial distribution and formation time of planetary embryos for a specific planetesimal surface density evolution. The total number of embryos is given as the number of orbital separations (Criterion II) within the possible area of embryo formation (Criterion I). Their spatial distribution is determined by their orbital separation, which is a function of the mutual Hill radii. This way the absolute orbital separation between embryos increases linearly with increasing distance to the star, leading to a logarithmic cumulative number of initial embryos (see Fig. \ref{Fig:Cumulative_number}).
\\
Due to the low number of embryos for early times, the mutual distance can differ strongly between the analytical model and the LIPAD runs. This behavior however would also occur if one attempts to compare two LIPAD runs with similar initial conditions, due to the chaotic nature of the N-body evolution. We can show that for a larger number of embryos, the orbital separation in the analytical model shows the same behavior as in the N-body simulations.
\subsection{Active number of embryos and mass in embryos}
\label{Sec:active_number}
\begin{figure*}[]
\centering
\begin{minipage}{.33\textwidth}
  \centering
  \includegraphics[width=1.0\linewidth]{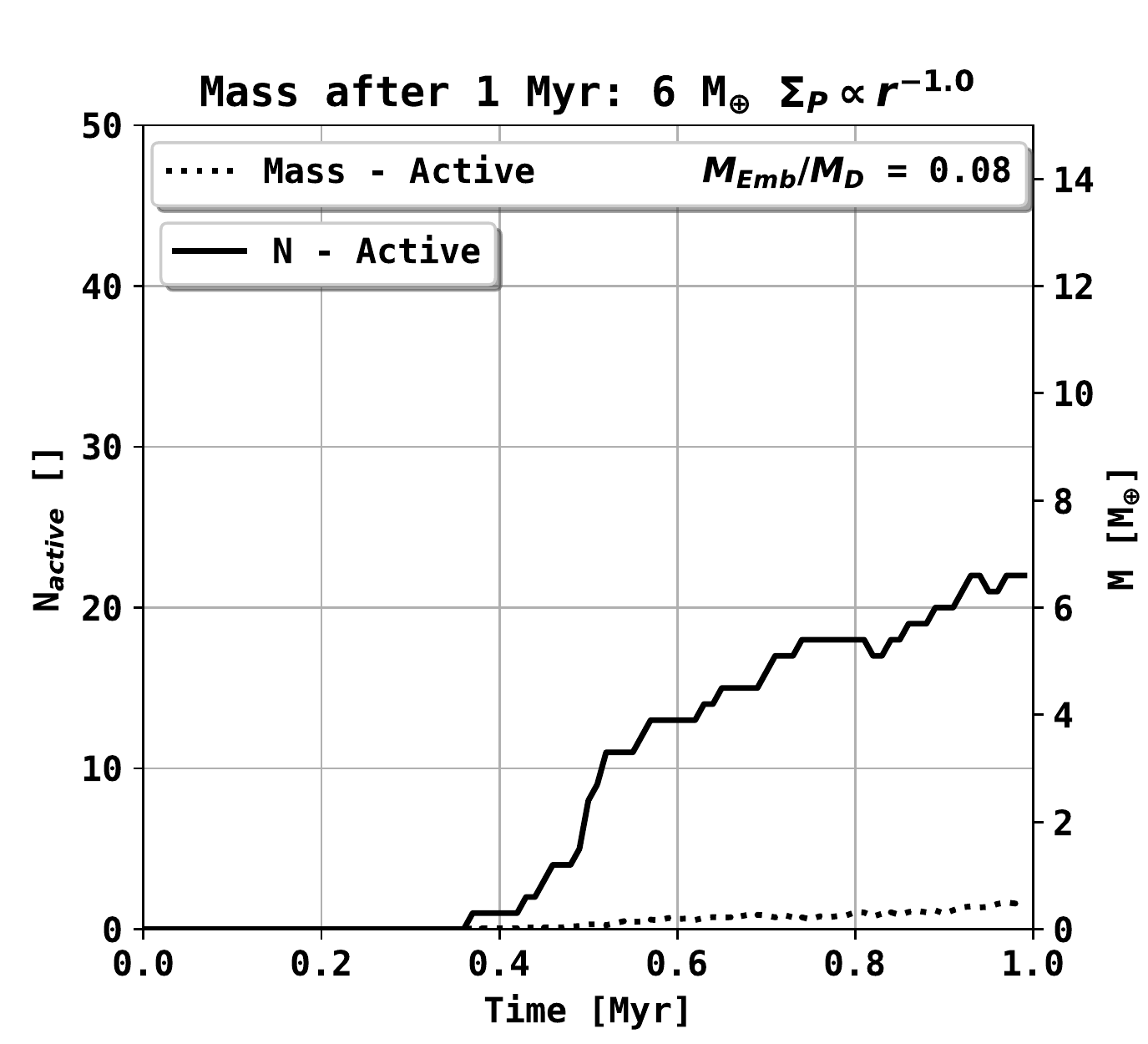}
\end{minipage}
\begin{minipage}{.33\textwidth}
  \includegraphics[width=1.0\linewidth]{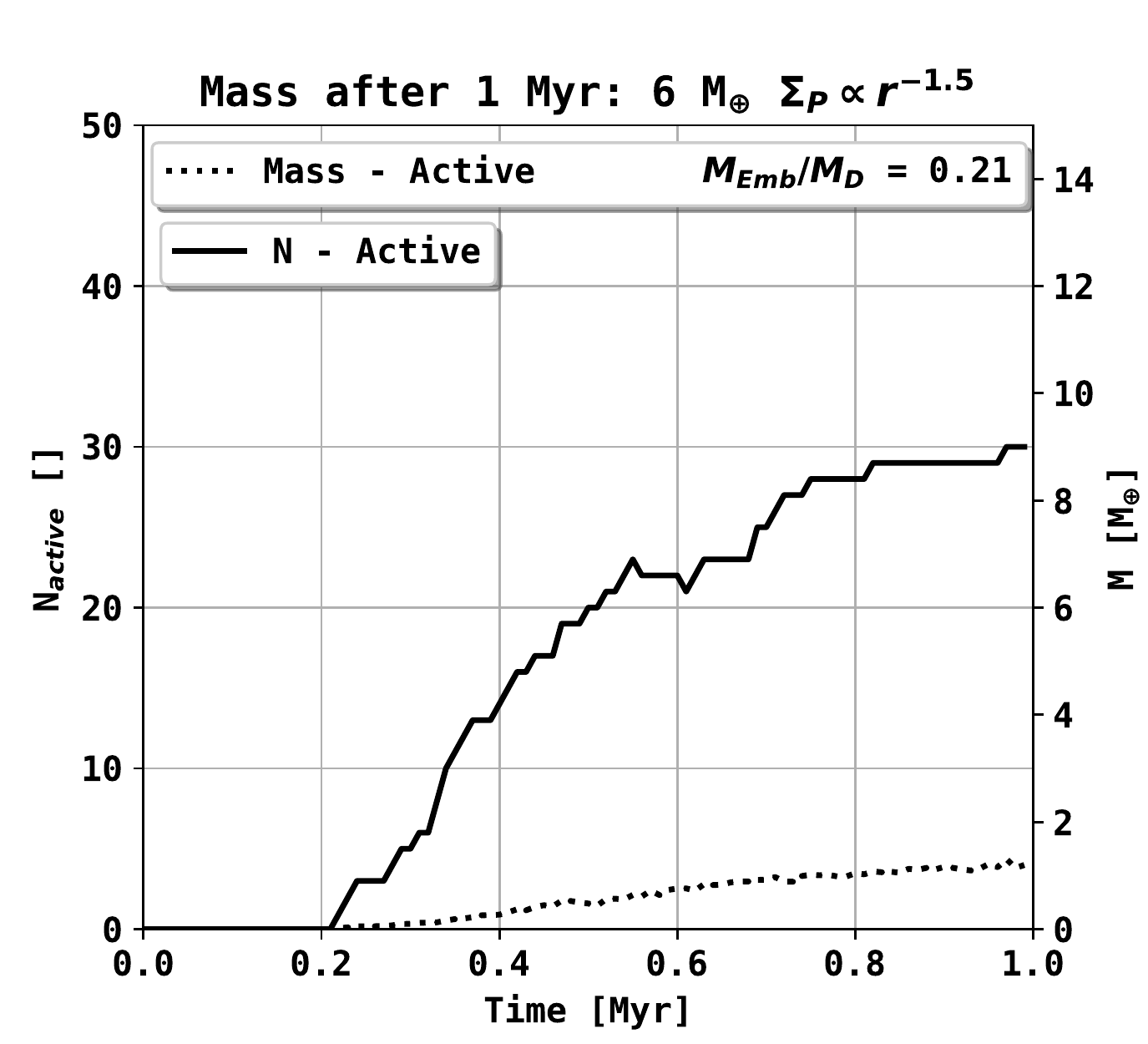}
\end{minipage}%
\begin{minipage}{.33\textwidth}
  \includegraphics[width=1.0\linewidth]{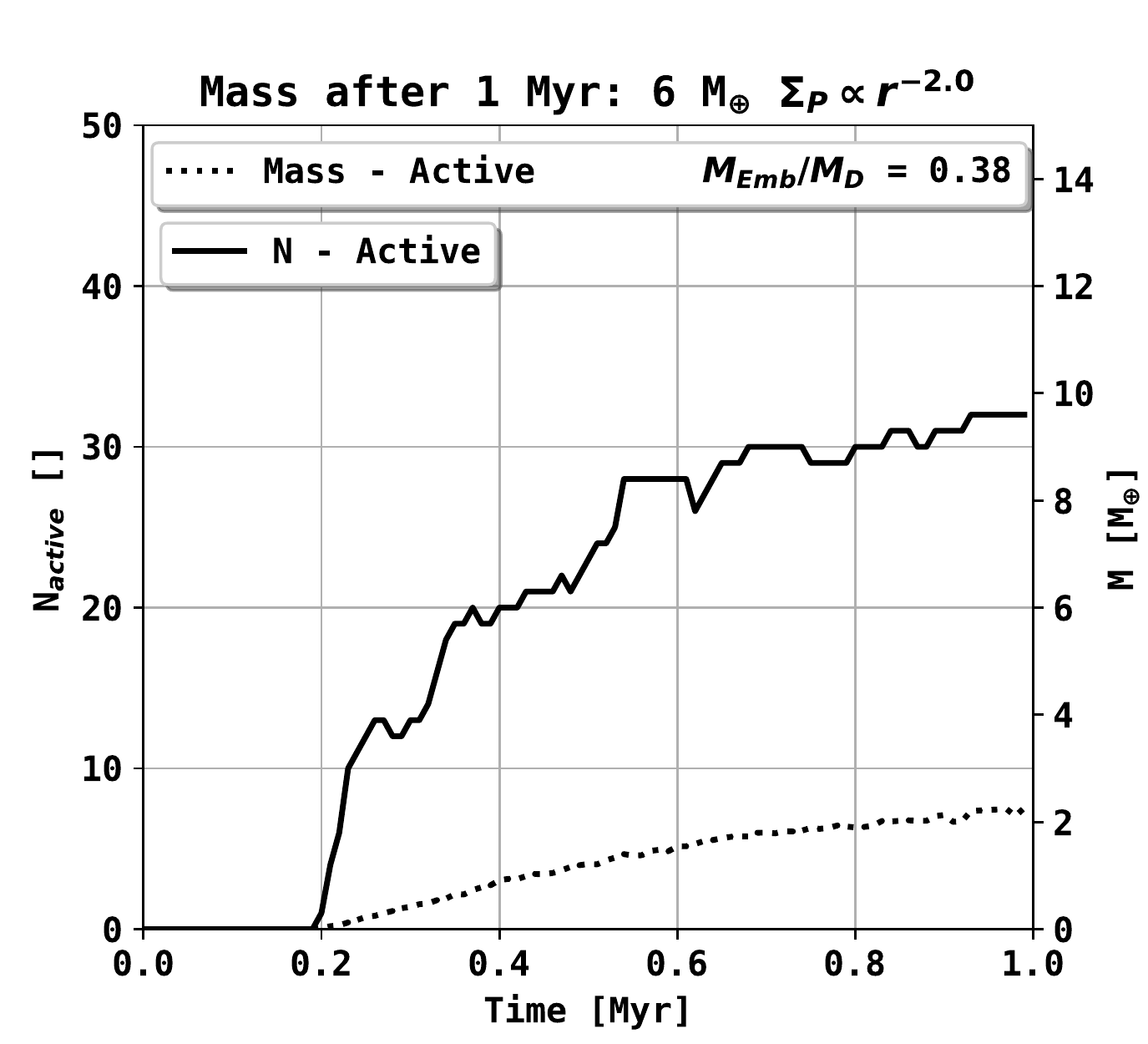}
\end{minipage}%
\\
\begin{minipage}{.33\textwidth}
  \centering
  \includegraphics[width=1.0\linewidth]{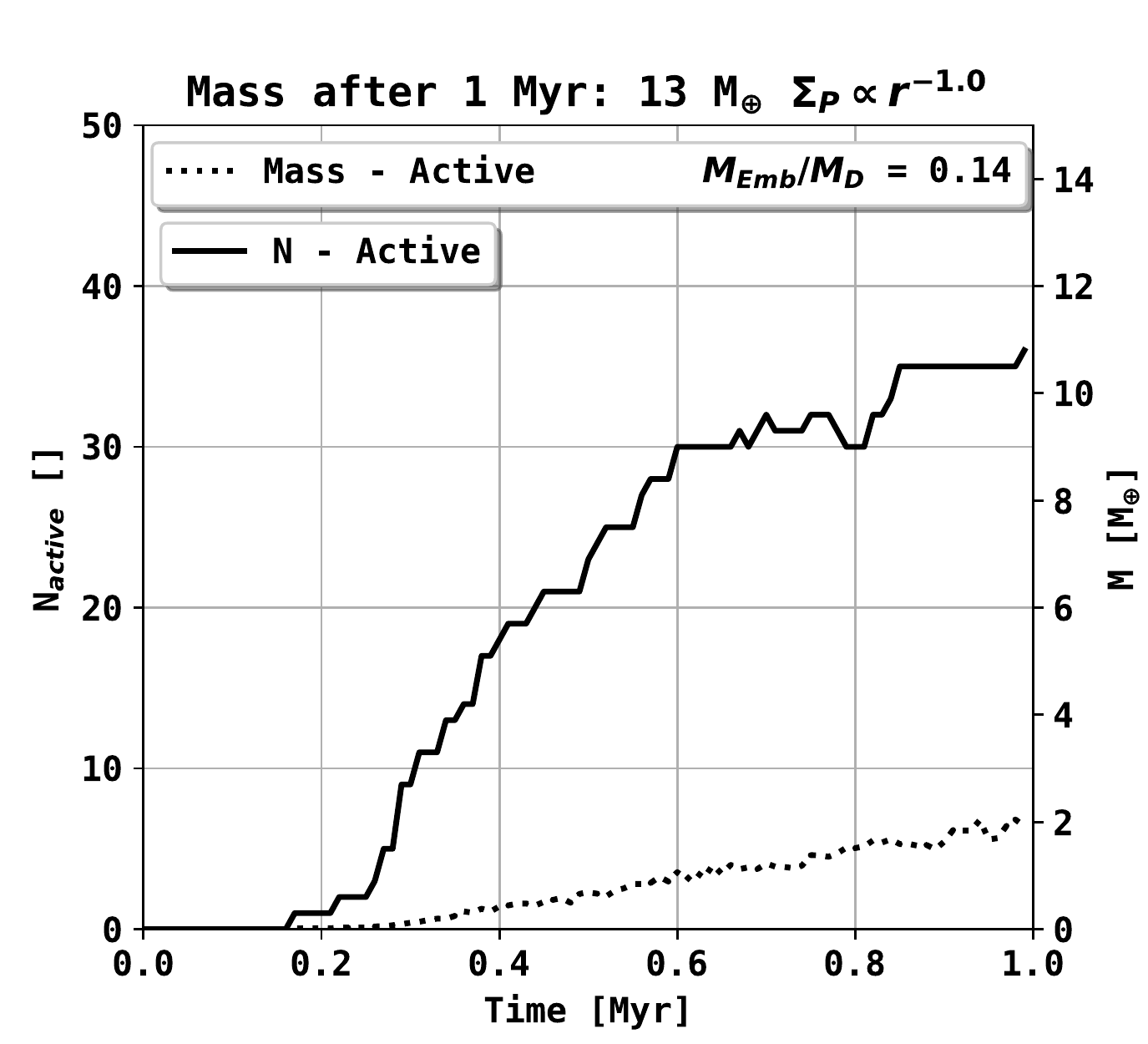}
\end{minipage}
\begin{minipage}{.33\textwidth}
  \includegraphics[width=1.0\linewidth]{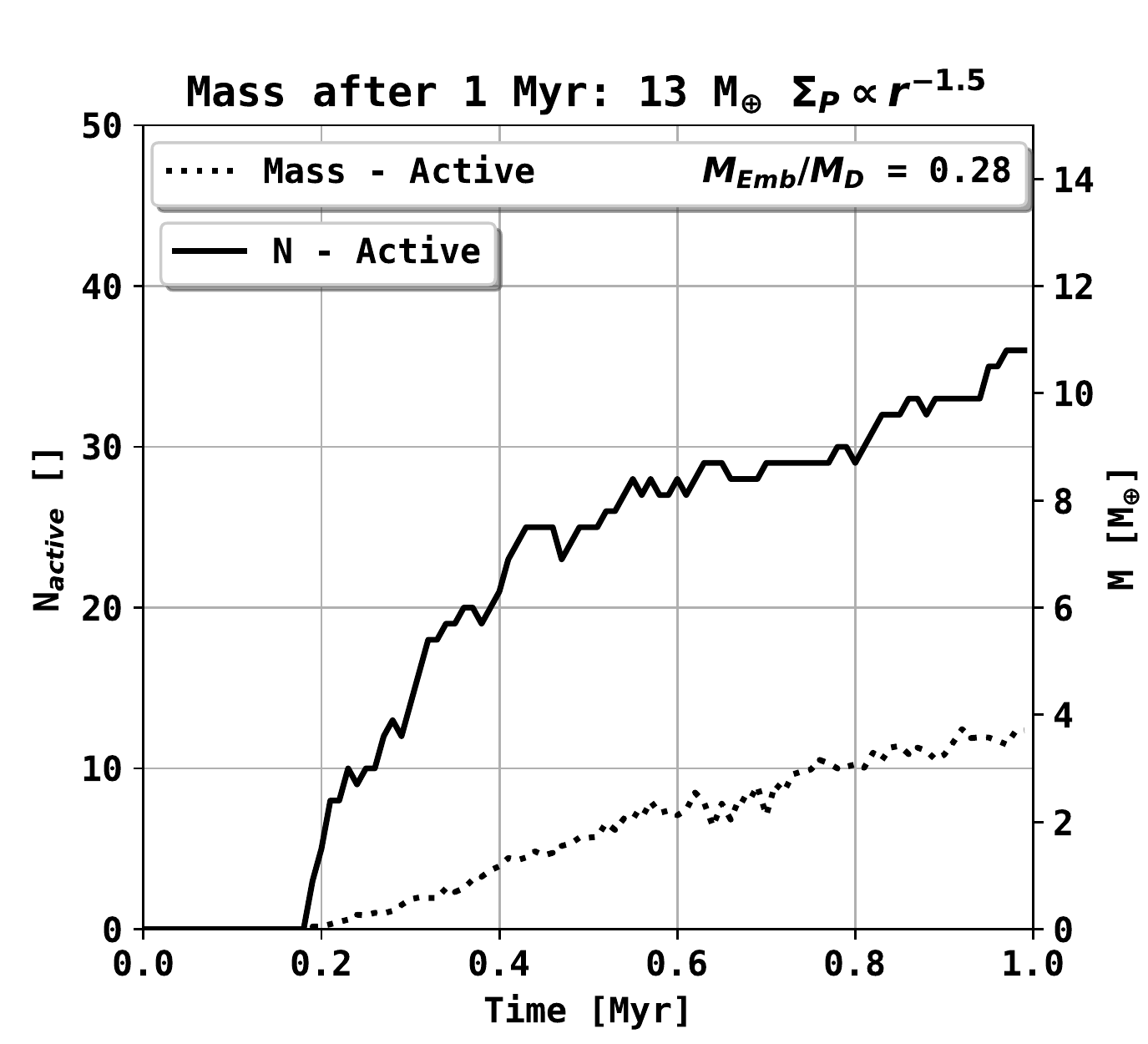}
\end{minipage}%
\begin{minipage}{.33\textwidth}
  \includegraphics[width=1.0\linewidth]{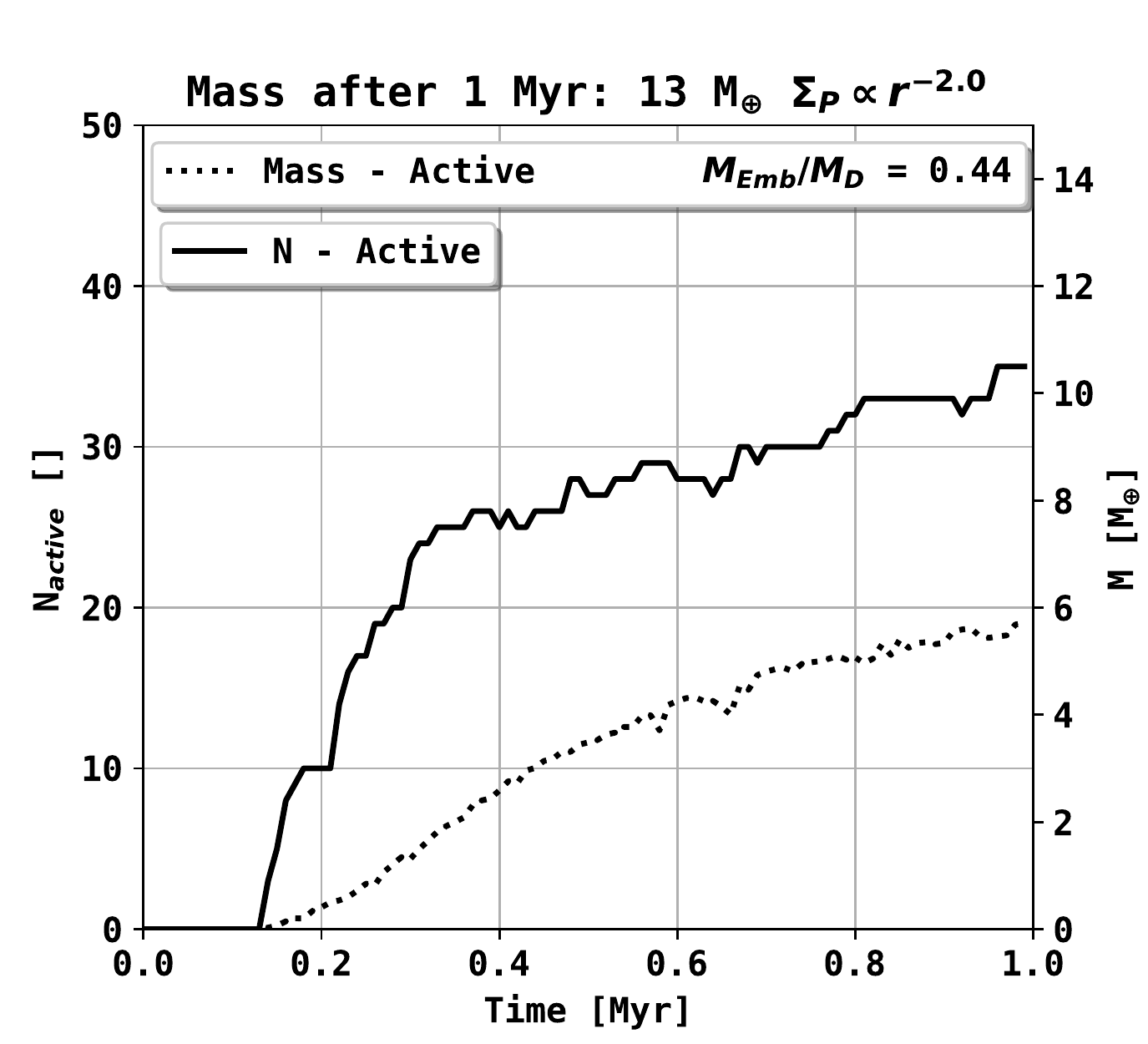}
\end{minipage}%
\\
\begin{minipage}{.33\textwidth}
  \centering
  \includegraphics[width=1.0\linewidth]{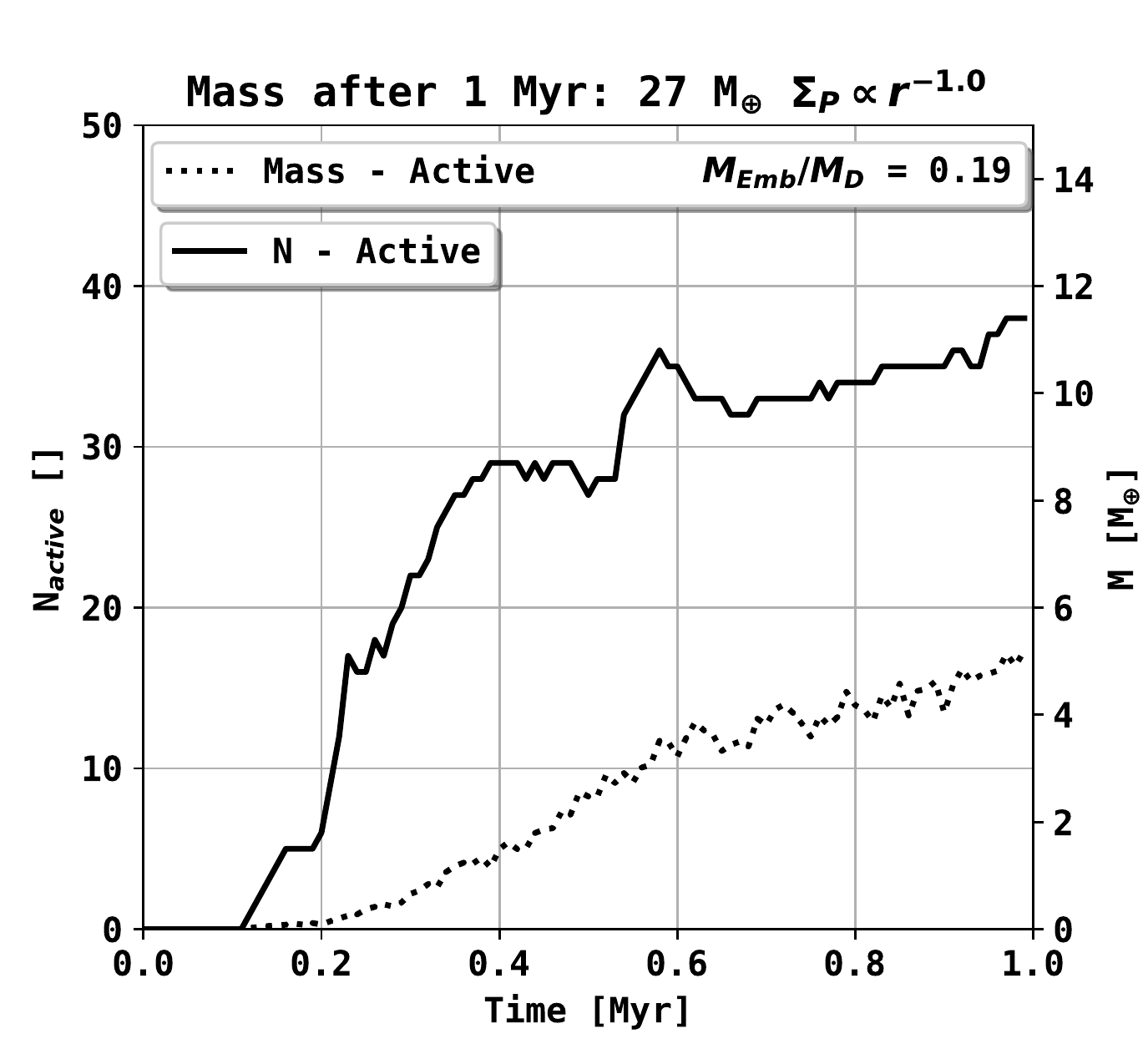}
\end{minipage}
\begin{minipage}{.33\textwidth}
  \includegraphics[width=1.0\linewidth]{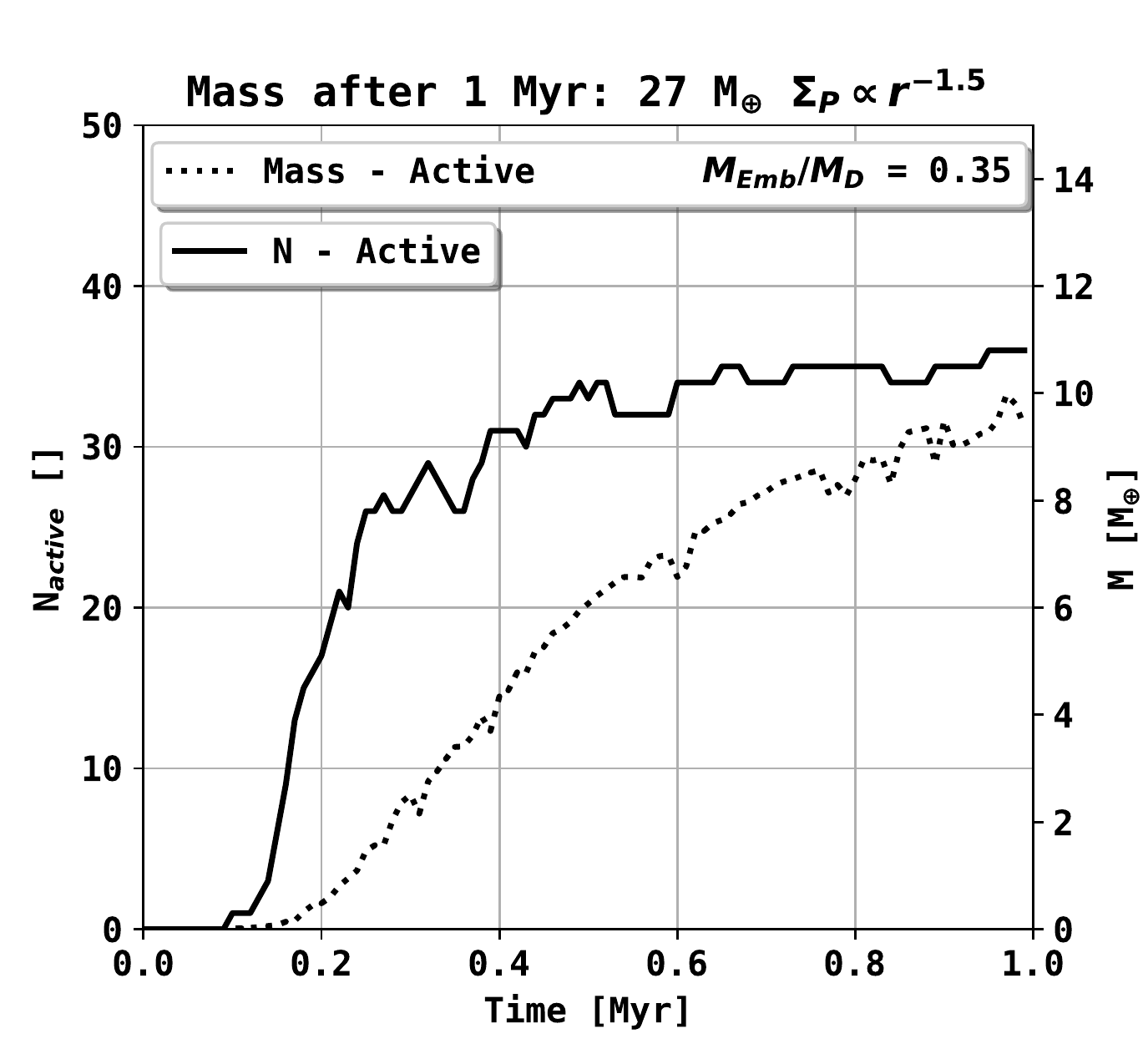}
\end{minipage}%
\begin{minipage}{.33\textwidth}
  \includegraphics[width=1.0\linewidth]{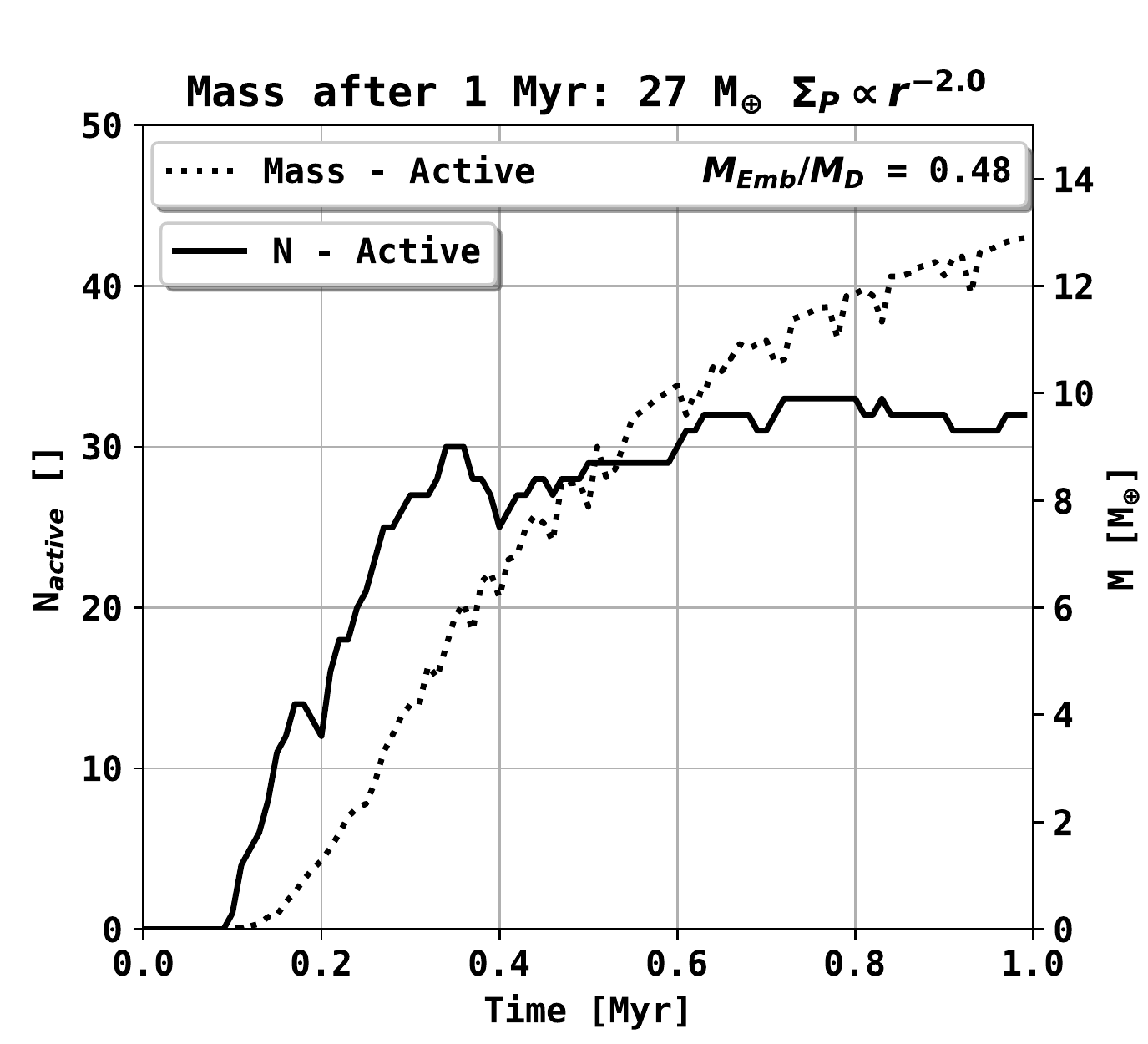}
\end{minipage}%
\caption{\small  Active number of planetary embryos and mass in planetary embryos over time for the LIPAD runs from Fig. \ref{Fig:Emb_form_LIPAD_6_ME} - Fig. \ref{Fig:Emb_form_LIPAD_27_ME}. We also show the fraction ($M_{Ebm}/M_{D}$) of mass in embryos over the final planetesimal mass that entered the disk after 1 Myrs. 
}
\label{Fig:Active_number}
\end{figure*}

Fig. \ref{Fig:Active_number} shows the number of active embryos over time, the total mass that is given in these embryos and the fraction of the total mass in embryos after 1 Myr ($M_{Emb}$) over the total mass in the system after 1 Myr ($M_{D}$). The number of active embryos after 1 Myr is between 30 and 40 embryos for 8 out of our 9 runs. Only the $6$\,$M_{\oplus}$ and $\Sigma_P \propto r^{-1.0}$ run contains less embryos ($N_{active}=$ 22) after 1 Myr. While the total number of active embryos seems insensitive to the total planetesimal mass or the planetesimal surface density profile, the same is not true for the total mass that is in planetary embryos after one million years.
\\
The total mass in embryos increases for steeper planetesimal surface density profiles and higher total masses after 1 Myr. The fraction of mass $M_{Emb}/M_{Disk}$ that is transformed into embryos increases for both higher masses and the slope of the planetesimal surface density. The number of embryos does not simply increase for more massive planetesimal disks in our runs. The reason being that the embryos that form grow larger in more massive disks. They thereby increase their orbital separation to the other embryos again. While larger planetesimal disk masses allow for a larger zone in which embryo formation is possible (Criterion I), the present embryos increase their orbital spacing due to their higher masses as well. 
\\
In the case of 27 $M_{\oplus}$ in planetesimals after 1 Myrs we can see that the number of embryos decreases slightly ($N_{active}=$ 38 for $\Sigma_P \propto r^{-1.0}$, $N_{active}=$ 36 for $\Sigma_P \propto r^{-1.5}$, $N_{active}=$ 32 for $\Sigma_P \propto r^{-2.0}$) for the steeper planetesimal surface density profiles but their mass increases drastically ($M_{Emb}\approx$5$\,$M$_{\oplus}$ for $\Sigma_P \propto r^{-1.0}$ , $M_{Emb}\approx$ 10$\,$M$_{\oplus}$ for $\Sigma_P \propto r^{-1.5}$ , $M_{Emb}\approx$ 13$\,$M$_{\oplus}$ for $\Sigma_P \propto r^{-2.0}$).
\section{Discussion}
\label{Sec:Discussion}
\subsection{Embryo formation - LIPAD}
\label{Subsec:embryo_formation_LIPAD}
Fig. \ref{Fig:Emb_form_LIPAD_6_ME} - Fig. \ref{Fig:Emb_form_LIPAD_27_ME} clearly show that embryo formation for every power law planetesimal surface density profile occurs from the inside out. This is an expected result due to the shorter growth time scales in the inner disk and the correspondingly higher densities in planetesimals. Even though the individual moment and location at which an embryo forms (black dots) appears to be stochastic, there is a pattern to be found in the embryo formation of the system. The red curve that marks Criterion I is well within the area of the initial embryos. The embryos individual locations, even though following the trend of the red line, appear chaotic. The exact location and time at which an object reaches the size of a planetary embryo appears stochastic due to the stochastic behavior of the N-body, but the analytic growth equations do well in constraining the zone of their individual formation. 
\\
Another effect that can be found is that embryos increase their orbital distance to other embryos when they grow in mass. This effect has already been found and discussed by \cite{kokubo1998oligarchic} and \cite{Kobayashi_2011} when studying the oligarchic growth of massive objects. In the general picture, initial embryos begin to form the earliest at closer distance to the star. 
\\
Furthermore the orbital separation of planetary embryos when expressed in terms of their Hill radii converges to a similar value in every setup studied, as can be seen in Fig \ref{Fig:Orbital_Seperation}. This directly results in a cumulative number of embryos that scales logarithmic with distance, as can be see in Fig. \ref{Fig:Cumulative_number}. As comparison we show the cumulative number of embryos that would form with the analytic model, in which the orbital separation is always expressed in terms of the Hill radius of the previously placed embryos. The cumulative number of embryos and the number of active embryos does not vary sensitively with the initial parameters (total mass and planetesimal surface density slope). The total mass that is converted into embryos however does depend strongly on the planetesimal surface density slope and the total mass in planetesimals, as it it shown in Fig. \ref{Fig:Active_number}. The number of active embryos even decreases slightly for higher disk masses and steeper planetesimal surface density profiles. As Fig. \ref{Fig:Emb_form_LIPAD_6_ME} - Fig. \ref{Fig:Emb_form_LIPAD_27_ME} show, the area in which planetary embryos form becomes larger for higher masses and steeper density profiles. Since their orbital separation increases for higher masses and since the mean orbital distance converges to the same number of Hill radii (Fig. \ref{Fig:Orbital_Seperation}), we see that the total number of embryos within 1 Myrs also does not sensitively abbreviate for different input parameters.
\label{Subsec:toy_model_emb_formation}
\subsection{Implications for pebble accretion}
\label{SubSec:Implications_pebble_acc}
While the effect of pebble accretion on the formation of planetarey embryos will be the main subject of our companion paper, we can already discuss some viable constraints here. It is notable to mention that the formation timescale of planetesimals is well within the formation timescales of the planetary embryos. This states that the formation of planetesimals continues to occur after planetary embryos have already formed from previously formed planetesimals. Since the growth rate of planetary embryos depends linearly on the local planetesimal surface density (Eq. \ref{eq:mass_growth}), the local formation of planetesimals has to be taken into account to model the growth timescales consistently. We conclude that since the formation of planetesimals requires a radial pebble flux, using our setup we can estimate first constraints on said pebble flux and subsequently on the possibility of continuous pebble accretion.
\\
Even though we do not take the accretion of pebbles onto planetesimals or planetary embryos into account in our simulations, we wish to highlight their importance in the general context of planetary growth, as already displayed by several studies like \cite{ormel2010effect}, \cite{bitsch2015growth} and \cite{Ndugu_2017} to mention just a few. The efficiency of pebble accretion directly depends on the local pebble flux at the location of an accreting body of sufficient mass. Since the formation of planetesimals $\Delta M_{disk}$ scales linearly with the local pebble flux, we can also derive from Fig. \ref{fig:formation_rate} that the pebble flux decreases drastically within the first $10^6$ years of the systems evolution. However, since we continue to form planetsimals well after the first embryos have formed, these embryos could grow by the remaining pebble flux that continues to form planetesimals as well. This indicates that the growth time scales of planetary embryos is a determining factor in defining the global efficiency of pebble accretion. There has to be a certain embryo size reached at first to effectively accrete pebbles. 
\\
Another crucial impact on the pebble flux evolution is the formation of planetesimals itself, since they form based on the disks evolution. The more planetesimals form, the earlier we also form planetary embryos, that could accrete pebbles. However, the more planetesimals one forms, the lower the pebble flux would become, due to the mass transfer into planetesimals. Even though the exact evolution of the pebble flux differs for every disk, the results of our study can already be used to apply first constraints on the magnitude of the pebble flux, based on the formation timescales of planetary embryos. Since $\sim 90 \%$ of our planetesimals form within 400ky of our setup, we conclude that the magnitude of the pebble flux has decreased significantly before
that time. Embryos that form after 400ky would therefore not be able to undergo significant pebble accretion in our model. 
\\
We find that in our setup most embryos that form within 400$\,$ky also form within 1$\,$au. The formation of further out embryos around 1.5-2.0$\,$au occurs well after 400$\,$ky. In conclusion, it is not possible for further out planetary embryos to undergo pebble accretion in our setup. This statement yields true if one assumes a power law distribution for the planetesimal surface density like the minimum mass solar nebula hypothesis, the dust profile of a viscous disk, or the pebble flux regulated planetesimal formation surface density profile.
\subsection{On the architecture of planetary systems}
\label{Subsec:architecture}
Following up on our findings from Sect. \ref{SubSec:Implications_pebble_acc} it is not too far-fetched to state that the architecture of planetary systems might very well be determined within the first few 100$\,$ky of their formation in terms of pebble accretion. Our study assumes power law density profiles for the planetesimal surface density and our results of inside out planetary embryo formation is a direct consequence of this. If one would assume abbreviations from the power law profile due to local substructures in the the disk, like e.g. around the iceline \citep{Drazkowska2017}, this picture might change. 
\\
The early formation of planetary embryos around the iceline could lead to the formation of cold giant planets via pebble accretion. The formation of those planets can then have major consequences to the subsequent evolution of the inner system. Assuming that outer planetary embryos form early enough to undergo significant pebble accretion, they could alter the evolution of the inner system drastically, as they would reduce the pebble flux that reaches the terrestrial planet region. Also the additional planetesimal formation itself will have strong consequences for the interior pebble flux. An early decrease in the pebble flux would also lead to a decrease in the formation of planetesimals in the terrestrial planet region. This would again effect the formation of planetary embryos and planetary growth. It becomes clear that the formation of planetesimals, the formation of planetary embryos and the evolution of the pebble flux are tightly connected within the first few 100$\,$ky of a circumstellar disk.
\\
Another scenario that might change the evolution of the system would be the stochastic formation of a planetesimal with an initial size much larger than 100$\,$km \citep{Johansen2007}. The formation of a significantly larger planetesimal in a reservoir of 100$\,$km planetesimals and pebbles could reduce the timescales of planetary embryo formation significantly. This could lead to the presence of planetary embryos at much larger distances within the lifetime of the pebble flux.
\subsection{Embryo formation - Analytic model}
The one dimensional analytic parameterized approach agrees well with the sophisticated N-body simulations in terms of the formation timescales of a lunar mass object and the total number of objects that reach this given size. In 2 out of our 9 runs, the deviation of the total number of embryos is below 5$\,\%$, in 4 out of 9 it is below 10$\,\%$ and in 8 out of 9 it is below 25$\,\%$. Only the 6$M_{\oplus}$, $\Sigma_P \propto r^{-1.0}$ run deviates stronger ($\approx 40\%$). The Hill criterion for the orbital separation of planetary embryos completely determines the number of planetary embryos without additional assumptions. Considering the time and location at which an object reaches the mass of a planetary embryo, we show that the analytic prescription does well in handling the analytic planetesimal surface density evolution (Sect. \ref{subsec:toy_model_comparison}). 
\\
It is worth mentioning that the N-body simulations require weeks (sometimes months) of computation time with the same planetesimal input, whereas the parameterized model takes merely seconds. While the N-body simulations clearly involve more complexity that allow for a more complete picture of the problem, the question on where, when and how many initial planetary embryos form is well reproduced with the analytic model. This makes the analytic approach well suited for other studies that aim for statistical properties in which computational time is a limiting factor, like e.g. planet population synthesis. 
\\
Even though our study focused on an area from 0.5$\,$au to 5$\,$au, the analytical model should also yield true at further locations and could be a valuable asset in considering planetary embryo formation in far out ring-like structures of circumstelar disks, as seen in ALMA observations. Other studies regarding planet formation via pebble accretion may use our findings to modify their initial conditions in terms of the available pebble flux, as explained in greater detail in Sect. \ref{SubSec:Implications_pebble_acc}.
\section{Summary $\&$ Outlook}
\label{Sec:Summary}
We study the spatial distribution and formation timescales of planetary embryos from an initial disk of gas and dust. For this purpose, we couple a one dimensional model for viscous disk evolution and planetesimal formation to the LIPAD code that studies the dynamical N-body evolution of the evolving planetesimal system. The size of an initial planetesimal is given as 100$\,$km in diameter and dynamically grows due to collisions with other planetesimals. We analyze the first million years of nine different systems in which we vary the total mass in planetesimals and their surface density profile. 
\\
In combination with analytic estimates on growth rates of planetesimals based on their local surface density, we derive an analytical model for planetary embryo formation. Our model does well in reproducing the spatial distribution and formation time of planetary embryos. We use their orbital separation as a free parameter that can be fit to match the N-body simulations. The model can be used in further studies (e.g. global models of planet formation, population synthesis etc. ) that use a planetesimal surface density description to consistently model the spatial distribution and formation time of planetary embryos. 
The main findings of planetary embryo formation based on pebble flux regulated planetesimal formation are:
\begin{itemize}
    \item Embryos form first in the innermost regions of planetesimal formation due to shorter growth time scales close to the star and higher planetesimal surface densities. 
    \\
    \item The innermost embryos (<1$\,$au) form well within the presence of an active pebble flux for most planetesimal disks, whereas the outer embryos(>2au) fail to do so in any disk studied.
    \\
    \item Higher planetesimal disk masses, or steeper planetesimal surface density profiles do not result in a higher number of active embryos, but in more massive embryos within a larger area.
\end{itemize}
We link the formation timescale of planetesimal formation and the evolution of the radial pebble flux to the formation timescale of lunar mass objects that formed by planetesimal collisions. In doing so we find crucial constraints for the possibility of pebble accretion as a planet formation process. These constraints need to be considered in studies that involve pebble accretion on planetary embryos, since as we show,  the presence of a planetary embryo and the presence of a pebble flux strongly depend on the radial distance of the embryo to the star. 
\\
It is shown that a power law planetesimal surface density profile cannot build planetary embryos at larger distances within the timescale of a radial pebble flux. This consequence arises from the interplay of pebble flux regulated planetesimal formation and the timescales involved to form planetary embryos from 100$\,$km sized bodies. The more planetesimals one forms, the earlier one forms planetary embryos, but the more planetesimals one forms, the less mass remains in pebbles. Vice versa, if one decreases the formation of planetesimals to maintain a higher pebble flux, the growth time scales for planetary embryos increase as a result of lower planetesimal surface densities.
\\
Future studies will include disk consistent pebble accretion in the N-body simulation to study the effect of an active pebble flux on the formation of planetary embryos. Another study that will follow our presented approach will study the formation of embryos in far out planetesimal rings that could result from pressure bumps during the disks evolution. 
\section*{Acknowledgements}
\bibliography{main.bib}
\end{document}